\documentclass[twocolumn]{aastex62}
\usepackage{mathrsfs}
\usepackage{color}
\usepackage{comment}
\usepackage{url}
\usepackage{graphicx}   % Including figure files
\usepackage{amsmath}    % Advanced maths commands
\usepackage{amssymb}    % Extra maths symbols
\usepackage{bm}         % Bold maths symbols, including upright Greek
\usepackage{array}
\usepackage{booktabs}
\usepackage{longtable}
\usepackage{supertabular}
\usepackage{threeparttable}
\usepackage{booktabs}
\usepackage{algorithm}
\usepackage{algorithmic}

\submitjournal{AAS (AJ)}
\begin{document}
%%%
\title{Determination of size, albedo and thermal inertia of 10 Vesta family asteroids with WISE/NEOWISE observations}
%%%%
%%
\correspondingauthor{Jianghui Ji}
\email{jijh@pmo.ac.cn}
%% Author information %%%
\author{Haoxuan Jiang}

\affil{CAS Key Laboratory of Planetary Sciences, Purple Mountain Observatory, Chinese Academy of Sciences, Nanjing 210008, China\\}
\affil{University of Science and Technology of China, Hefei 230026, China\\}
\author{Jianghui Ji}
\affil{CAS Key Laboratory of Planetary Sciences, Purple Mountain Observatory, Chinese Academy of Sciences, Nanjing 210008, China\\}
\affil{CAS Center for Excellence in Comparative Planetology, Hefei 230026, China\\}
\author{Liangliang Yu}
\affil{State Key Laboratory of Lunar and Planetary Science, Macau University of Science and Technology, Macau, China\\}

%% Mark off the abstract in the ``abstract'' environment.
\begin{abstract}
\noindent
In this work, we investigate the size, thermal inertia, surface roughness and geometric albedo of 10 Vesta family asteroids by using the Advanced Thermophysical Model (ATPM), based on the thermal infrared data acquired by mainly NASA's Wide-field Infrared Survey Explorer (WISE).  Here we show that the average thermal inertia and geometric albedo of the investigated Vesta family members are 42 $\rm J m^{-2} s^{-1/2} K^{-1}$ and 0.314, respectively, where the derived effective diameters are less than 10 km.  Moreover, the family members have a relatively low roughness fraction on their surfaces. The similarity in thermal inertia and geometric albedo among the V-type Vesta family member may reveal their close connection in the origin and evolution. As the fragments of the cratering event of Vesta, the family members may have undergone similar evolution process, thereby leading to very close thermal properties.
Finally, we estimate their regolith grain sizes with different volume filling factors.
\end{abstract}


\section{Introduction}
An asteroid family is usually supposed to be formed from the fragmentation of a parent body in the main-belt. The family members may share similar composition and physical characteristics with their parent body. To identify and make a study of the asteroid families, one of the classical methods is to evaluate the proper elements and characterize the distribution of asteroids in proper element space by applying a clustering algorithm (e.g.,the Hierarchical Clustering Method (HCM)) \citep{zapp1990}. In addition, \citet{nesv2015} provided an asteroid family catalogue that contains 122 families calculated from synthetic proper elements. The Vesta family, as one of the largest asteroid population, consists of over 15,000 members and locates in the inner region of the main-belt with the proper orbital elements: $2.26\leq a_p \leq 2.48$ AU, $0.075\leq e_p \leq 0.122$, and $5.6^\circ \leq i_p \leq 7.9^\circ$, where $a_p$, $e_p$ and $i_p$ are the proper elements of semi-major axis, eccentricity and inclination, respectively \citep{zapp1995}. Taxonomically, basaltic asteroids are classified as V-type asteroids that have a photometric, visible-wavelength spectral and other observational relationships with (4) Vesta \citep{harder2014}. For example, their optical spectrums are similar with that of (4) Vesta that displays a strong absorption band attributed to pyroxene centered near $9000~\rm \mathring{A}$ \citep{binzel1993}. In the Vesta family, most of the members are believed to be V-type asteroids. But some V-type asteroids have been discovered outside the Vesta family recently, which may indicate the presence of multiple basaltic asteroids in the early solar system \citep{lic2017}. The Vesta family was inferred to be originated from (4) Vesta through a catastrophic impact event approximately 1 Gyr ago \citep{marz1996}.  \citet{carr2005} further investigated the dynamical evolution of the V-type asteroid outside the Vesta family and showed the possibility that the members of the Vesta family migrated via Yarkovsky effect and nonlinear secular resonance. \citet{hasegawa2014} investigated the rotational rates of 59 V-type asteroids in the inner main belt region and showed that the rotation rate distribution is non-Maxwellian, this may be caused by the long-term Yarkovsky-O'Keefe-Radzievskii-Paddack (YORP) effect which can change the direction of asteroids' spin axis and rotation periods \citep{delbo2015}. Additionally, by numerically integrating the orbits of 6600 Vesta fragments over a timescale of 2 Gyr, \citet{nesv2008} demonstrated that a large number of family members can evolve out of the family borders defined by clustering algorithms and constrained the age of this family to be older than 1 Gyr. Also, \citet{bottke2005} derived the cratering event may be occurred in the last 3.5 Gyr. According to Dawn spacecraft's observation, the two largest impact craters on Vesta were estimated to be formed about 1 Gyr ago \citep{Marchi2012}. Such early formation event of the family could provide sufficient time to yield the rotational distribution obtained by \citet{hasegawa2014} under the influence of YORP effect. Moreover, visible and infrared spectroscopic investigations imply that (4) Vesta may be the parent body of near-Earth V-type asteroids \citep{mig1997} and Howardite-Eucrite-Diogenite meteorites (HEDs) \citep{cruikshank1991,mig1997,burb2001}. The HEDs are believed to come from the melted basaltic magma ocean crystallization on the large asteroids \citep{sanctis2012,mandler2013}. \citet{ful2012} performed the irradiation experiments in laboratory on HED meteorites to simulate space weathering on Vesta and Vesta family asteroids by using different ions. Their experimental results indicate that space weathering effect can give rise to the spectral differences between (4)Vesta and other V-type bodies.

Vesta is known as one of the most frequently observed bodies \citep{reddy2013,hasegawa2014b}. \citet{tho1997} explored the pole orientation, size and shape of Vesta by using images from Hubble Space Telescope (HST). The Hubble's observations unveil an amazing impact crater with a diameter 460 km, being supportive of the collision site \citep{tho1997}. In 2011, the spacecraft Dawn arrived at Vesta and further discovered that the giant basin observed by HST is, as a fact, composed of two overlapping huge impact craters, Rheasilvia (500 km) and Veneneia (400 km), respectively, which their excavation was found to be sufficient to supply the materials of the Vesta family asteroids and HEDs \citep{schenk2012}.  The Rheasilva crater appears to be younger and overlies the Veneneia.  Moreover, further study by Dawn mission revealed a wide variety of albedo on the surface of Vesta \citep{reddy2012} and a deduced core with a diameter $107\sim112$ km, indicating sufficient internal melting to segregate iron \citep{rus2012}.

As aforementioned, V-type asteroids and HEDs do provide key clues to formation and evolution scenario for the main-belt asteroids as well as essential information of the early stage of our Solar system, from a viewpoint of their similar orbits and the spectral properties. Therefore, the primary objective of this work is to investigate the thermophysical characteristics of thermal inertia, roughness fraction, geometric albedo and effective diameter etc., to have a better understanding of such kind of Vesta family members.  This can help us establish the relation between Vesta family asteroids and other main-belt asteroids (MBAs) from a new perspective. In fact, thermal inertia plays an important role in determining the resistance of temperature variation over the asteroid surface, which is associated with surface temperature and materials. As a result of the major fragments of the parent body or the impactor, although they may have similar features in orbital evolution or spectral feature, each member of the Vesta family can have a distinguished appearance in size, surface topography and roughness due to surface evolution over secular timescale in space, thereby causing diverse thermal inertia on the asteroid's surface. Moreover, the geometric albedo does hold significant information of the asteroidal composition. Therefore, by comparing thermal inertia and geometric albedo of the family members with those of the parent body, we can have a deep understanding of origin of the Vesta family.

In this work, we extensively investigate the thermal properties for 10 Vesta family asteroids whose polyhedron shape models and thermal infrared observational data can be acquired, by using the Advanced Thermophysical Model (ATPM) \citep{roz2011,yu2017,jiang2019}. Moreover, we derive the thermophysical parameters for the Vesta fmaily asteroids on the basis of ATPM and the mid-infrared data and further explore the correlation of various thermal parameters, to provide implications of the impact history on (4) Vesta. Furthermore, we explore the homology of these Vesta family asteroids by comparing their thermal parameters with those of (4) Vesta. The 10 family members, are (63) Ausonia, (556) Phyllis, (1906) Neaf, (2511) Patterson, (3281) Maupertuis, (5111) Jacllif, (7001) Neother\footnote{According to the orbital database at AstDyS node (http://hamilton.dm.unipi.it/astdys/), we take (7001) Neother as a Vesta family asteroid, because the proper orbital elements of (7001) Neother are within the range of those of the Vesta family \citep{zapp1995}. The other 9 Vesta family asteroids here are simply adopted from the Vesta family list \citep{nesv2015}.}, (9158) Plate, (12088) Macalintal, and (15032) Alexlevin. Shape models for them can be obtained from the Database of Asteroid Models from Inversion Techniques (DAMIT)\footnote{https://astro.troja.mff.cuni.cz/projects/asteroids3D/web.php}, whereas thermal infrared observations can be acquired from  WISE/NEOWISE, IRAS and AKARI. Table \ref{tableshape} summarizes the detailed parameters of the targets under study.

The structure of this paper is as follows. Section \ref{thermalmodel} gives a brief description on modelling of thermal process as well as the convex shape models, mid-infrared observations and ATPM. The radiometric results for each Vesta family asteroids and their analysis are presented in Section \ref{resultanalysis}. Section \ref{discu} summarizes the discussions on the relationship of thermal inertia, effective diameter and geometric albedo and the evaluation of regolith grain size. Section \ref{conclu} gives the conclusions.

%%% Section 2.1
\section{THERMAL MODELING}\label{thermalmodel}
\subsection{Shape Model}
\begin{figure*}
        \includegraphics[width=18cm, height=5.4cm]{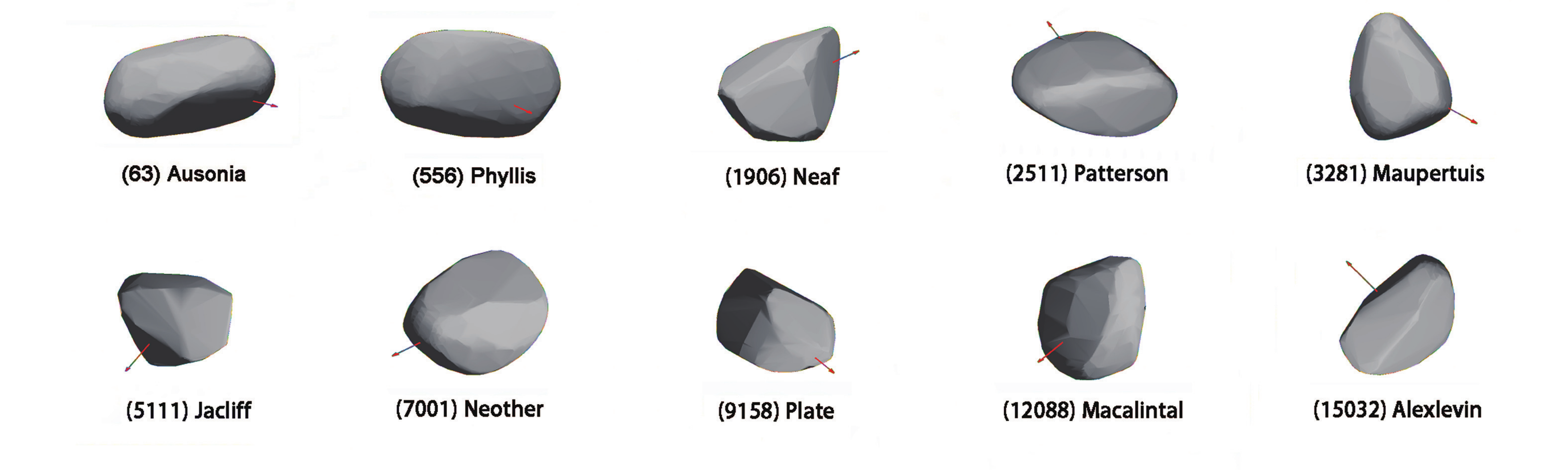}
    \caption{3D convex shape model of the Vesta family asteroids (from DAMIT) in this work, where red arrows illustrate the direction of spin orientation for each asteroid.}
    \label{shapefigure}
\end{figure*}

As mentioned above, here we adopt 3D convex shape models for 10 Vesta family asteroids \citep{kaa2001} from DAMIT.  In ATPM, the asteroids are considered to be composed of N triangular facets, hence we employ a fractional coverage of hemispherical craters to describe the surface roughness of asteroids, where each crater is assumed to be composed of M smaller triangular sub-facets \citep{roz2011}. Table \ref{tableshape} lists the parameters of the asteroids' shape model, which includes the number of facets, number of vertices and pole orientations. As can be seen in Fig. \ref{shapefigure}, the shape models for 10 Vesta family asteroids are plotted, where the red arrow represents the spin axis of each asteroid.

\makeatletter\def\@captype{table}\makeatother
\begin{table*}
        \centering
        \caption{Orbital and physical characteristics of the investigated asteroids and the parameters in this work (JD = 2458600.5, from Minor Planet Center).}
        \label{tableshape}
        %\begin{threeparttable}
        \begin{tabular}{lccccccccccc} % four columns, alignment for each
                \hline
                Asteroid & $a$(AU)&$e$ &$i$ ($^\circ$)&$P_{\rm orb}$(yr)&$Abs_{\rm mag}$& $N_{\rm facets}$&$N_{\rm vertices}$&Orientation$(^ \circ)$&$P_{\rm rot}$(h)&Spectral type\\
                \hline
                (63)  Ausonia     &2.395 &0.1269 &5.7756  &3.71 &7.55 &3192 & 1598 & (120, -15)&9.282 &Sa,Sw\\
                (556) Phyllis     &2.464 &0.1037 &5.2461  &3.87 &9.56 &3192 &1598 &(209, 41)&4.923 &S\\
                (1906) Neaf       &2.375 &0.1345 &6.4707  &3.66 &12.7 &2028 &1016 & (72, -70) &11.010 &V\\
                (2511) Patterson  &2.298 &0.1037 &8.0479  &3.48 &12.7 &1144 &574 & (194, 51) &4.141 &V\\
                (3281) Maupertuis &2.350 &0.0975 &5.9904  &3.60 &12.7 &1092 &548 &(231, -74) &6.730 &-\\
                (5111) Jacliff    &2.355 &0.1264 &5.8043  &3.61 &12.7 &1144 &574 &(259, -45) &2.839 &R,V\\
                (7001) Neother    &2.378 &0.1508 &7.0250  &3.67 &13.3 &2040 &1022 &(13, -66)&9.581 &-\\
            (9158) Plate      &2.300 &0.1509 &7.6954  &3.49 &13.6 &1140 &572 &(119,-52)&5.165 &$\rm SQ_{\rm p}$\\
            (12088) Macalintal&2.355 &0.0739 &6.2439  &3.61 &14.0 &1144 &574 &(265, 50) &3.342 &V\\
            (15032) Alexlevin &2.373 &0.1179 &5.5082  &3.66 &14.5 &1144 &574 &(353,-46)&4.405 &V\\
                \hline
        \end{tabular}
         \begin{tablenotes}
    \footnotesize
    \item[]Note: $a$,$e$,$i$ represents the semi-major axis, eccentricity and inclination, respectively. $P_{\rm orb}$ is the orbital period and $Abs_{\rm mag}$ is the absolute magnitude. $N_{\rm facets}$ and $N_{\rm vertices}$ describes the number of shape facets and vertices in the shape models. $P_{\rm rot}$ is the rotation period.
    \end{tablenotes}
        %\end{threeparttable}
\end{table*}

\subsection{Observations}
In this work, thermal infrared data are obtained from three space-based telescopes: WISE/NEOWISE, Infrared Astronomical Satellite (IRAS), and AKARI. For example, WISE surveyed the sky in 4 wavebands (3.4, 4.6, 12.0 and 22.0 $\rm \mu m$ noted as W1, W2, W3 and W4, respectively) until the solid hydrogen cryostat (which was utilized to cool down W3 and W4 bands) was depleted on September 30, 2010. Thereafter the satellite continued to work at W1 and W2 bands, known as NEOWISE. In this situation, we can download the data from two source tables of WISE archive (http://irsa.ipac.caltech.edu/applications/wise/), WISE All-Sky Single Exposure (L1b) and NEOWISE-R Single Exposure (L1b). Here we should emphasize that the surface temperature of MBAs is relatively lower than that of NEAs, as a result the data in shorter wavelengths (e.g., W1) can include a large percentage of reflected sunlight. As will be discussed in the following section, W1 band contains roughly 90\% of reflected sunlight in the observations, indicating that the thermal portion is merely comparable to the uncertainty of the entire observed flux. For this reason, we do not adopt W1 band data of the Vesta family asteroids in our fitting. In the target searching, we employ  a  Moving Object Search with a search cone radius of 1". Similar to that of \citet{masiero2011} and \citet{grav2012}, all artifact identification $\rm CC_{FLAG}$ other than 0 and $p$ is rejected, where 0 indicates no evidence of known artifacts found whereas $p$ means that an artifact may be present. Additionally, the modified Julian date needs to be within 4s of the epochs given in MPC.

Subsequently, we follow the method described by \citet{wright2010} to convert the magnitudes into fluxes and the color correction factors of 1.3448, 1.0006 and 0.9833 for W2 $\sim$ W4 bands. Since the observed flux is proportional to the cross-section area of the asteroid in the direction of the observer, therefore the thermal light curve of an asteroid should not have an amplitude that may exceed a certain value \citep{jiang2019}. According to this point, we further screen the data set for each asteroid and the thermal light curves will be discussed later in this work. Table. \ref{numobs} reports the number of observations for W2 $\sim$ W4 wavelength of WISE, the range of phase angle, heliocentric distance and distance from asteroid in reference to observer. For detailed WISE/NEOWISE observations for each asteroid are summarized in the Appendix. The observational uncertainties here are set to be 10\% for all Vesta family asteroids. Here the observations from AKARI and IRAS are only applied to the fitting for (63) Ausonia and (556) Phyllis, which are not given in the table.

\subsection{Advanced Thermophysical Model with Reflected Sunlight}
Here ATPM accepts global shape models in the triangular facet formalism and adopts a hemispherical crater to represent the roughness surface. In order to constrain the thermal properties such as thermal inertia, roughness fraction and geometric albedo, we need to compute the temperature distribution over the asteroid's surface. For each shape facet, the temperature $T$ can be determined by solving the 1D heat conduction function
\begin{equation}
    \frac{\partial T}{\partial t}=\frac{\kappa}{\rho C}\frac{\partial^2 T}{\partial x^2},
        \label{heatchonduct}
\end{equation}
with specific boundary condition in which shadowing effect, multiple-sunlight scattering and re-absorption of thermal radiation are taken into consideration \citep{roz2011}. Here, $\kappa$, $\rho$ and $C$ represents the thermal conductivity, surface density, and heat capacity. The method to simplify and solve the heat-conduction equation are described in \citet{spe1989} and \citet{lag1996}. Once the temperature distribution is ascertained, we can use the Planck function to evaluate theoretical thermal emission of each facet, thus the total theoretical thermal emission of an asteroid can be written as
\begin{equation}
\begin{split}
 F_{\rm thermal} = (1-f_{\rm r})\sum_{i}^{N}\pi\epsilon B(\lambda, T_i)A_iv_i \\ +f_{\rm r}\sum_{i}^{N}\sum_{j}^{M}\pi\epsilon B(\lambda, T_{ij})A_{ij}v_{ij},
\end{split}
\end{equation}
where $ f_{\rm r}$ is the roughness fraction, $\epsilon$ is the monochromatic emissivity at wavelength $\lambda$ which is assumed to be 0.9. For facet $i$ and sub-facet $ij$, $A$ and $v$ denote the area and the view factor, respectively. $B$ is the Planck function, described by
\begin{equation}
B(\lambda, T) = \frac{2hc^2}{\lambda^5}\frac{1}{e^{hc/ \lambda k_{\rm B}T}-1},
 \label{vf}
\end{equation}
and the view factor $v$ is defined as
\begin{equation}
\begin{split}
 v_i = s_i\frac{\vec{n}_{i}\cdot\vec{n}_{\rm obs}}{\pi d_{\rm ao}^{2}},
 \label{vf}
\end{split}
\end{equation}
where $s_{\rm i}$ indicates whether facet $i$ can be seen by the observer, $\vec{n}_{\rm i}$ and $\vec{n}_{\rm obs}$ represents the facet normal and the vector pointing to the observer respectively, and $d_{\rm ao}$ is the distance between asteroid and observer.

In addition, as pointed out by \citet{myhrovold2018}, thermal infrared observations of shorter wavelengths (e.g., W1 and W2) are dominated by reflected sunlight. It is necessary to remove the effects of the reflected part when we use these observations in shorter wavelengths. Hence, we further deal with the reflected sunlight contained in the observed flux by using the method described in \citet{jiang2019}. We treat each facet $i$ and sub-facet $ij$ as Lambertian surface, and the reflect part can expressed as
\begin{equation}
%\begin{split}
F_{\rm ref\_i}(\lambda) = B(\lambda,5778)\frac{R_{\rm sun}^2}{r_{\rm helio}^2}\cdot A_{\rm B}\cdot \psi_i \cdot A_i\cdot v_i,
%\end{split}
\end{equation}
\begin{equation}
%\begin{split}
F_{\rm ref\_ij}(\lambda) = B(\lambda,5778)\frac{R_{\rm sun}^2}{r_{\rm helio}^2}\cdot A_{\rm B}\cdot\psi_{ij}\cdot A_{ij} \cdot v_{ij},
%\end{split}
\end{equation}
where $B(\lambda, T)$ is the Planck equation in the temperature of the Sun, $R_{\rm sun}$ the radius of the Sun, $r_{\rm helio}$  the heliocentric distance of the asteroid, $A_{\rm B}$  the bond albedo and $\psi$ the sine value of the solar incidence angle. $S$ and $v$ are the area and view factor, respectively. The entire reflected sunlight portion will be given by
\begin{equation}
\begin{split}
\label{refl}
F_{\rm ref} =(1-f_{\rm r})\sum_{i}^{N}F_{\rm ref\_i}+f_{\rm r}\sum_{i}^{N} \sum_{j}^{M}F_{\rm ref\_ij}.
\end{split}
\end{equation}
According to our sunlight reflection model, we assess the reflected sunlight contained in W1 $\sim$ W4 observations, which it covers  $\sim 90\%$ at W1,  $30 \%\sim 50 \%$ at W2 and can be negligible at W3 and W4. Therefore, as described above, W1 band is not adopted in this work. The reflected sunlight contributes a significant part in W2 observation, but the proportion is no more than 50\%. As will be discussed in the following section, we account for an overall contribution of thermal emission and reflected sunlight to fit the observations. In addition, WISE only surveyed the sky for roughly 9 months in 2010, being suggestive of the observations of Main-Belt asteroids in W3 and W4 simply covering a very narrow range of solar phase angle. In comparison, the utilization of W2 data can provide diverse observational solar phases, wavelengths, as well as observational numbers, which make the fitting process more reliable. Thus, we utilize the observations from W2 band but we also take reflected sunlight into consideration in our fitting.

\subsection{Thermal-infrared fitting}
Heat conduction into and outwards the asteroids' sub-surface material can lead to a certain thermal memory, named as thermal inertia. In particular,  thermal inertia is defined as $\Gamma$=$\sqrt{\kappa \rho c}$, where $\kappa$, $\rho$, $c$ have the same meaning as in Equation~(\ref{heatchonduct}). As a matter of fact, the thermal inertia plays a very important role of governing the heat conduction process and inducing the non-zero night-side temperature of an asteroid. Moreover, this thermal parameter can result in the surface temperature to peak at the afternoon side of an asteroid, thereby causing the diurnal Yarkovsky effect \citep{delbo2015}. In order to derive a best-fitting thermal emission with the observed fluxes, we set the initial thermal inertia in the range $\rm \Gamma$=$0\sim 300$ $\rm J m^{-2} s^{-1/2} K^{-1}$ at equally spaced steps of 10 $\rm J m^{-2} s^{-1/2} K^{-1}$ in search for  best-fitting value. Other parameters, such as pole orientation, absolute magnitude, rotation period are listed in Table \ref{tableshape}. A bolometric and spectral emissivity of 0.9 is assumed for all wavelengths in the fitting procedure. On the other hand, as shown in Table. \ref{numobs}, for each Vesta family asteroids, the solar phase angle only covers a very small range, therefore it brings about the difficulty in placing constraints on thermal inertia and roughness fraction at the same time. In general, the roughness fraction of main-belt asteroids could be usually small, thus we assume a priori roughness for these Vesta family asteroids to be $0.1 \sim 0.5$.  Hence, for each wavelength, we obtain three free parameters, i.e.,  thermal inertia $\Gamma$, geometric albedo $p_{\rm v}$ and the rotation phase $\phi$. In fact, the effective diameter  $D_{\rm eff}$ is in connection with the geometric albedo via
\begin{equation}
\begin{split}
D_{\rm eff} = 1329 \times \frac{10^{-0.2H_{\rm v}}}{\sqrt{p_{\rm v}}},
\label{deffpv}
\end{split}
\end{equation}
\\ \\
\par \noindent
where $H_{\rm v}$ is the absolute magnitude. For each asteroid, the entire theoretical flux $F_{\rm m}$ can be written as
\begin{equation}
\begin{split}
F_{\rm m} = F_{\rm thermal} + F_{\rm ref},
\end{split}
\end{equation}
where $F_{\rm thermal}$ is the total theoretical thermal emission, $F_{\rm ref}$ represents the reflected sunlight. Then we compare $F_{\rm m}$ with the observations, and we adopt the minimum $\chi^2$ fitting defined by \citet{press2007}
\begin{equation}
\label{chi2r}
\begin{split}
\chi_{\rm r}^2 = \frac{1}{n-3}\sum_{i=1}^{n}\left[\frac{F_{\rm m}(p_{\rm \nu}, \Gamma, f_{\rm r}, \lambda, \phi)-F_{\rm obs}(\lambda)}{\sigma_{\rm \lambda}} \right]^2,
\end{split}
\end{equation}
where $n$ is the observation number, and $\sigma_{\lambda}$ is the observation uncertainty. In the following, we will detailedly report our results for 10 Vesta family asteroids.

\begin{table*}
        %\small
        \centering
        \caption{WISE/NEOWISE observation numbers at W2-W4 wavebands and observation geometry}
        \label{numobs}
%       \renewcommand\tabcolsep{2.5pt}
        %\begin{threeparttable}
        \begin{tabular}{lcccccc} % four columns, alignment for each
                \hline
                Asteroid & $N_{\rm W2}$ &$N_{\rm W3}$&$N_{\rm W4}$&$\alpha(^\circ)$&$r_{\rm helio}$ (AU)& $r_{\rm obs}$ (AU)\\
                \hline
                (63) Ausonia &47 &17 &17 & 20.88 - 28.76 & $ 2.113 \sim 2.694$ & $1.637 \sim 2.495$\\
                (556) Phyllis &75 &30 &29 & 21.74 - 25.78 &$2.215 \sim 2.718$ &$1.764 \sim 2.539$ \\
                (1906) Neaf &30 &16 &16 &25.01 - 29.30 &$2.077 \sim 2.337$ &$1.618 \sim 2.048$ \\
                (2511) Patterson&0 &11 &13 &22.79 - 28.37 &$2.093 \sim 2.534$ & $1.712 \sim 2.244$ \\
                (3281) Maupertuis &18 &13 &14 & 24.61 - 29.65 &$2.132 \sim 2.194$ & $1.803 \sim 1.856$ \\
                (5111) Jacliff &28 &11 &8 &25.04 - 25.07 &$2.055 \sim 2.359$ & $1.703 \sim 2.057$ \\
                (7001) Neother &18 &11 &11&25.33 - 29.32 &$2.058 \sim 2.348$ & $1.579 \sim 2.045$\\
            (9158) Plate &16 &16 &22 & 25.69 - 25.73 &$2.258 \sim 2.265$ & $1.934 \sim 1.986$ \\
            (12088) Macalintal &0 &10 &5 &24.41 - 24.42 &$2.459 \sim 2.460$ & $2.246 \sim 2.262$ \\
            (15032) Alexlevin  &16 &12 &14 &27.30 - 27.35 &$2.140 \sim 2.142$ & $1.793 \sim 1.180$\\
                \hline
        \end{tabular}
                 \begin{tablenotes}
    \footnotesize
    \item[]Note: $N$ is the number of observations for each wavelength, $\alpha$ denotes the range of solar phase for each asteroid, $r_{\rm helio}$ and $r_{\rm obs}$ represent the heliocentric distance and the distance between the asteroid and the observer, respectively.
    \end{tablenotes}

\end{table*}

\section{RESULTS}\label{resultanalysis}
\subsection{(63) Ausonia}
\begin{figure}
        \includegraphics[width=9cm,height=4.8cm]{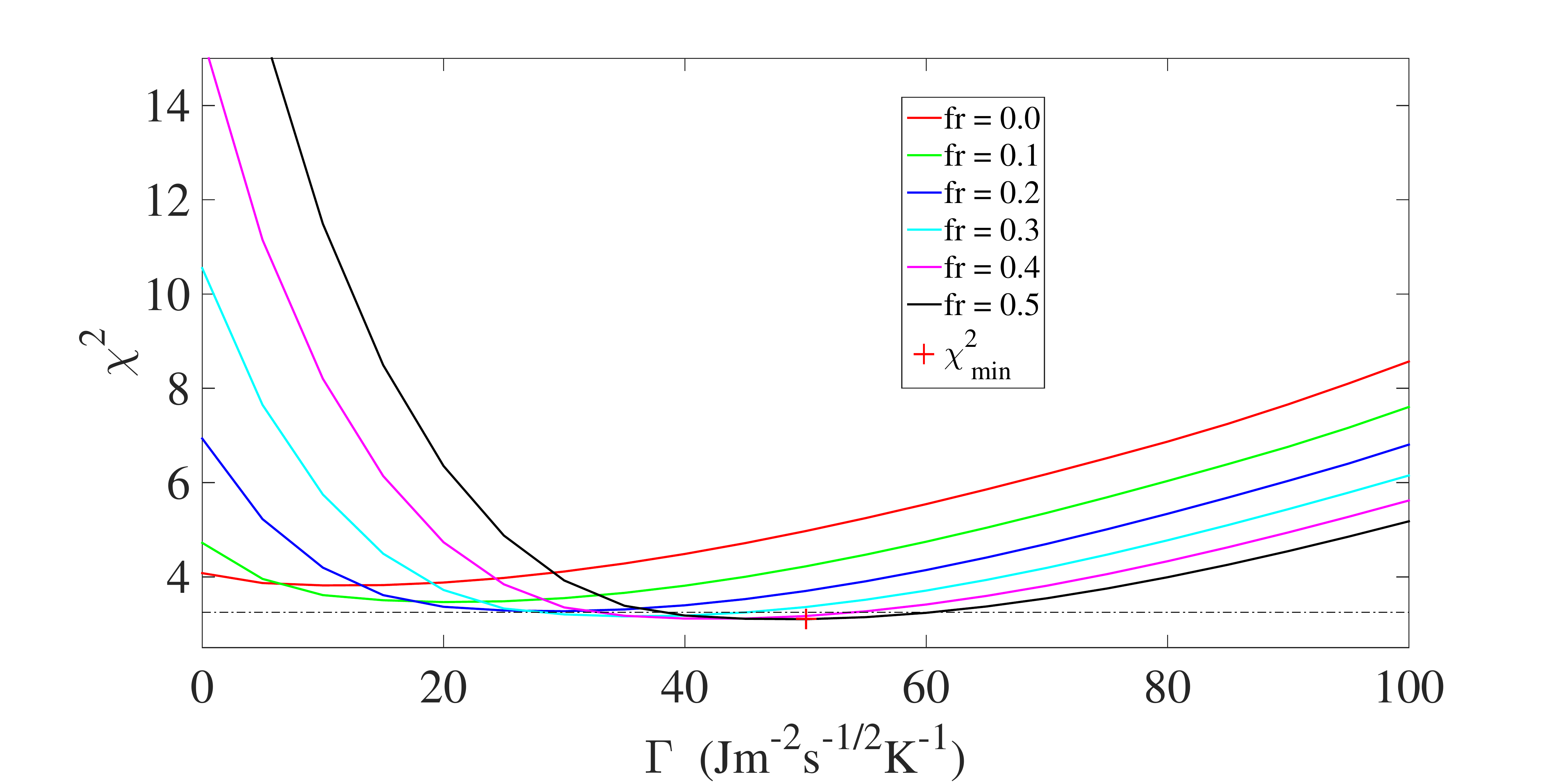}
    \caption{$\Gamma -\chi^2$ profile fit to observation of (63) Ausonia. The solid lines with each color indicate the roughness ranging from $0.0 \sim 0.5$ during the fitting process.}
    \label{chi2ga63}
\end{figure}

Asteroid (63) Ausonia is the largest Vesta family member with a diameter of roughly 100 $\rm km$. In the Tholen classification, (63) Ausonia is a stony S-type asteroid \citep{tholen1984}, and in the SMASSII classification, this asteroid is classified as Sa type \citep{bus_binzel2002}, while in the Bus-Demeo taxonomic it is an Sw subtype \citep{Demeo2009}.  \citet{tanga2003} estimated the overall shape, spin orientation, angular size of (63) Ausonia using the observations from Fine Guidance Sensors (FGS) of Hubble Space Telescope (HST). They derived an effective diameter of 87 km for this asteroid, which was smaller than the IRAS diameter (103 km) \citep{tedesco2004} and that of \citet{masiero2012} (116 km). In this work, we adopt 97 observations from IRAS ($3 \times 12  \rm \mu m$, $3 \times 25 \rm \mu m$, $3 \times 60 \rm \mu m$, and $1 \times 100 \rm \mu m$), AKARI \citep{usui2011_akari,alilagoa2018_akari} ($4 \times 9 \rm \mu m$, $2 \times 18 \rm \mu m$) and WISE/NEOWISE ($47 \times 4.6 \rm \mu m$, $17 \times 12.0 \rm \mu m$ and $17 \times 22.0 \rm \mu m$) to explore the thermal parameters for (63) Ausonia. Fig.  \ref{chi2ga63} illustrates the $\Gamma -\chi^2$ profile of (63) Ausonia, where the minimum value $\chi^2$  is related to the thermal inertia  $50_{-24}^{+12}$ $\rm J m^{-2} s^{-1/2} K^{-1}$ and the roughness fraction $0.5_{-0.3}^{+0.0}$.  The horizontal line represents the $3-\sigma$ range of $\Gamma$. Furthermore, we derive the effective diameter  $94.595_{-2.483}^{+2.343} ~ \rm km$, and the geometric albedo is then evaluated to be $0.189^{-0.009}_{+0.010}$ for this asteroid.

\begin{figure}
        \includegraphics[width=9cm,height=20cm]{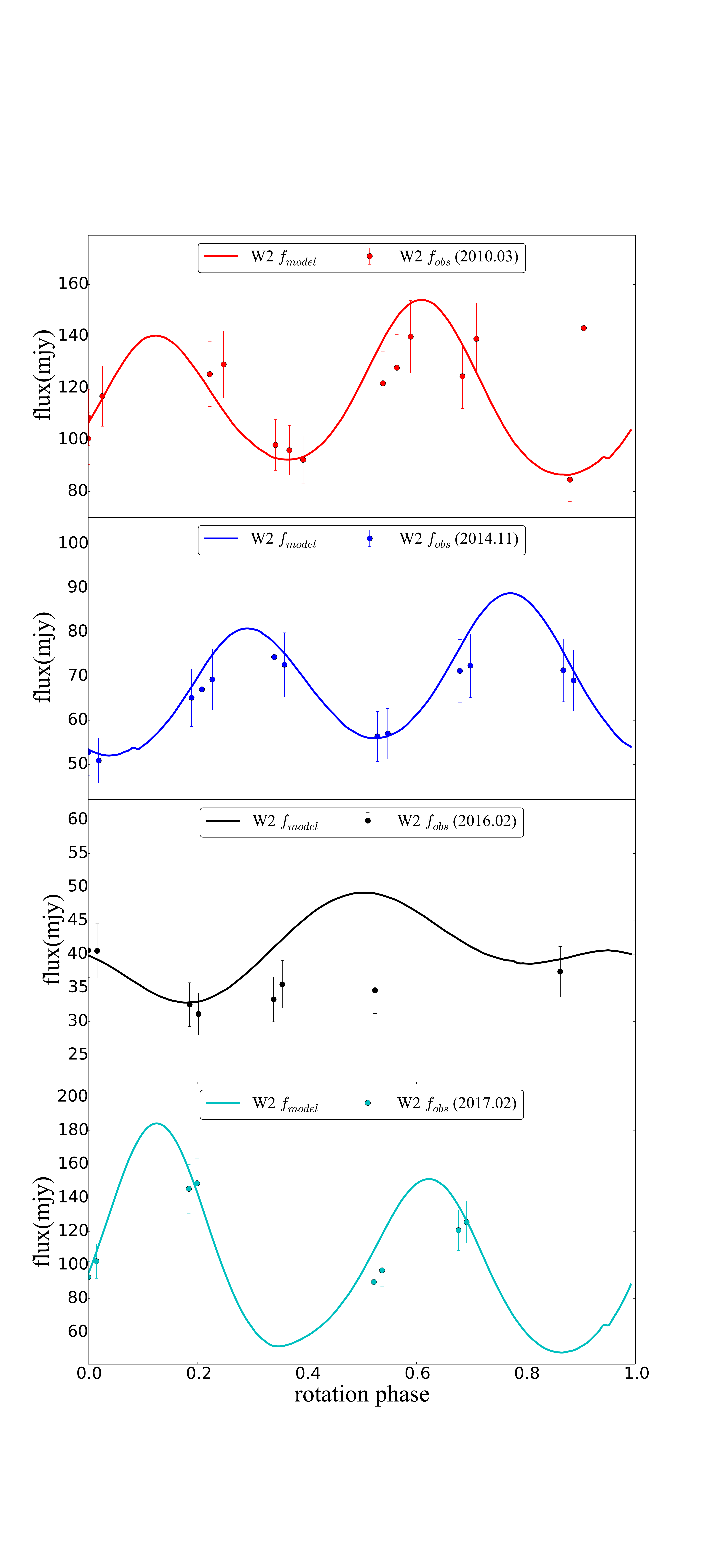}
    \caption{Thermal light curves of (63) Ausonia at W2 band.}
    \label{w2thli_curve63}
\end{figure}

\begin{figure}
        \includegraphics[width=9cm,height=10cm]{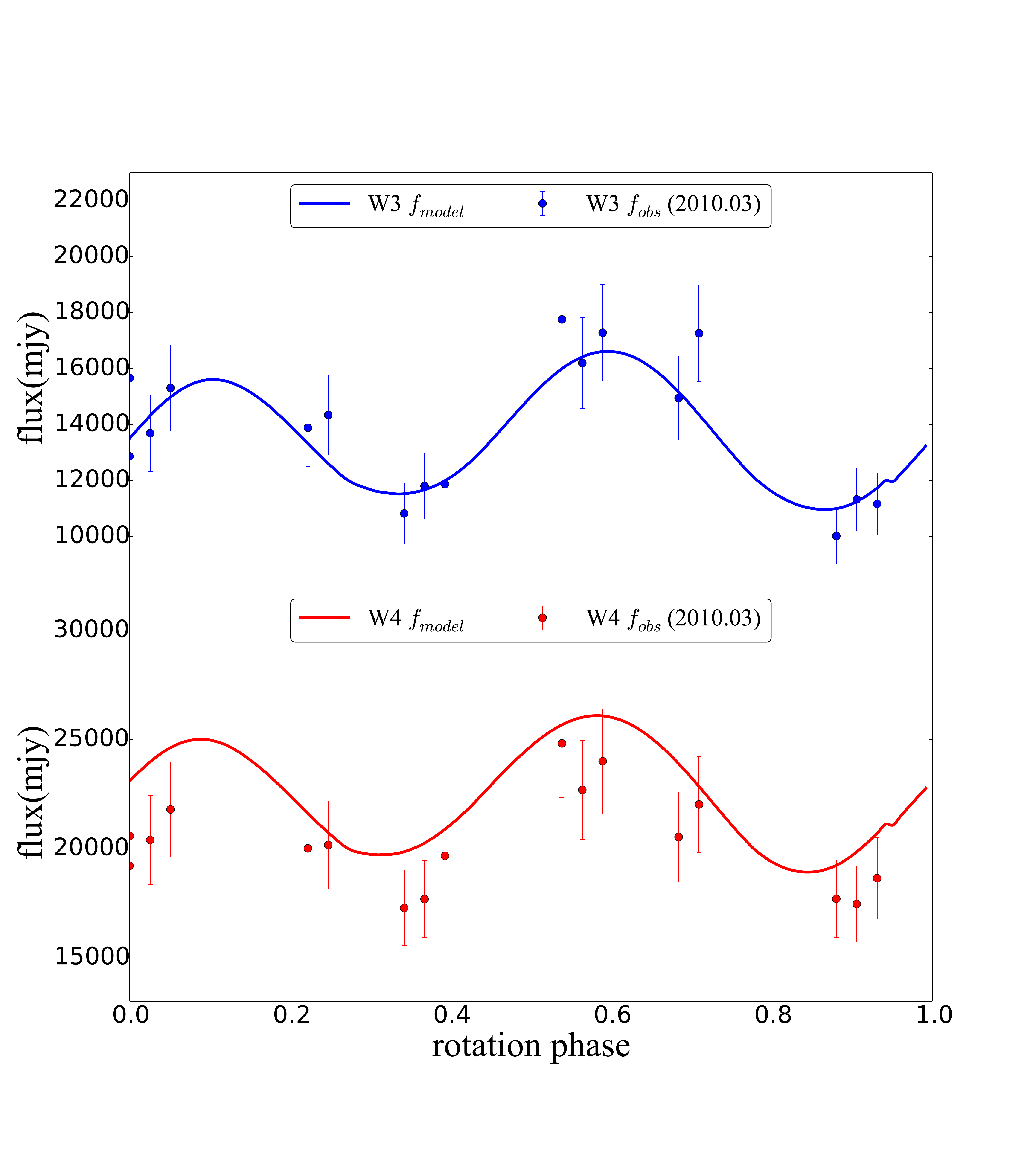}
    \caption{Thermal light curves of (63) Ausonia at W3 and W4 bands.}
    \label{w34thli_curve63}
\end{figure}

\begin{figure}
        \includegraphics[width=9cm,height=6cm]{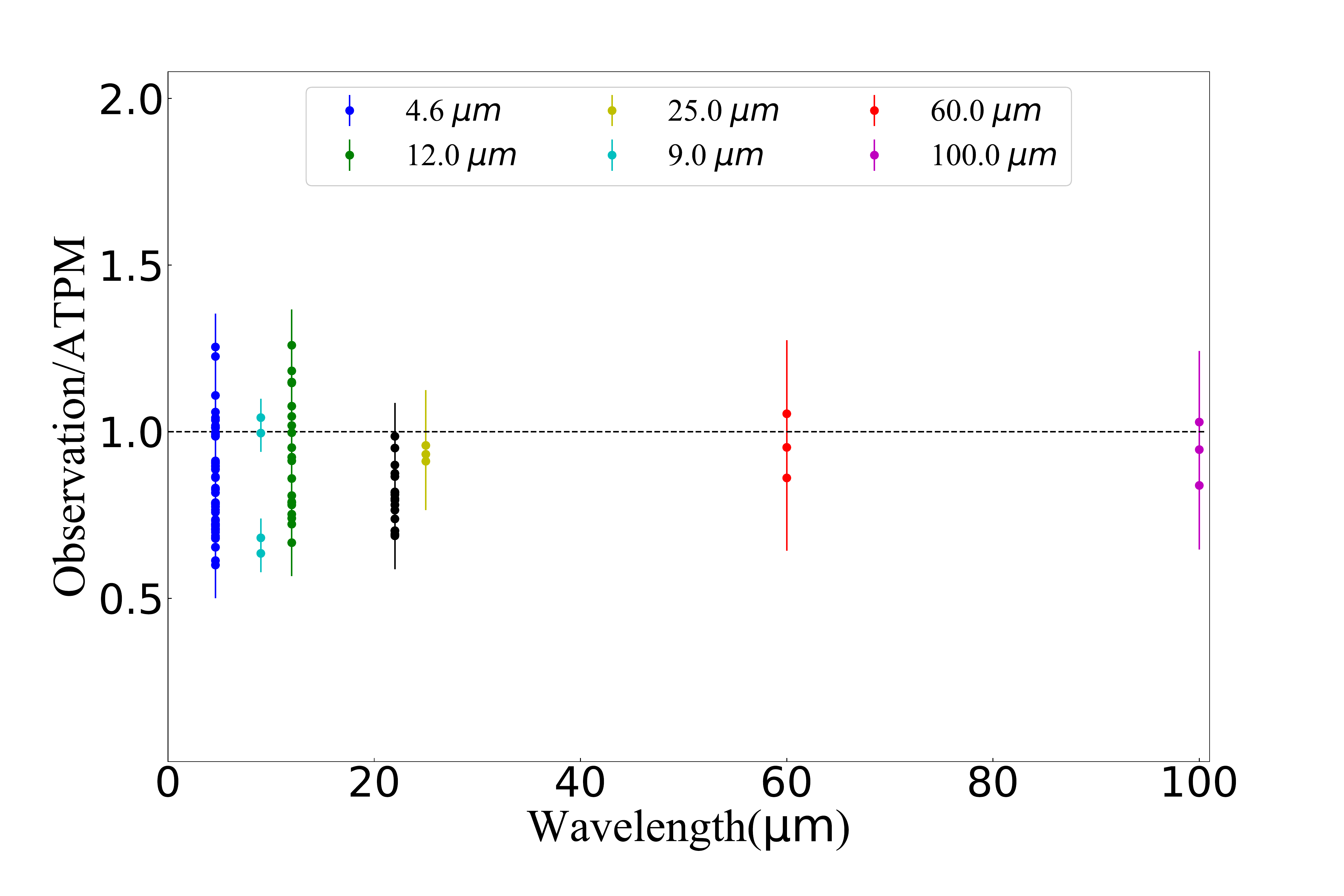}
    \caption{The observation/ATPM ratios as a function of wavelengths for (63) Ausonia}
    \label{fmobsratio63}
\end{figure}

To examine the best-fitting parameters for (63) Ausonia, here we follow the method described in \citet{yu2017} and \citet{jiang2019} to plot the theoretical thermal light curves of (63) Ausonia compared with the mid-infrared observations. As shown in Fig. \ref{w2thli_curve63}, the thermal flux from  ATPM offers a good matching with the data at W2 band for each of four separate epoch. In addition, Fig. \ref{w34thli_curve63} exhibits a similar behaviour of the thermal light curves with the observations at W3 and W4 bands, respectively. In order to examine the reliability of our derived results, we again calculate the ratio of theoretical flux obtained by ATPM and the observational flux at diverse wavelengths.  Fig. \ref{fmobsratio63} displays the observation/ATPM ratios as a function of wavelengths for (63) Ausonia at each wavelength by distinguished colors, and the values of the ratio fluctuate about 1 for each wavelength, being indicative of a reliable outcome for (63) Ausonia.

\subsection{(556) Phyllis}
\begin{figure}
 \includegraphics[width=9cm,height=4.8cm]{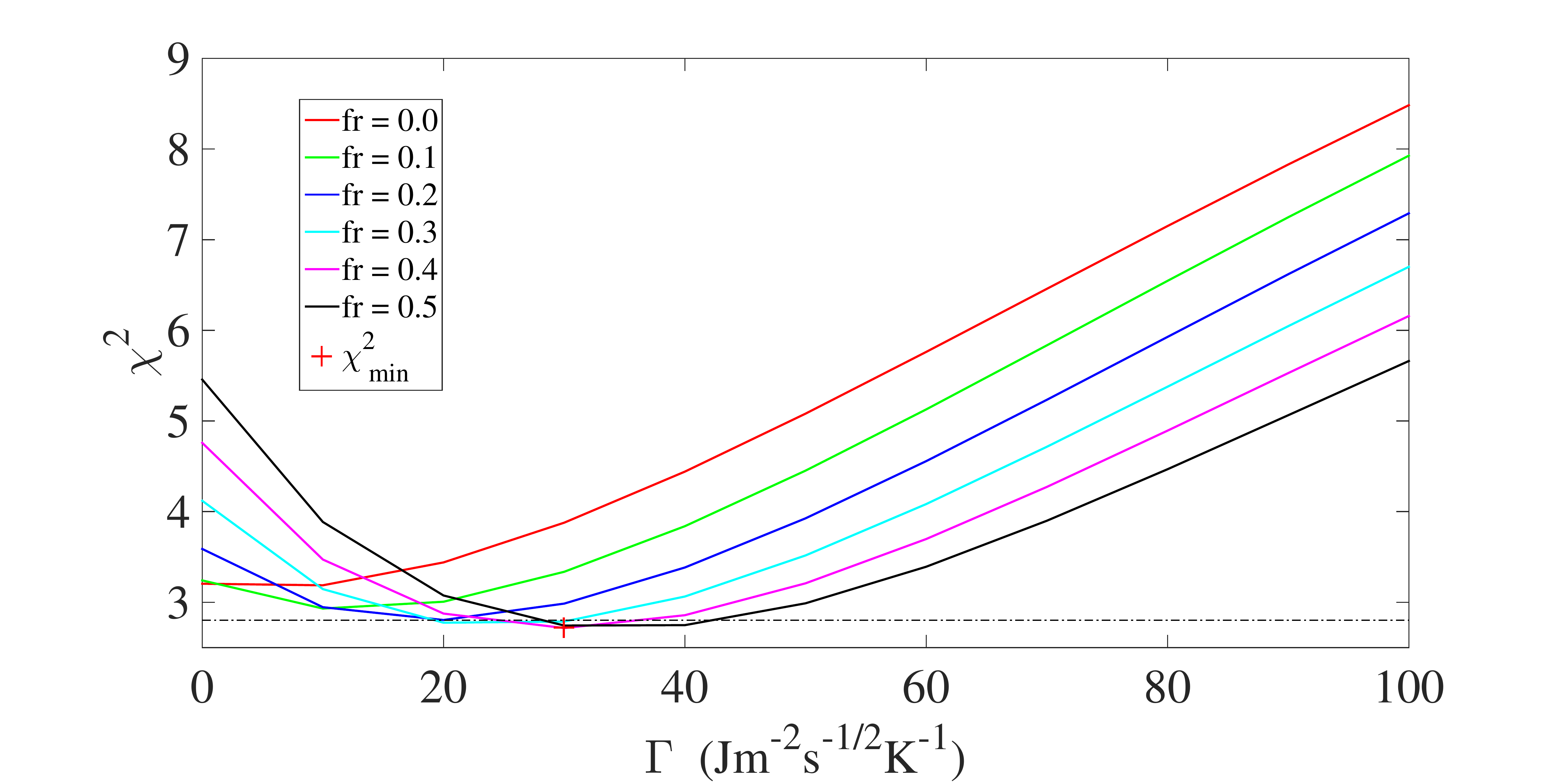}
    \caption{$\Gamma -\chi^2$ profile fit to observation of (556) Phyllis.}
    \label{chi2ga556}
\end{figure}

\begin{figure}
        \includegraphics[width=9cm,height=20cm]{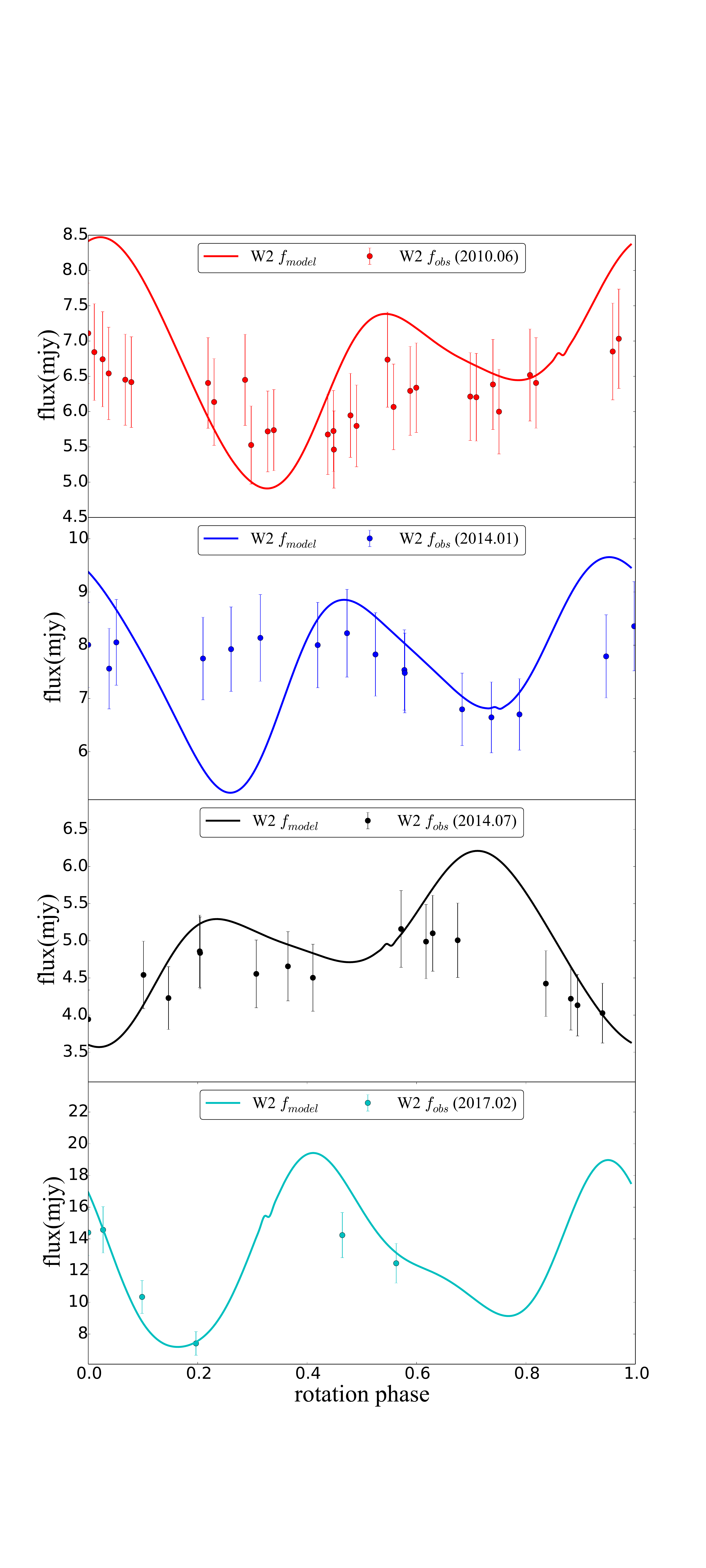}
    \caption{Thermal light curves of  (556) Phyllis at W2 band.}
    \label{w2thli556}
\end{figure}

\begin{figure}
        \includegraphics[width=9cm,height=10cm]{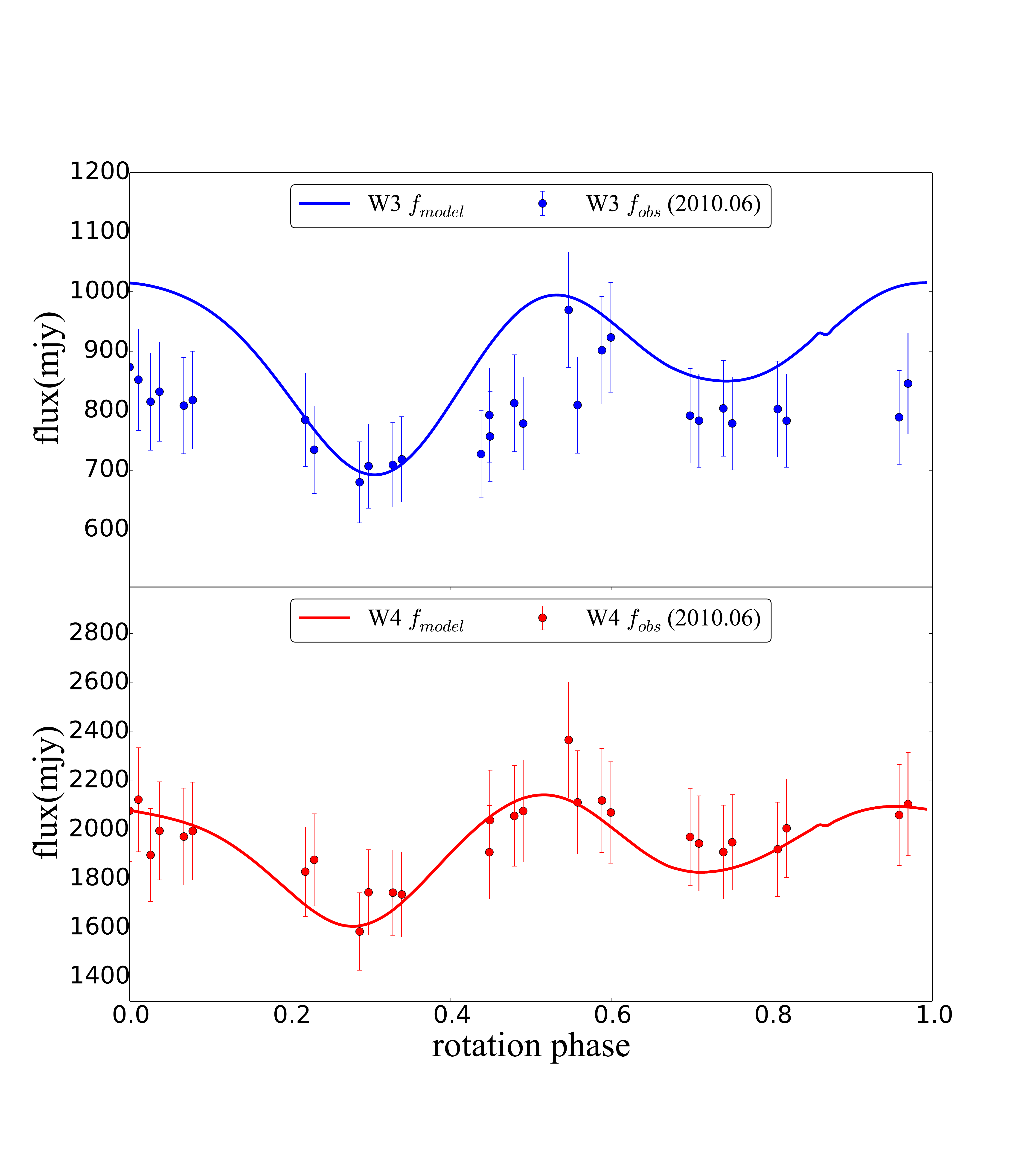}
    \caption{Thermal light curves of  (556) Phyllis at W3 and W4 bands.}
    \label{w34thli556}
\end{figure}

\begin{figure}
        \includegraphics[width=9cm,height=6cm]{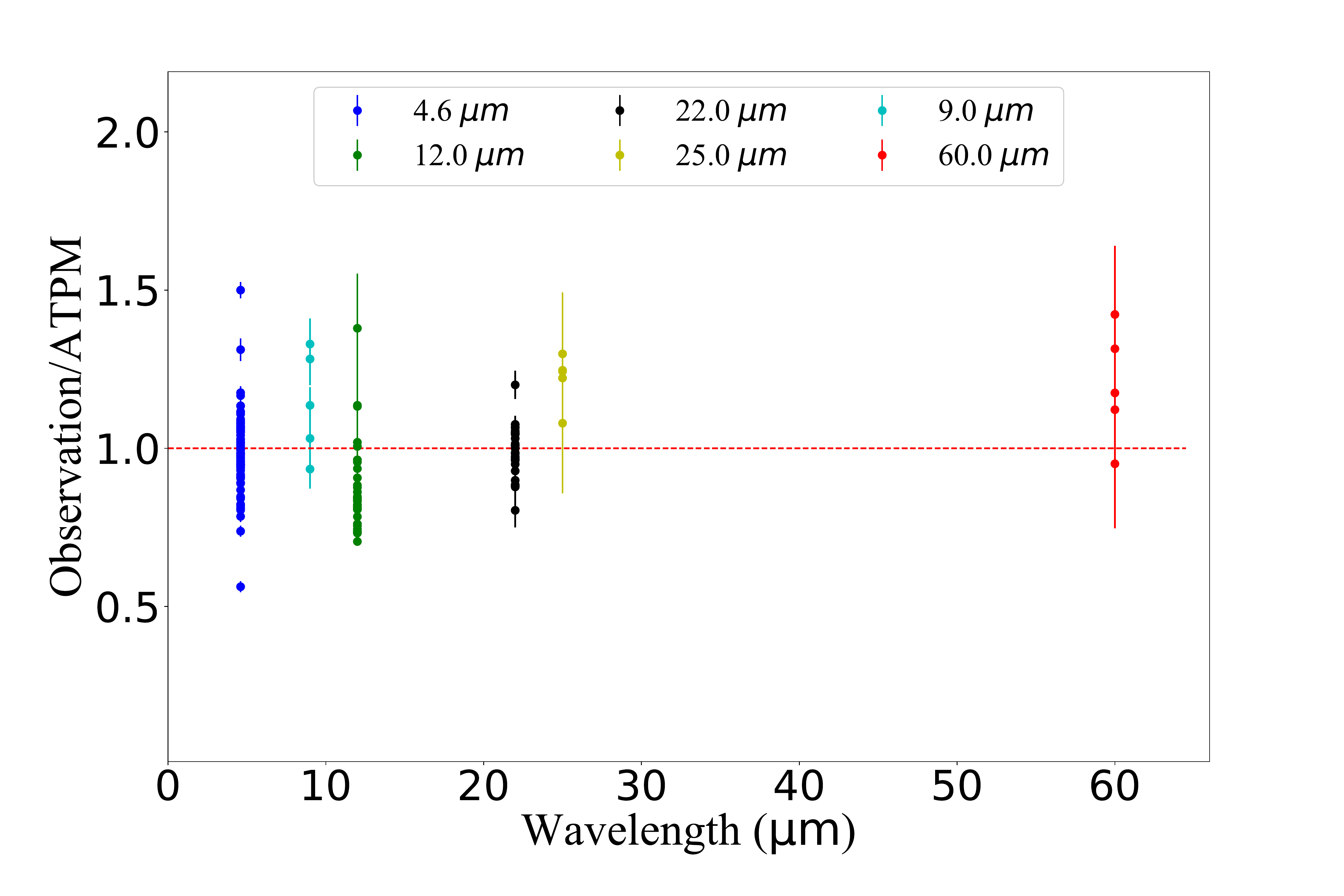}
    \caption{The observation/ATPM ratios as a function of wavelengths for (556) Phyllis.}
    \label{fmobsratio556}
\end{figure}

(556) Phyllis is also taxonomically classified as an S-type \citep{tholen1984,bus_binzel2002} asteroid with a diameter 36.28 km, and a geometric albedo  0.201 \citep{masiero2014}. The first photometric observations and optical light curves of (556) Phyllis were performed by \citet{zapp1983}, and they derived a rotation period   $4.28 \pm 0.002$ h. \citet{Marciniak2007} observed this asteroid for five distinct observation epochs in 1998, 2000, 2002, 2004 and 2005-2006, respectively. They updated a rotation period $4.293 \pm 0.001$ h and provided two resolved pole orientations  $(18^\circ, 54^\circ)$ and $(209^\circ, 41^\circ)$. In the present study, we adopt the pole orientation of the former from that of \citet{Marciniak2007}. For the observations, we include the thermal data from IRAS ($5 \times 12  \rm \mu m$, $5 \times 25 \rm \mu m$ and $5 \times 60 \rm \mu m$), AKARI ($5 \times 9 \rm \mu m$, $4 \times 18 \rm \mu m$) and WISE/NEOWISE ($75 \times 4.6 \rm \mu m$, $30 \times 12.0 \rm \mu m$ and $29 \times 22.0 \rm \mu m$), where the number of entire observations is 130. We derive an effective diameter of $35.600_{-0.901}^{+0.883}$ km with the geometric albedo $0.209_{-0.010}^{+0.011}$ by considering the absolute magnitude 9.56. As shown in Fig. \ref{chi2ga556}, the thermal inertia and roughness fraction are constrained to be $30_{-11}^{+12}$ $\rm J m^{-2} s^{-1/2} K^{-1}$ and $0.40_{-0.20}^{+0.10}$, respectively, with respect to a minimum $\chi^2$ value 2.715. WISE/NEOWISE observed this asteroid on eight different epochs. However, for several epochs, the number of data points are too small to fully reflect the asteroid's varied flux with rotational phases. Moreover, Fig. \ref{w2thli556} gives the results of thermal light curves for 4 different epochs at W2, while \ref{w34thli556} reveals the outcomes of W3 and W4. Figs. \ref{w2thli556} and \ref{w34thli556} both demonstrate that the theoretical model fits the observations well. Further evidence can be provided from Fig. \ref{fmobsratio556}, which shows the Observation/ATPM ratio of (556) Phyllis for each wavelength.

\subsection{(1906) Neaf}

\begin{figure}
        \includegraphics[width=9cm,height=4.8cm]{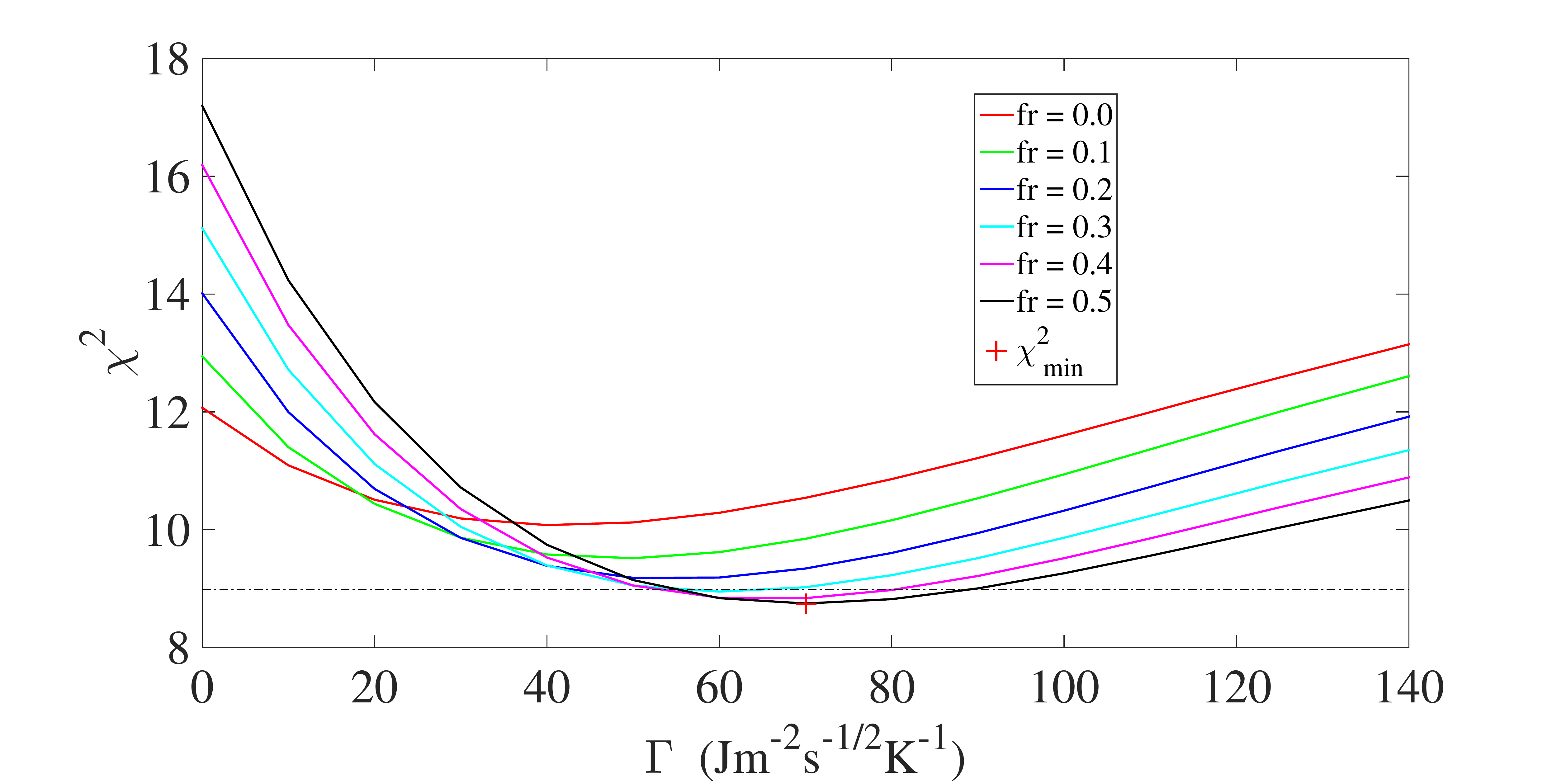}
    \caption{$\Gamma -\chi^2$ profile fit to observation of (1906) Neaf.}
    \label{chi2ga1906}
\end{figure}

\begin{figure}
        \includegraphics[width=9cm,height=10cm]{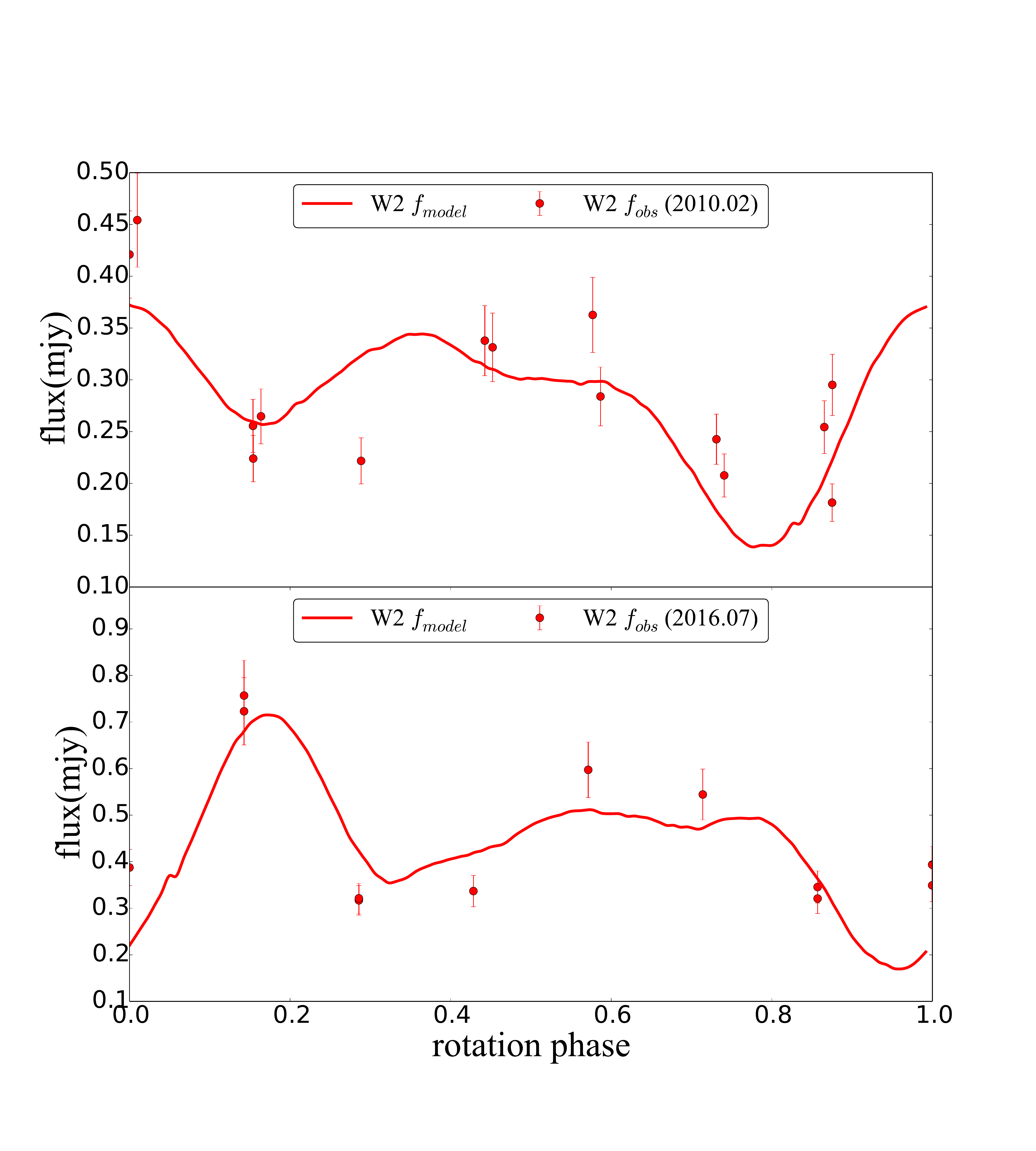}
    \caption{Thermal light curves of (1906) Neaf at W2 band for two epochs.}
    \label{w2thli1906}
\end{figure}

\begin{figure}
        \includegraphics[width=9cm,height=10cm]{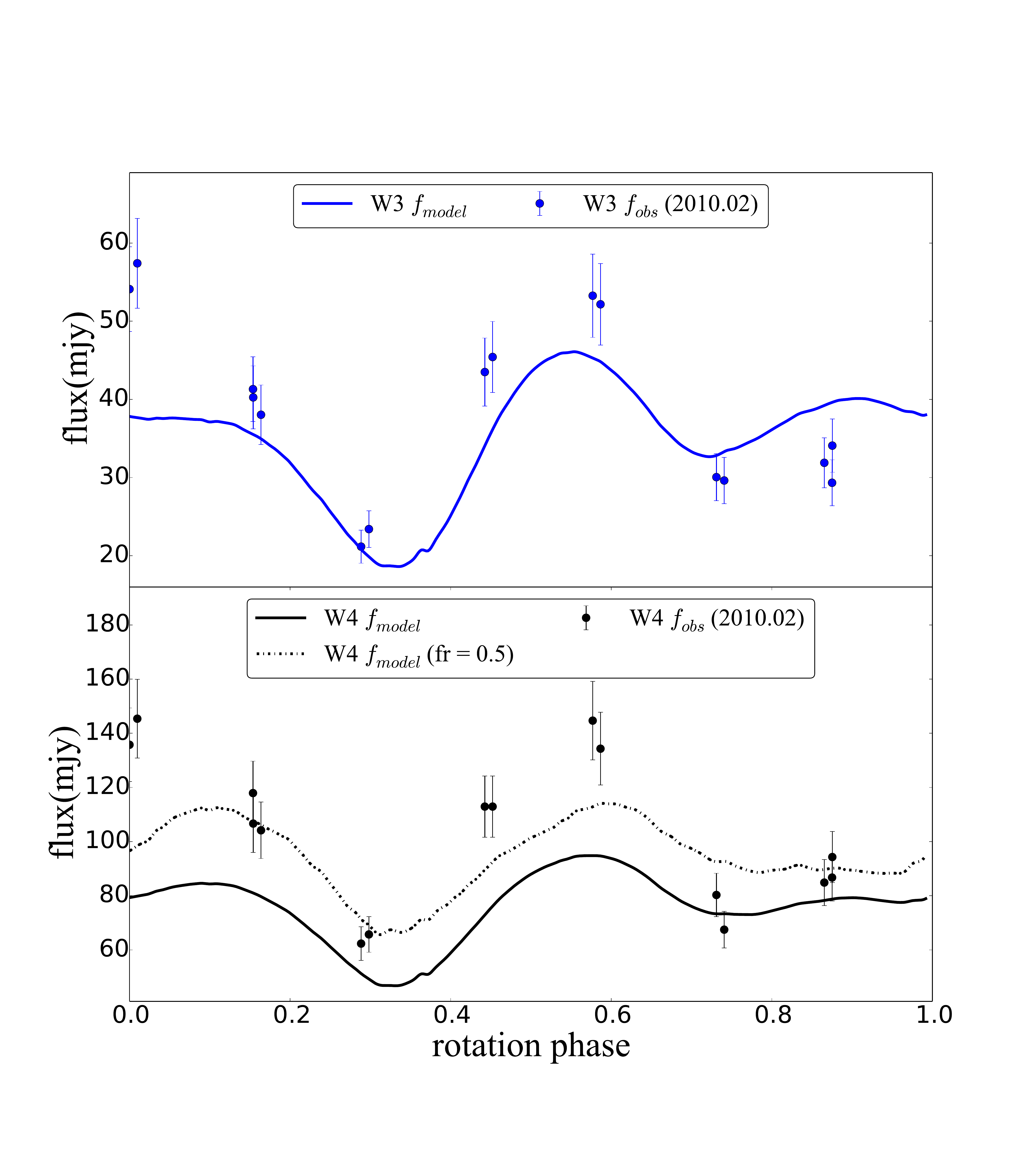}
    \caption{Thermal light curves of (1906) Neaf at W3 and W4 band.}
    \label{w34thli1906}
\end{figure}

\begin{figure}
        \includegraphics[width=9cm,height=6cm]{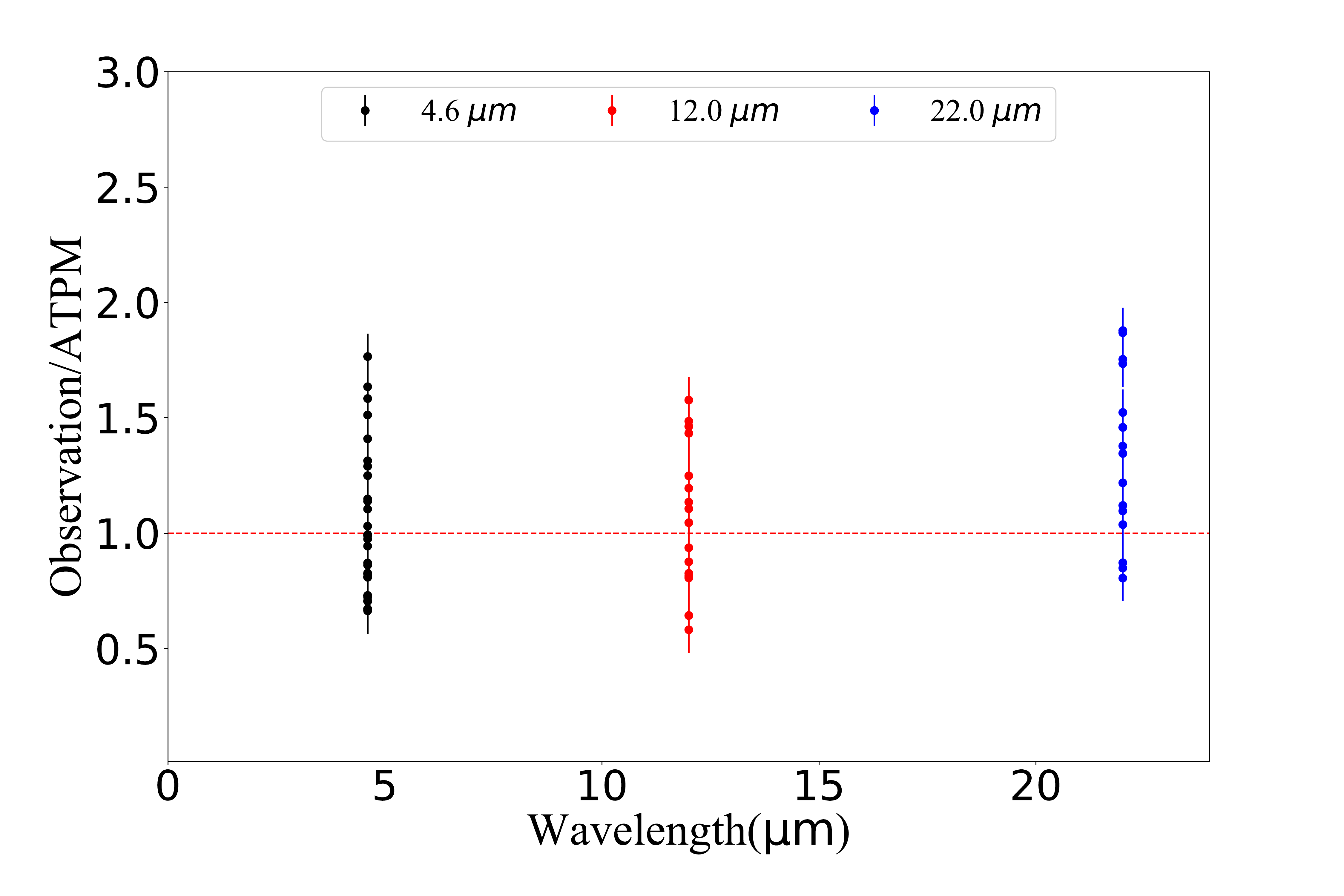}
    \caption{The ratio of Observation/ATPM for (1906) Neaf.}
    \label{obsfmratio1906}
\end{figure}

Asteroid (1906) Neaf, provisional designation 1972 RC, is a V-type asteroid \citep{xu1995_icarus} in the inner main-belt region. It orbits the Sun at a heliocentric distance $2.1 \sim 2.7$ AU every 3.66 yr. \citet{masiero2014} derived its diameter $7.923 \pm 0.09$ km and a geometric albedo $0.234 \pm 0.052$, based on the observations from WISE/NEOWISE. Similarly, we utilize  the observations from WISE/NEOWISE, which are combined with ATPM to derive its thermal properties. Here we entirely use 57 WISE observations at 3 separate epochs (30 in W2, 16 in W3 and 16 in W4). We obtain the effective diameter of this asteroid to be $7.561_{-0.443}^{+0.449}$ km and geometric albedo $0.257_{-0.028}^{+0.033}$, and these results agree with the findings of \citet{masiero2014}. According to $\Gamma - \chi^2$ profile in  Fig. \ref{chi2ga1906}, the thermal inertia and roughness fraction are confined to be $70_{-16}^{+19}$ $\rm J m^{-2} s^{-1/2} K^{-1}$ and $0.5_{-0.2}^{+0.0}$, respectively.  From the result of $\Gamma$ and $f_{\rm r}$, we plot the 3-bands thermal light curves in Figs. \ref{w2thli1906} and  \ref{w34thli1906}. Our computed thermal fluxes reasonably fit the WISE observation with $\chi_{\rm min}^{2}=$ 7.095. This may be the reason that the W4 theoretical flux do not agree well with the observations according to Fig. \ref{w34thli1906}, but the fluctuation trends of the thermal light curves are consistent with the observations. In the lower panel of Fig. \ref{w34thli1906}, we again provide an additional thermal curve at W4 with a high roughness $f_{\rm r} = 0.5$ (marked by the black dashed line) in comparison with the case of low roughness, which leads to a better-fitting solution at W4 band. With the help of all data, the best-fitting thermal inertia is evaluated to be approximately 70 $\rm J m^{-2} s^{-1/2} K^{-1}$  with respect to a low roughness fraction. The ratio of observed flux and theoretical flux are plotted in Fig. \ref{obsfmratio1906}.

\subsection{(2511) Patterson}

\begin{figure}
        \includegraphics[width=9cm,height=4.8cm]{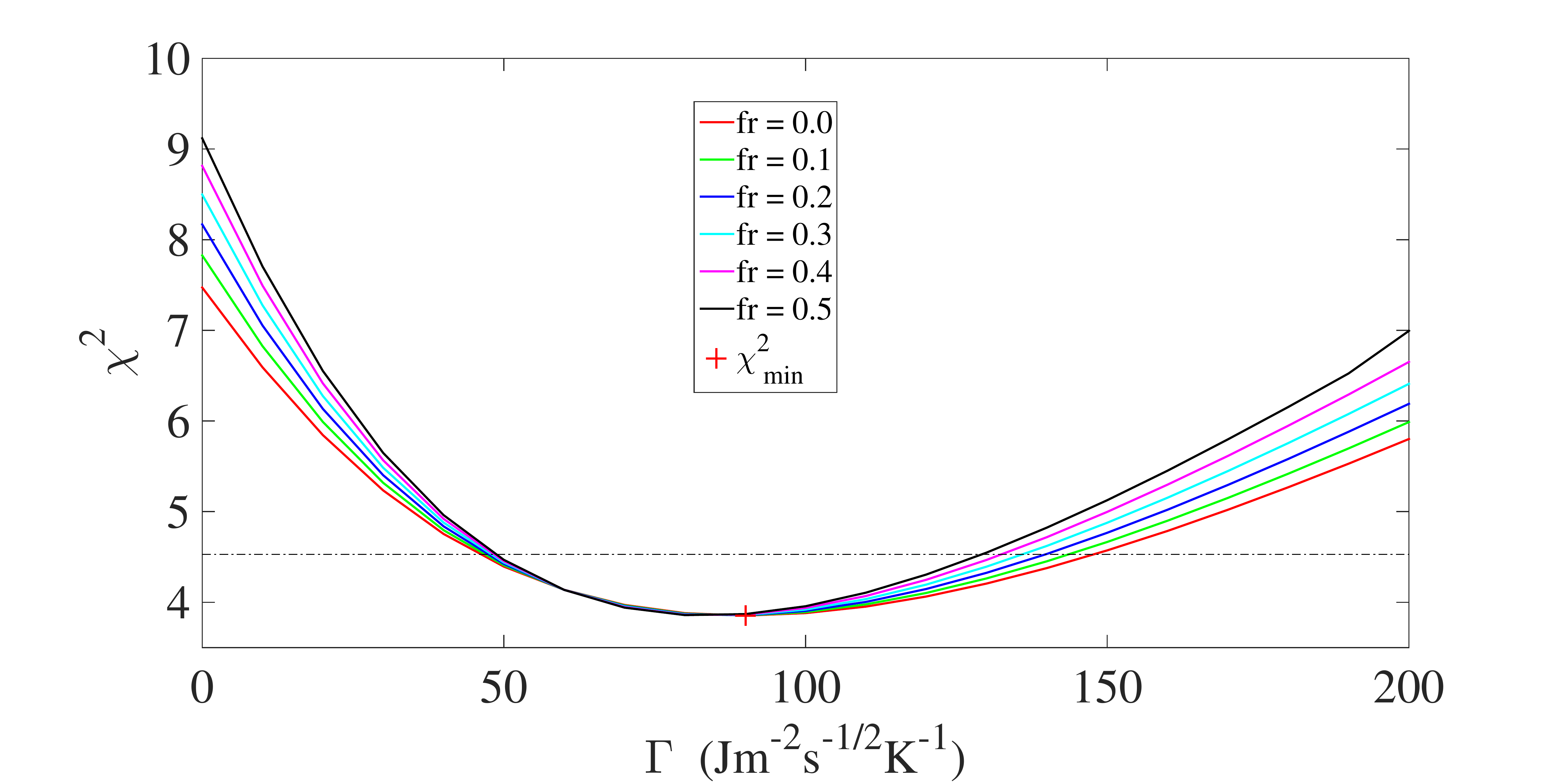}
    \caption{$\Gamma -\chi^2$ profile of (2511) Patterson.}
    \label{chi2ga2511}
\end{figure}

\begin{figure}
        \includegraphics[width=9cm,height=10cm]{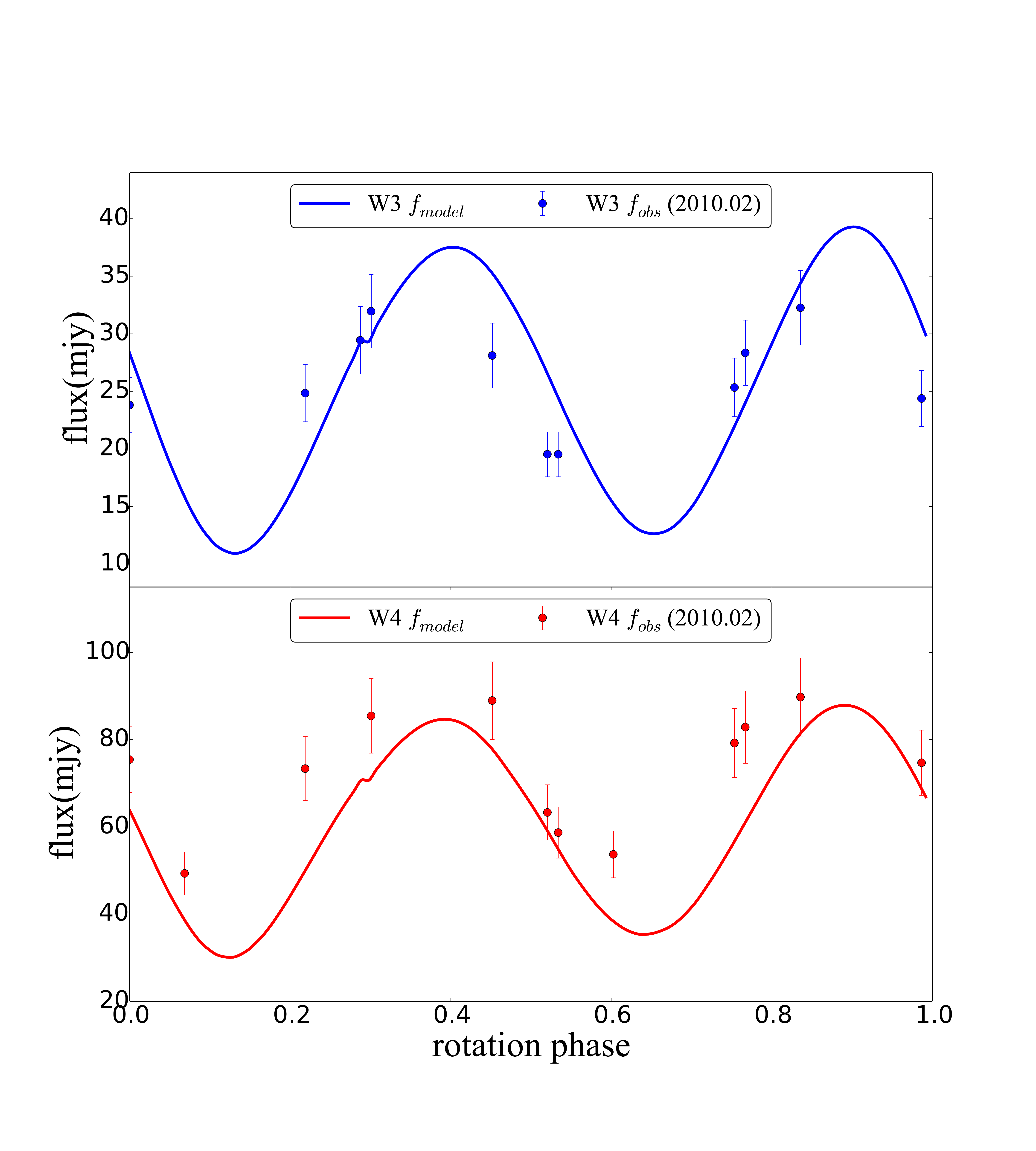}
    \caption{W3 and W4 thermal light curves of (2511) Patterson }
    \label{w34thli2511}
\end{figure}

\begin{figure}
        \includegraphics[width=9cm,height=5.5cm]{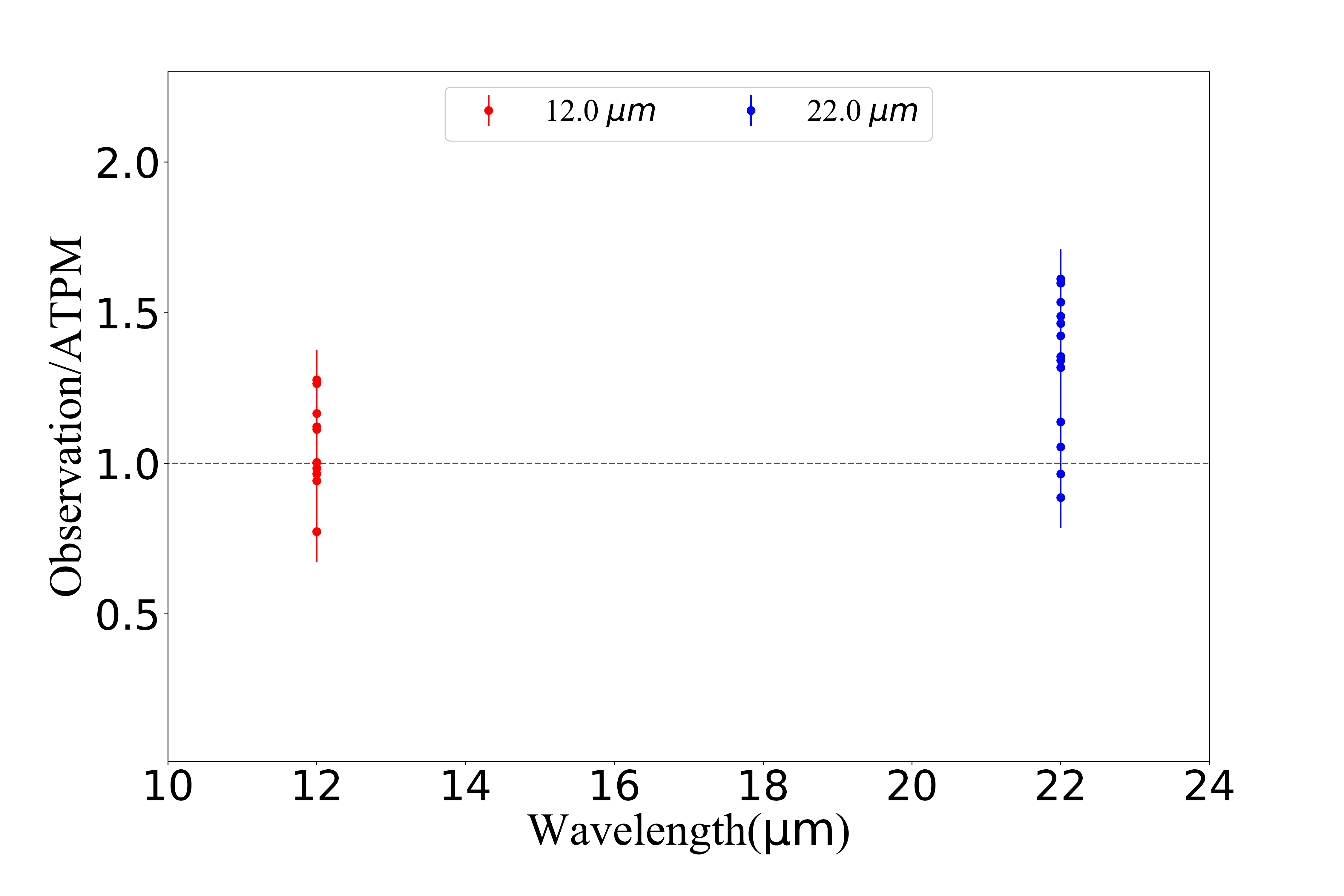}
    \caption{The ratio of Observation/ATPM for (2511) Patterson.}
    \label{fmobsratio2511}
\end{figure}

Asteroid (2511) Patterson is a V-type asteroid \citep{bus_binzel2002} that was discovered by the Palomar Observatory in 1980. It has a semi-major axis 2.298 AU, an eccentricity 0.104, and an orbital inclination $8.046^{\circ}$. \citet{durech2017} derived the shape model of (2511) Patterson by using the sparse-in-time photometry from the Lowell Observatory photometry database and WISE observations, which can be obtained from DAMIT. They further presented the pole orientation $(194^\circ, 51^\circ)$. Using the NEATM (Near Earth Asteroid Thermal Model), \citet{masiero2011} derived the geometric albedo and effective diameter to be $0.287 \pm 0.039$ and $7.849 \pm 0.174$ km, respectively. In this work we use the ATPM and combined with 24 WISE/NEOWISE observations ($11 \times 12.0  \rm \mu m$ and $13 \times 22.0  \rm \mu m$) to determine the thermal characteristics of (2511) Patterson. As shown in Fig. \ref{chi2ga2511}, the minimum $\chi^2$  corresponds to thermal inertia  $90_{-43}^{+58}$ $\rm J m^{-2} s^{-1/2} K^{-1}$, and the roughness fraction can be constrained to be $0.0_{-0.0}^{+0.50}$. In addition, the geometric albedo is estimated to be $0.180_{-0.034}^{+0.055}$, which is smaller than that of \citet{masiero2011}, and thus the effective diameter is $9.034_{-1.128}^{+0.997}$ km. To examine our results, we plot the observation/ATPM ratio and thermal light curves for each waveband in Fig. \ref{fmobsratio2511} and Fig. \ref{w34thli2511}.  The solid curves in Fig. \ref{w34thli2511}  are modeled with $\Gamma=90$ $\rm J m^{-2} s^{-1/2} K^{-1}$ and $f_{\rm r} = 0.0$. The model seems to slightly overestimate the W4 data, but the fit to the WISE light curve seems to be reliable for all wavelengths.

\subsection{(3281) Maupertuis}

\begin{figure}
        \includegraphics[width=9cm,height=4.8cm]{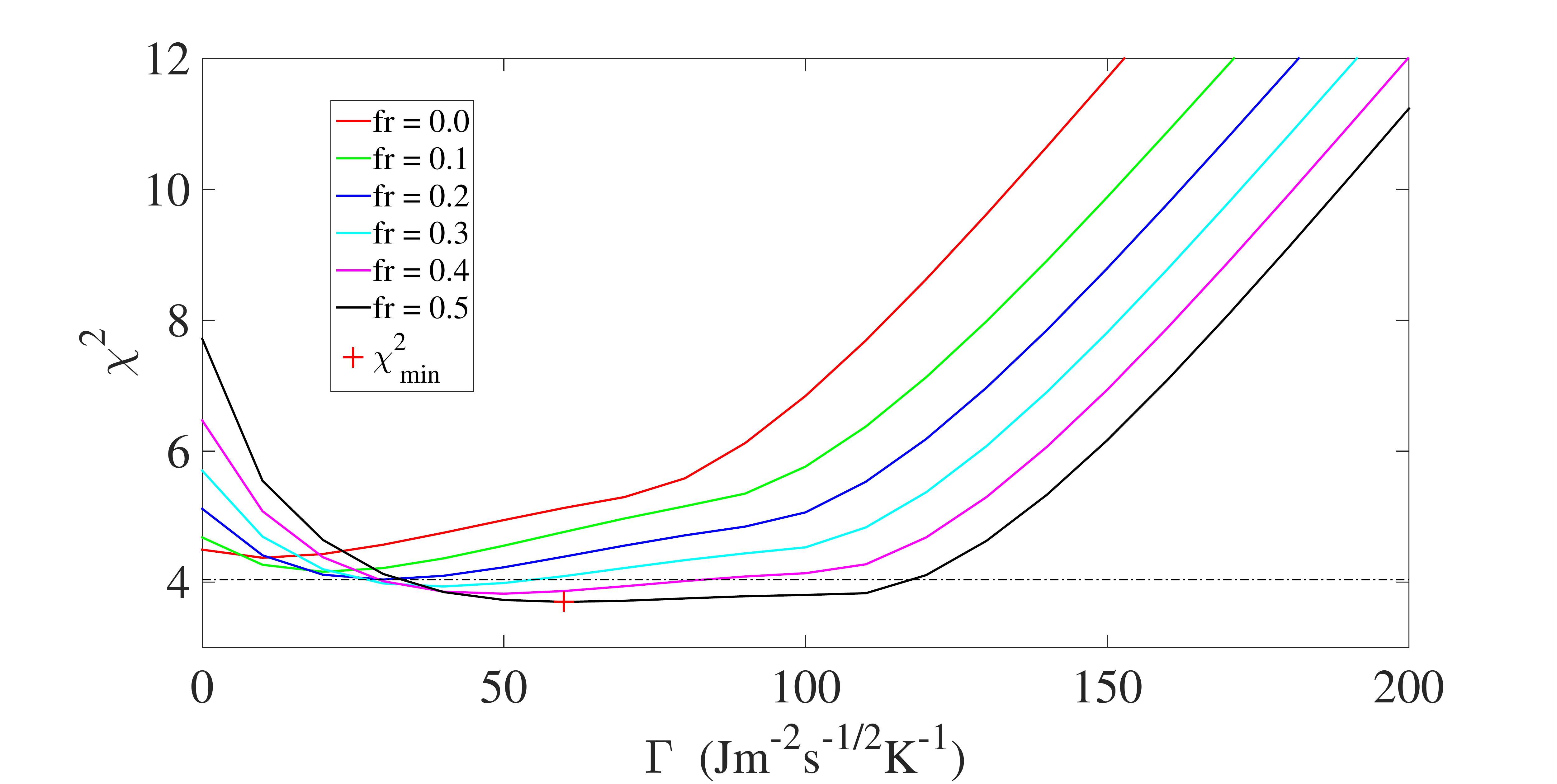}
    \caption{$\Gamma -\chi^2$ profile of (3281) Maupertuis.}
    \label{chi2ga3281}
\end{figure}

\begin{figure}
        \includegraphics[width=9cm,height=15cm]{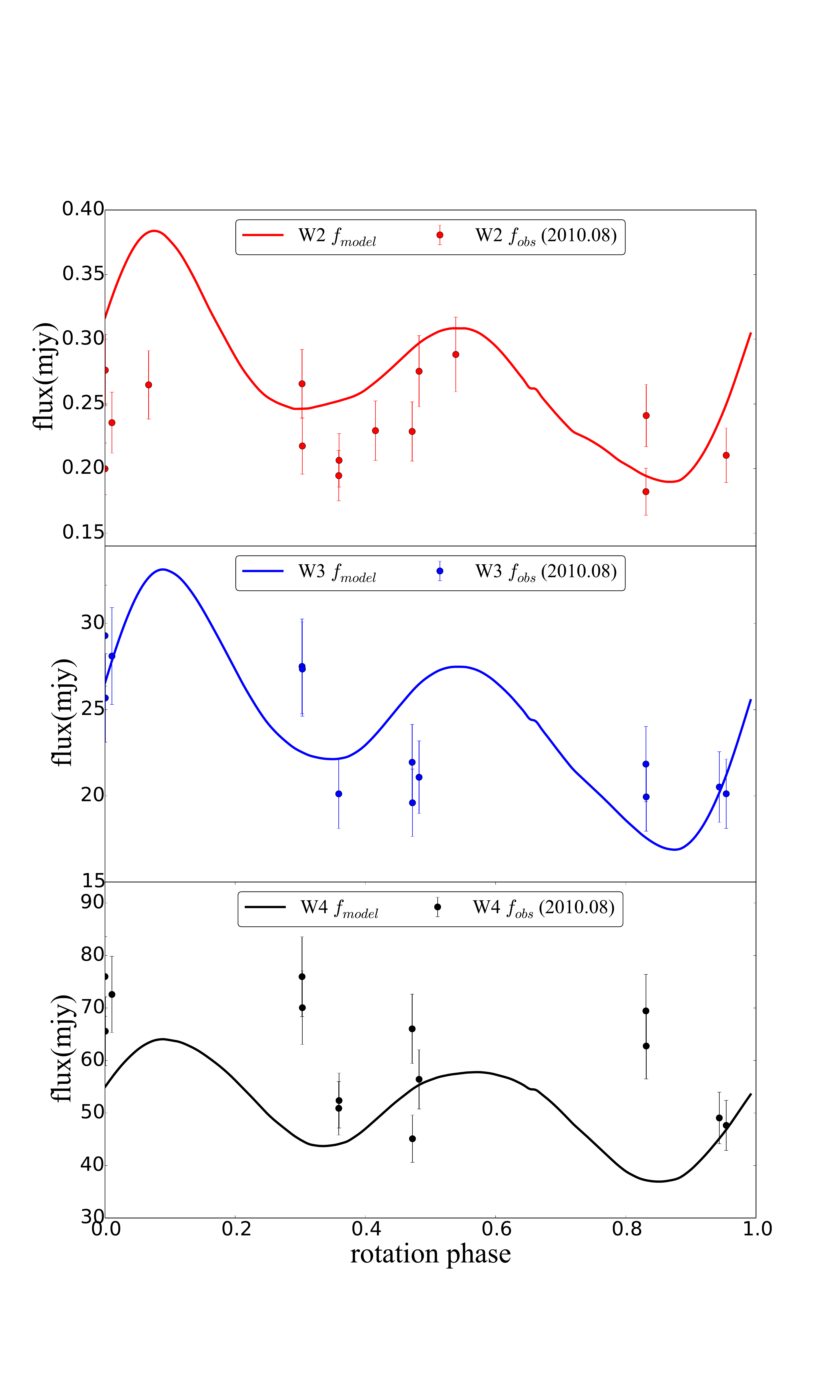}
    \caption{W2, W3 and W4 thermal light curves of (3281) Maupertuis.}
    \label{w234thli3281}
\end{figure}

\begin{figure}
        \includegraphics[width=9cm,height=6cm]{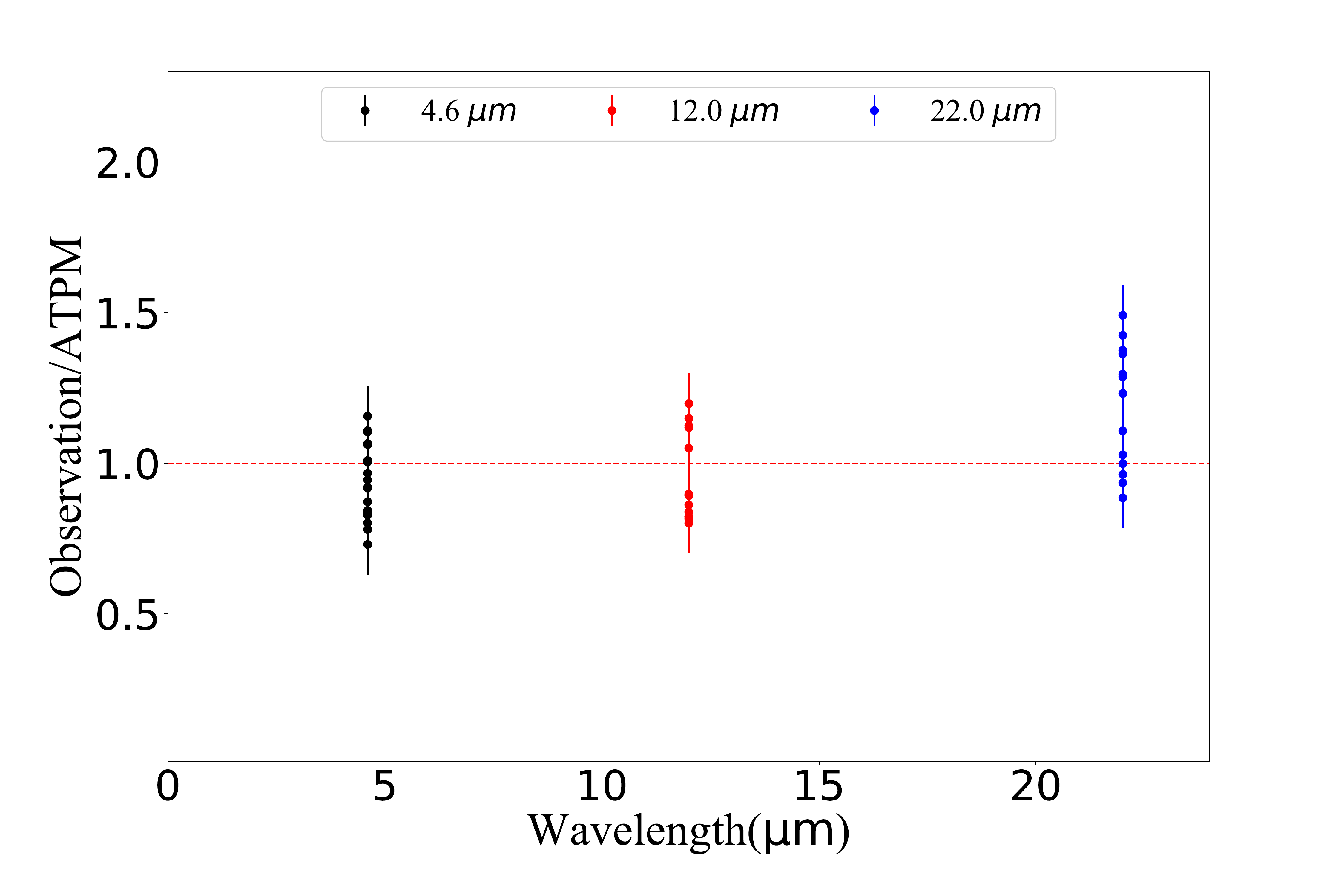}
    \caption{The ratio of Observation/ATPM for (3281) Maupertuis at three wavelengths.}
    \label{fmobsratio3281}
\end{figure}

Asteroid (3281) Maupertuis is a Vesta family member that orbits the Sun at a distance $2.121 \sim 2.579$ AU every 3.6 years, and has an absolute magnitude 12.9. The spectral type of this asteroid is not determined yet. Using the Lowell photometric data and light curve inversion method, \citet{durech2016} derived the shape model of this asteroid and the sidereal period $6.72894 \pm 0.00001$ h. Two pole orientations of $(231^\circ, -74^\circ)$ and $(62^\circ, -66^\circ)$ were given in their work. \citet{masiero2014} showed that (3281) Maupertuis has a geometric albedo $0.489 \pm 0.02$ and the effective diameter $5.482 \pm 0.043$ km. This result was close to that of \citet{mainzer2011}. For this asteroid, we employ 45 thermal data from WISE/NEOWISE at two separate epochs ($18 \times 4.6 \rm \mu m$, $13 \times 13.0 \rm \mu m$ and $14 \times 22.0 \rm \mu m$) in the fitting.  For (3281) Maupertuis, we find that its diameter is constrained to be $5.509_{-0.270}^{+0.447}$ km with a geometric albedo  $0.484_{-0.074}^{+0.051}$. The value of the geometric albedo is a bit high for a main-belt asteroid, but as a fragment of asteroid (4) Vesta, it is consistent with a wide range of $p_{\rm v}$ on (4) Vesta's surface. Fig. \ref{chi2ga3281} exhibits the best-fitting values of the thermal inertia  $60_{-31}^{+58}$ $\rm J m^{-2} s^{-1/2} K^{-1}$ and the roughness $0.50_{-0.30}^{+0.00}$.  As can be seen in Fig. \ref{w234thli3281}, the model seems to overestimate W4 data. Fig. \ref{fmobsratio3281} shows the ratio of Observation/ATPM for (3281) Maupertuis with respect to $\Gamma = 60$ $\rm J m^{-2} s^{-1/2} K^{-1}$ and $f_{\rm r} = 0.5$. However, the values of $\Gamma$ and $f_{\rm r}$ correspond to the minimum value of $\chi^2$.

\subsection{(5111) Jacliff}

\begin{figure}
        \includegraphics[width=9cm,height=4.8cm]{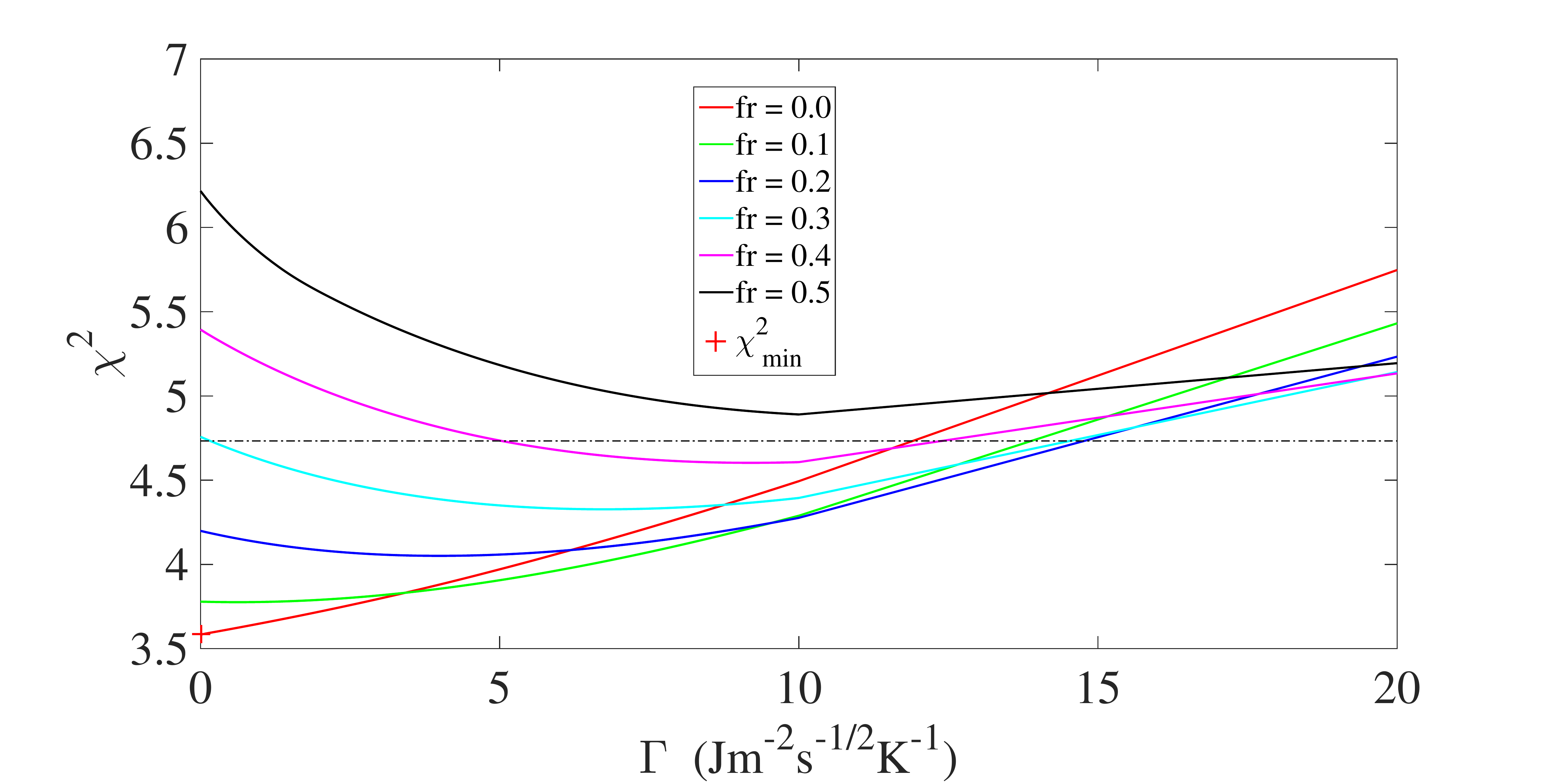}
    \caption{$\Gamma -\chi^2$ profile of (5111) Jacliff.}
    \label{chi2ga5111}
\end{figure}

\begin{figure}
        \includegraphics[width=9cm,height=10cm]{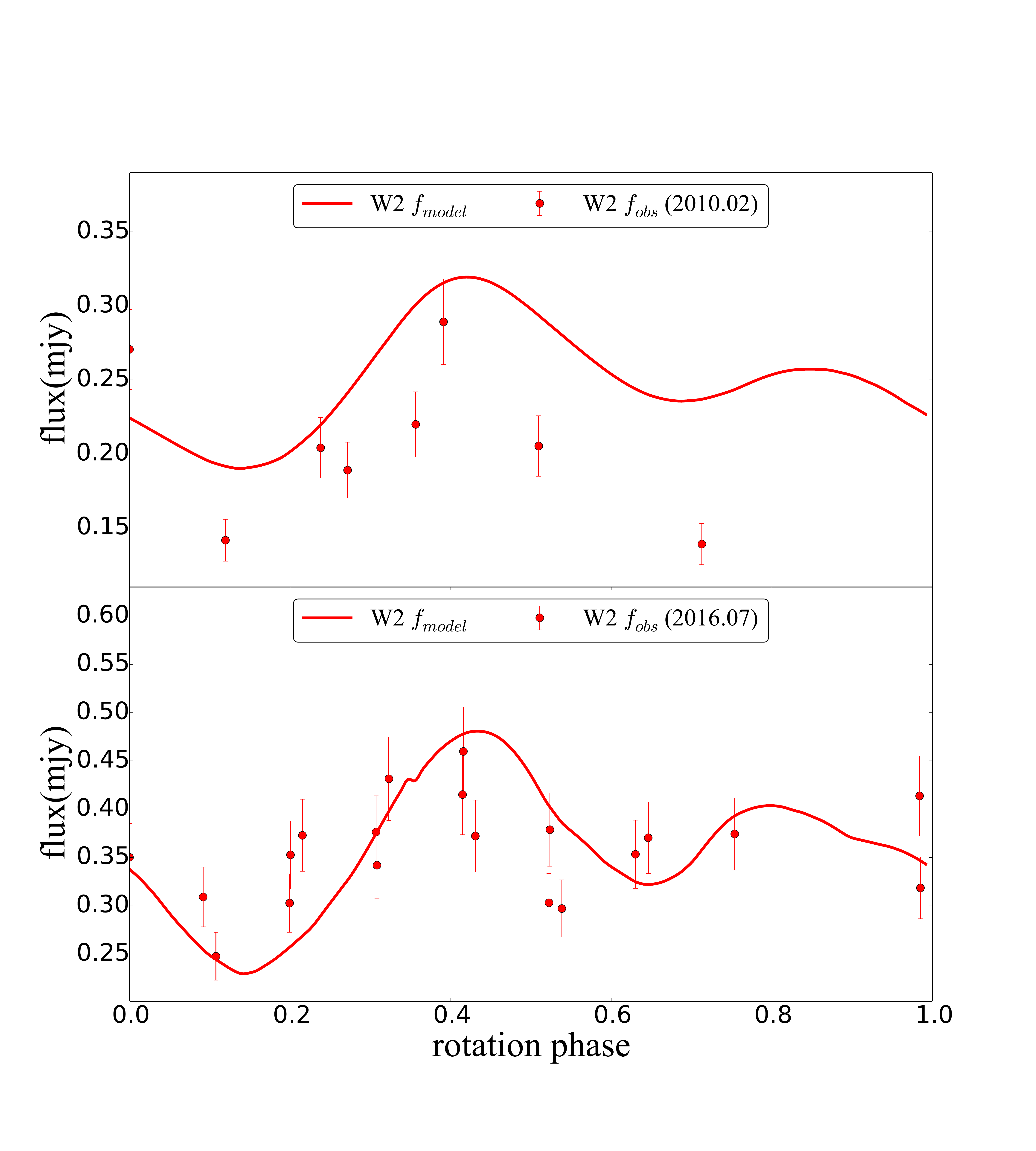}
    \caption{W2 thermal light curves of (5111) Jacliff at two epochs.}
    \label{w2thli5111}
\end{figure}

\begin{figure}
        \includegraphics[width=9cm,height=10cm]{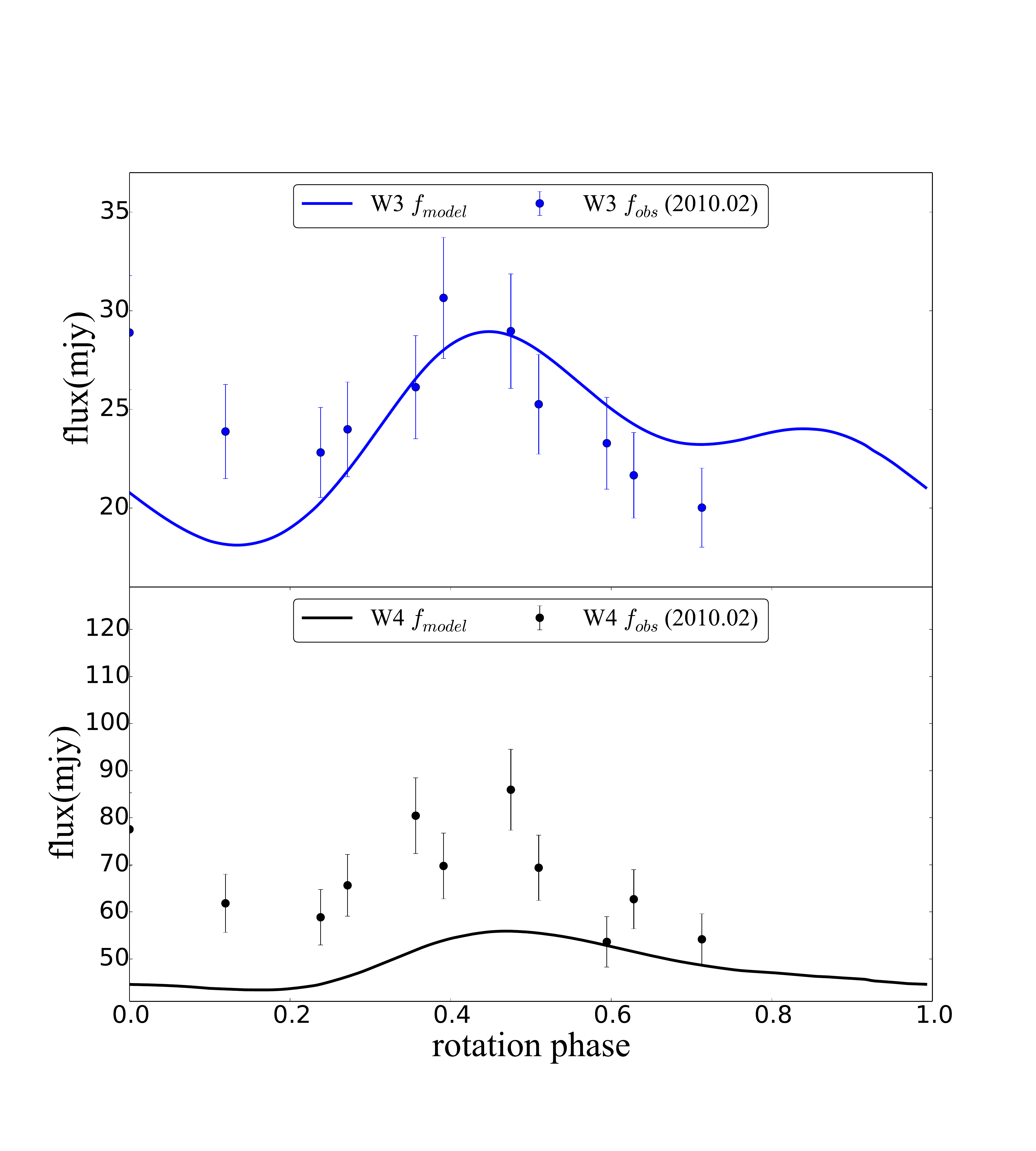}
    \caption{W3 and W4 thermal light curves of (5111) Jacliff.}
    \label{w34thli5111}
\end{figure}

\begin{figure}
        \includegraphics[width=9cm,height=6cm]{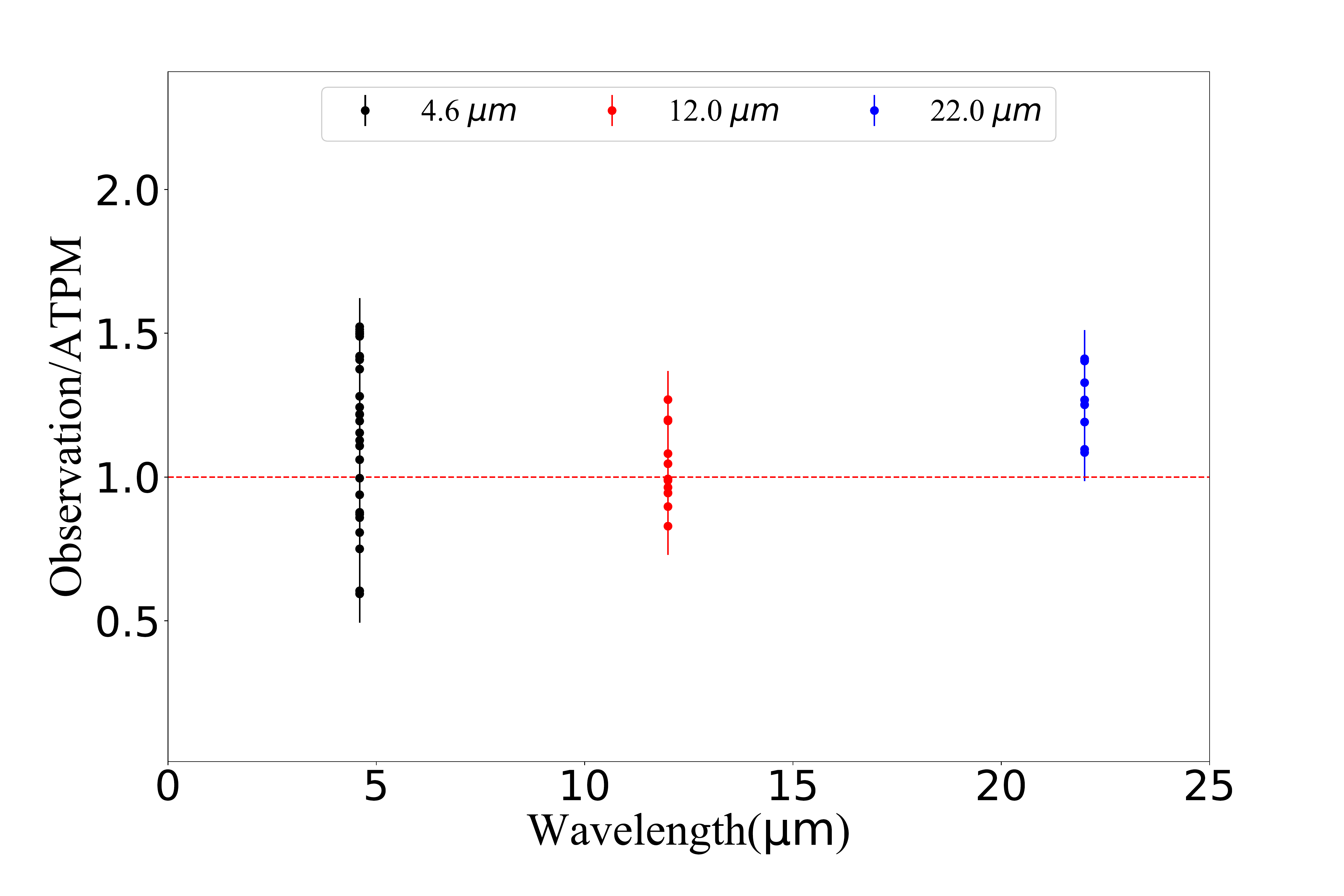}
    \caption{Observation/ATPM ratio of (5111) Jacliff at three wavebands.}
    \label{fmobsratio5111}
\end{figure}

Asteroid (5111) Jacliff, known as provisional designation 1987 SE24, orbits the Sun once every 3.61 yr. In the SMASSII classification, (5111) Jacliff is a R-type asteroid \citep{bus_binzel2002}, while in the Bus-Demeo taxonomy, the asteroid is classified as a V-type asteroid \citep{Demeo2009}. Moreover, \citet{Moskovitz2010} compared the near-infrared ($0.7-2.5~\mu m$) spectra of this asteroid with the laboratory spectra of HED meteorites, and showed that it is expected to be a V-type asteroid. By using all available disk-integrated optical data as input for the convex input method, \citet{hanus2016} derived the 3D shape model for (5111) Jacliff, and the pole orientation and rotation period were derived to be $(259^\circ, -45^\circ)$ and 2.840 h. In our study, we employ 47 WISE/NEOWISE observations ($28 \times 4.6 \rm \mu m$, $11 \times 12.0 \rm \mu m$ and $8 \times 22.0 \rm \mu m$) to derive thermal parameters of (5111) Jacliff. In the fitting process, when we set the step width of thermal inertia to be 10 $\rm J m^{-2} s^{-1/2} K^{-1}$, we can finally obtain a best-fitting value for $\Gamma$ to be 0 $\rm J m^{-2} s^{-1/2} K^{-1}$. Thus, to derive a more accurate value of $\Gamma$, we again set the step width of thermal inertia to be 0.1 $\rm J m^{-2} s^{-1/2} K^{-1}$ to perform additional fittings with observations. However, as shown in Fig. \ref{chi2ga5111}, the derived thermal inertia still remains $0_{-0}^{+15}$ $\rm J m^{-2} s^{-1/2} K^{-1}$ with a corresponding $\chi_{\rm min}^2$ 3.583 and a roughness fraction $0.00_{-0.00}^{+0.40}$. Here it should be emphasized that the derived thermal inertia is given in $\rm 3-\sigma$ confidence level. Although the $\chi_{\rm min}^2$ is related to thermal inertia of 0 $\rm J m^{-2} s^{-1/2} K^{-1}$, it does not mean the value of $\Gamma$ should be zero, but suggests that the probability of thermal inertia between $0\sim15$ $\rm J m^{-2} s^{-1/2} K^{-1}$ is about 99.7\%. We derive the effective diameter $5.302_{-0.397}^{+0.237}$ km, which produces a geometric albedo $0.523_{-0.044}^{+0.088}$. The results of $p_{\rm v}$ and $D_{\rm eff}$ are slightly different from those of \citet{masiero2014} from NEATM. Figs. \ref{w2thli5111} and \ref{w34thli5111} show the thermal light curves at W2, W3 and W4, respectively, and Fig. \ref{fmobsratio5111} displays the ratio of observed flux and theoretical flux.

\subsection{(7001) Neother}

\begin{figure}
        \includegraphics[width=9cm,height=4.8cm]{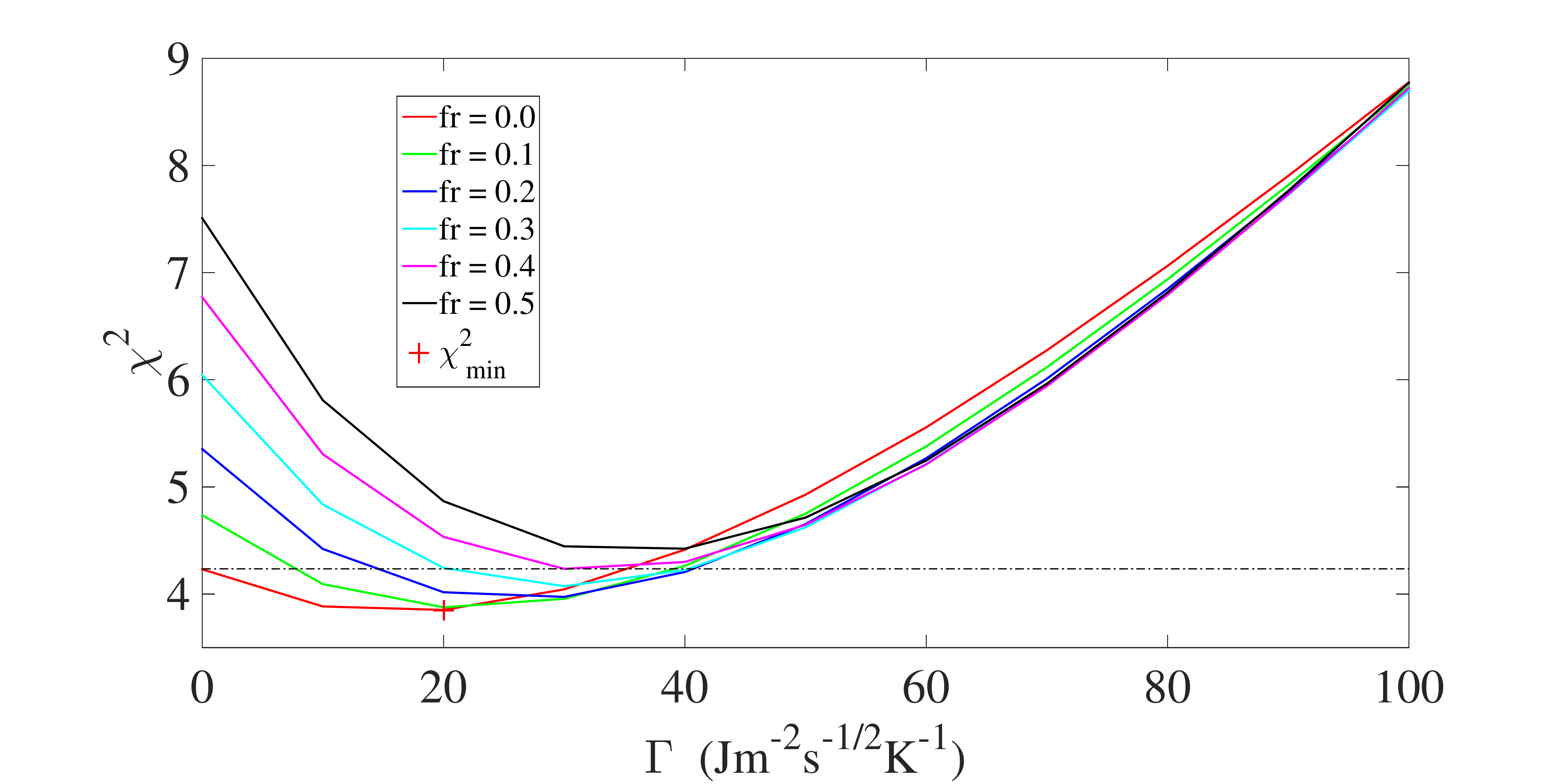}
    \caption{$\Gamma -\chi^2$ profile of (7001) Neother.}
    \label{chi2ga7001}
\end{figure}

\begin{figure}
        \includegraphics[width=9cm,height=15cm]{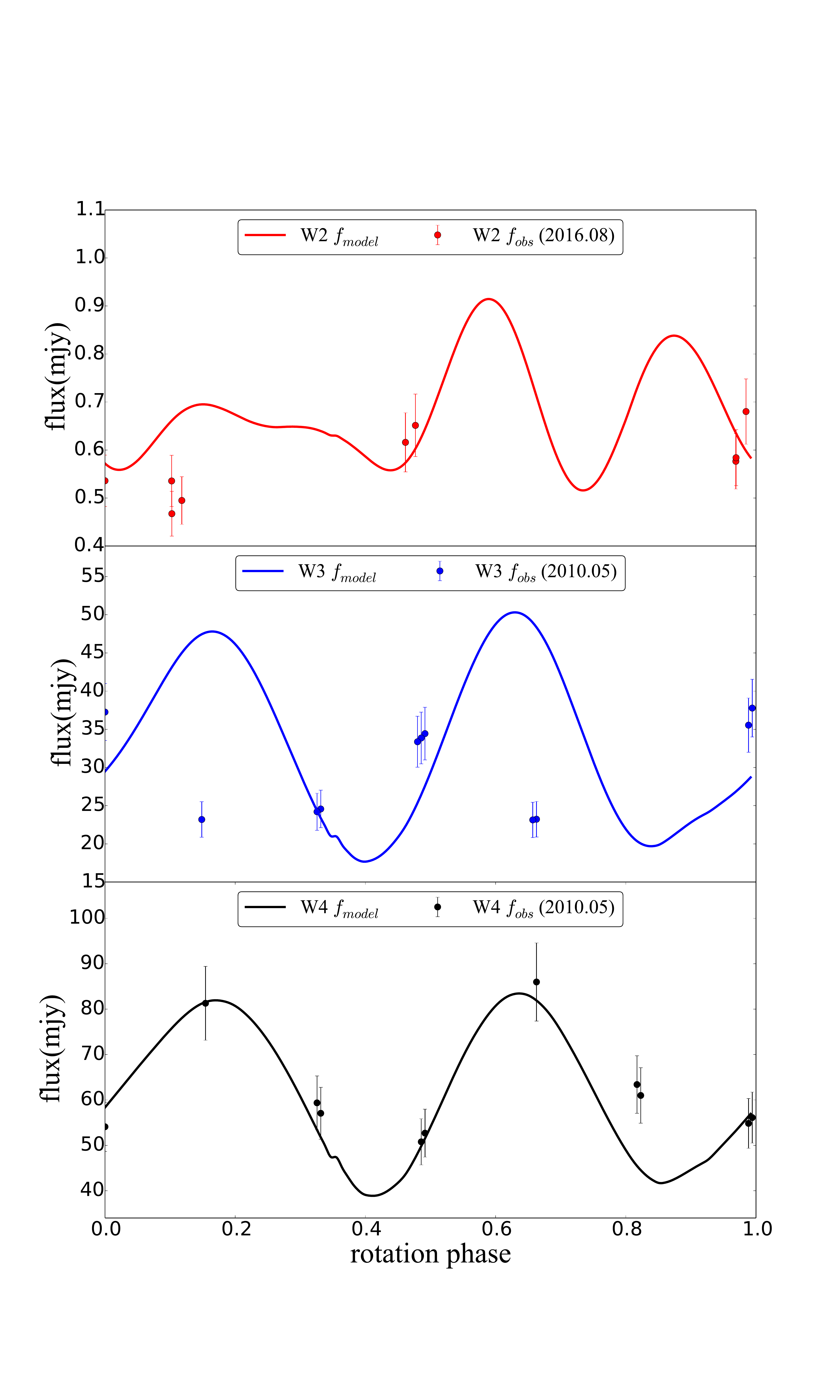}
    \caption{W2, W3 and W4 thermal light curves of (7001) Neother.}
    \label{w234thli7001}
\end{figure}

\begin{figure}
        \includegraphics[width=9.5cm,height=6cm]{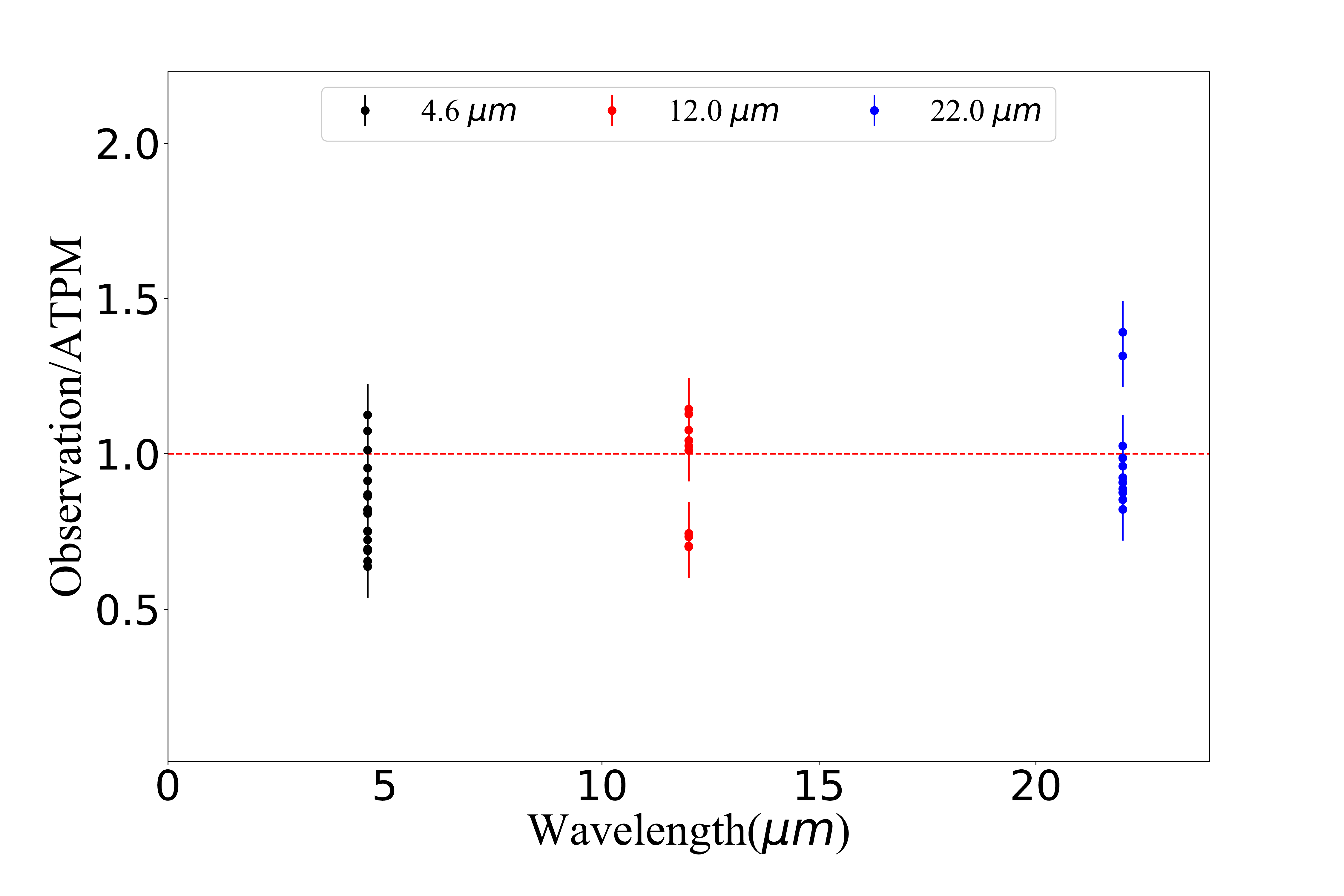}
    \caption{Observation/ATPM ratio of (7001) Neother.}
    \label{fmobsratio7001}
\end{figure}

Asteroid (7001) Neother orbits the Sun with the orbital period of 3.67 years. It has a perihelion distance of 2.023 AU and an aphelion distance  2.739 AU. Until now, the spectral type of this asteroid remains unknown.   \citet{warner2009} provided the rotation period 9.581 hours. Using 40 WISE observations ($18 \times 4.6 \rm \mu m$, $11\times 12.0 \rm \mu m$ and $11\times 22.0 \rm \mu m$) and ATPM,  we derive the thermal properties of (7001) Neother, i.e., the thermal inertia $\Gamma = 20_{-20}^{+21}$ $\rm J m^{-2} s^{-1/2} K^{-1}$, roughness fraction $f_r = 0.00_{-0.00}^{+0.40}$, geometric albedo $p_{\rm v} = 0.241_{-0.013}^{+0.034}$ and effective diameter $D_{\rm eff} = 5.923_{-0.378}^{+0.167}$ km. The outcomes of $p_{\rm v}$ and $D_{\rm eff}$ are close to those of \citet{masiero2011}, where $p_{\rm v} = 0.216 \pm 0.022$ and $D_{\rm eff} = 6.122 \pm 0.073$, respectively. The $f_{\rm r} - \Gamma$ and Observation/ATPM ratio are plotted in Figs. \ref{chi2ga7001} and \ref{fmobsratio7001} with a minimum  $\chi^2_{min}$ value 3.851. With the aid of the outcomes of $\Gamma$ and $f_{\rm r}$, we offer the thermal light curves for (7001) Neother at W2, W3 and W4 wavelengths (see Fig. \ref{w234thli7001}) with respect to two epochs of 2010.05.03 and 2016.08.31, respectively, indicating that the theoretical results from fitting accord with the observations.

\subsection{(9158) Plate}

\begin{figure}
        \includegraphics[width=9cm,height=4.8cm]{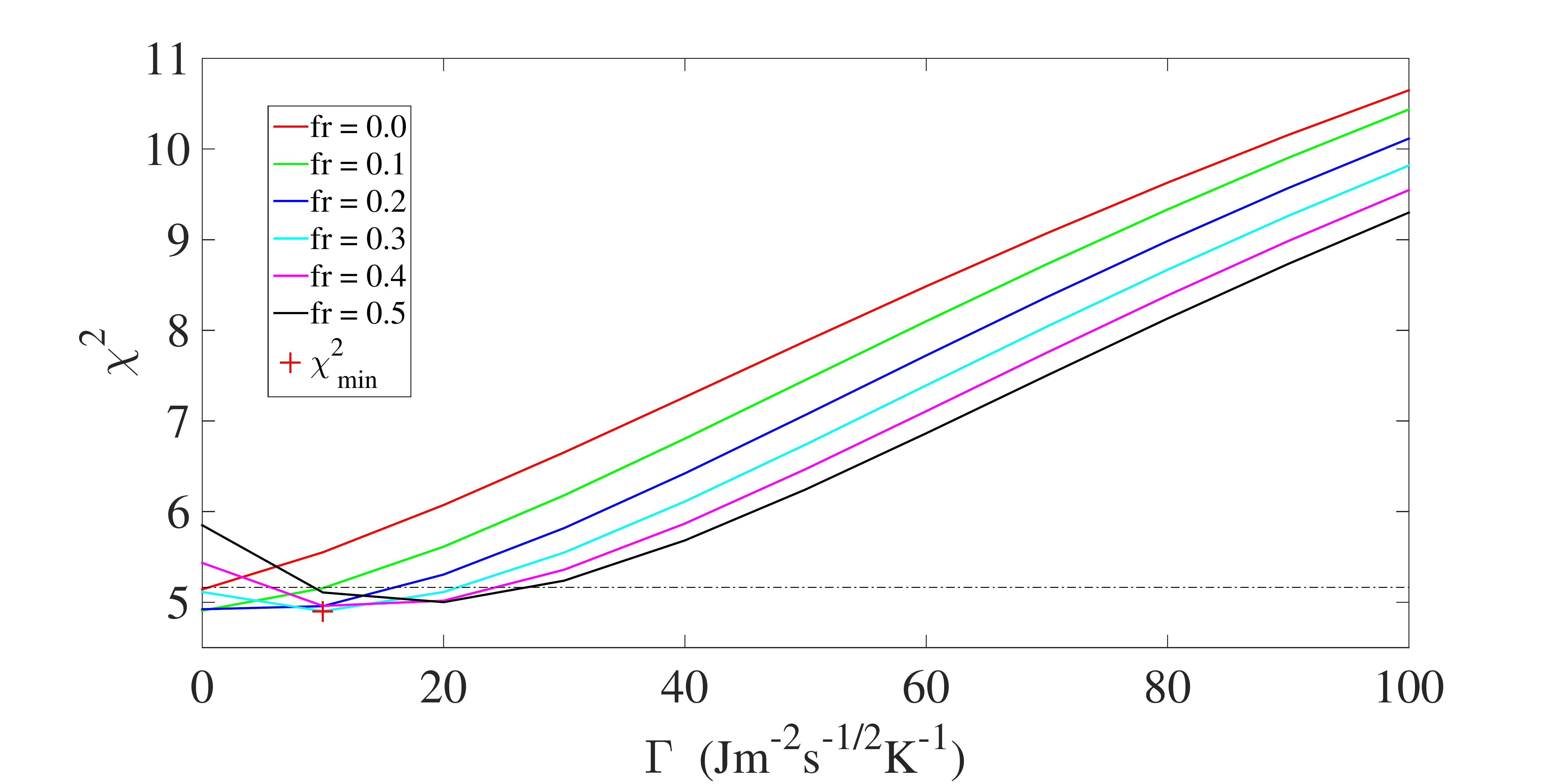}
    \caption{$\Gamma -\chi^2$ profile of (9158) Plate.}
    \label{chi2ga9158}
\end{figure}

\begin{figure}
        \includegraphics[width=9cm,height=12cm]{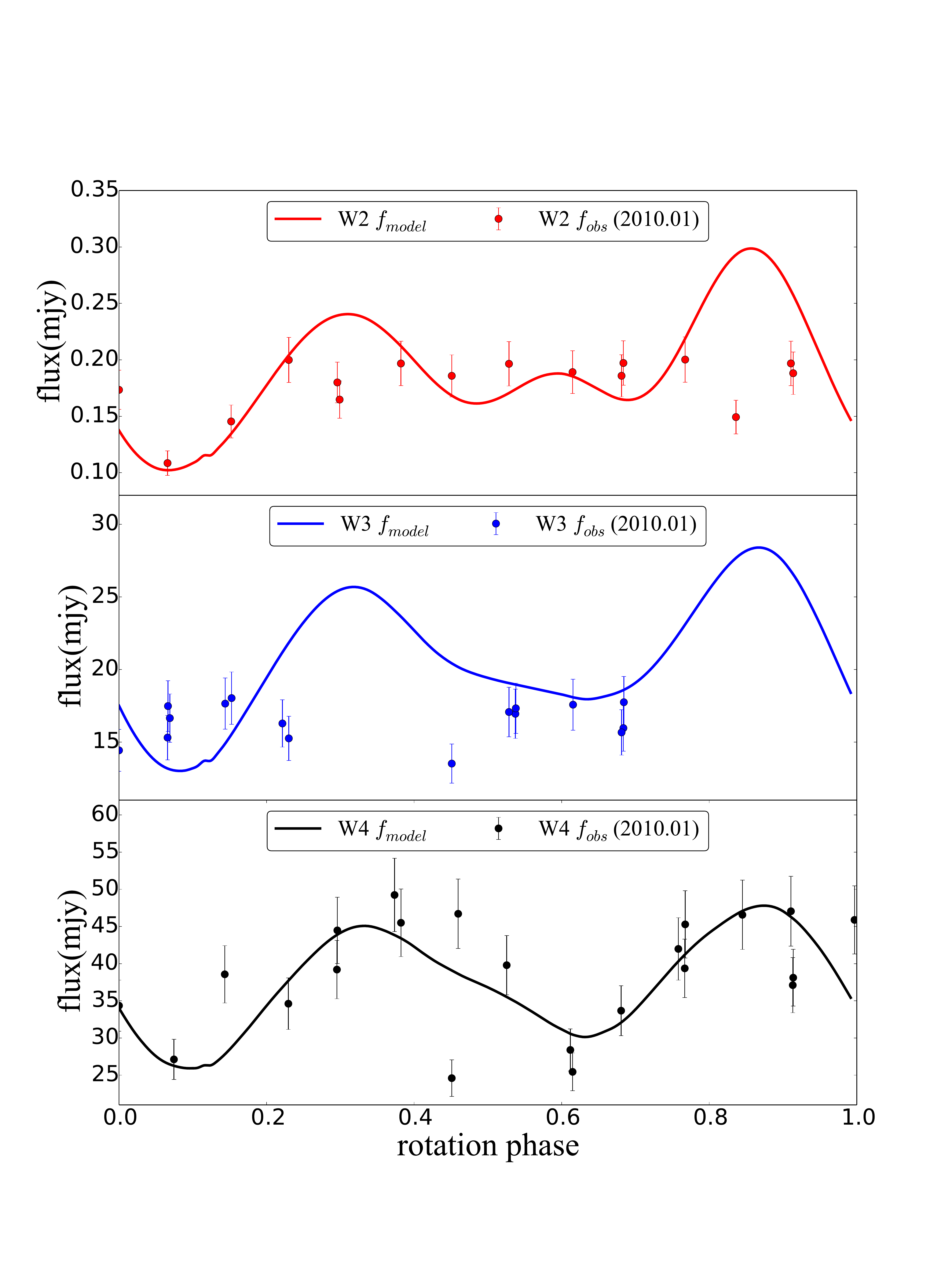}
    \caption{W2, W3 and W4 thermal light curves of (9158) Plate.}
    \label{w234thli9158}
\end{figure}

\begin{figure}
        \includegraphics[width=9.5cm,height=6cm]{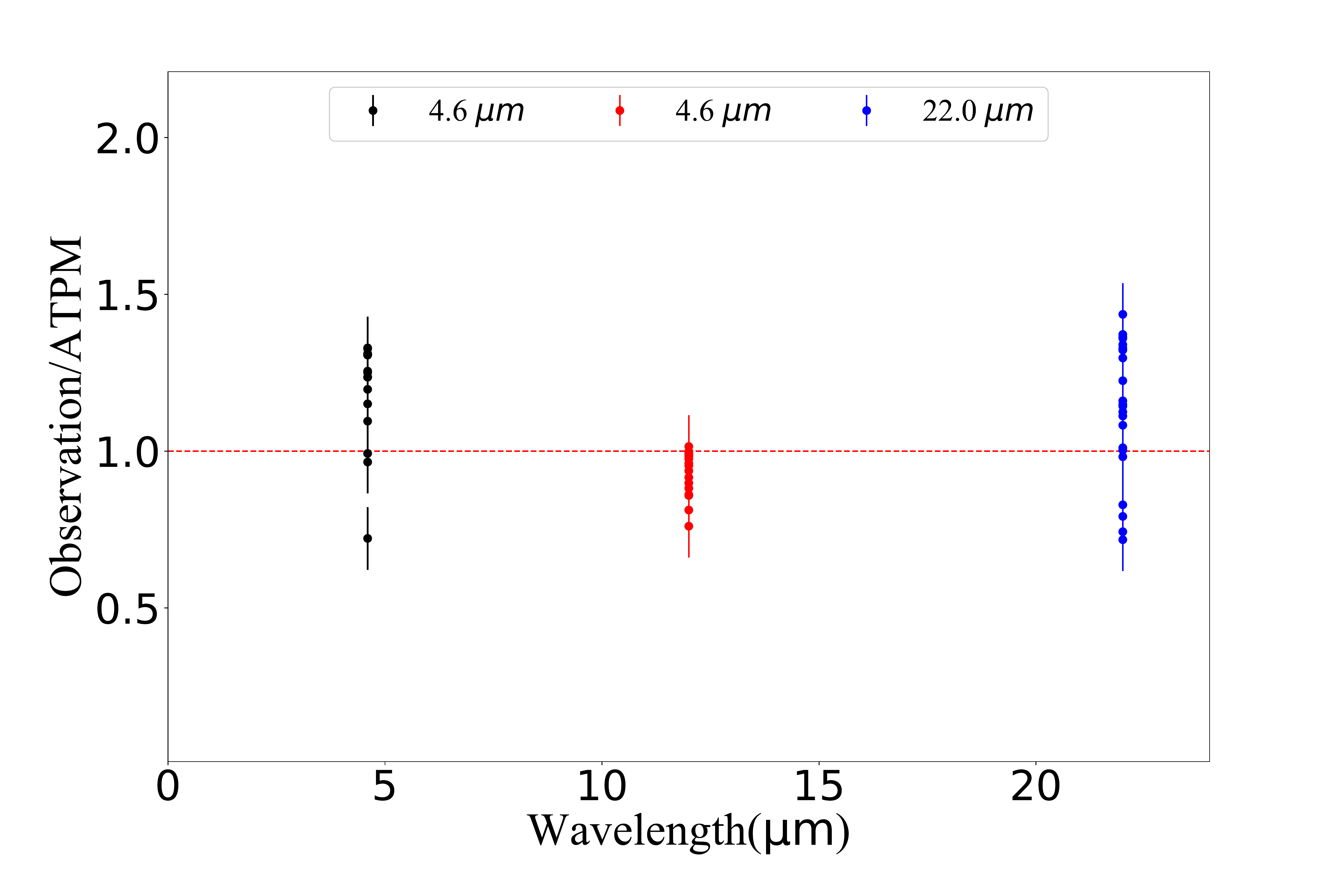}
    \caption{Observation/ATPM ratio of (9158) Plate at W2, W3, and W4 wavebands.}
    \label{fmobsratio9158}
\end{figure}

Asteroid (9158) Plate was discovered in 1984 and has an orbital period of 3.49 yr. According to the SDSS-based taxonomic classification developed by \citet{Carvano2010}, it is an $\rm SQ_{\rm p}$ asteroid. Using the combined data from Lowell Photometric Database and WISE, \citet{durech2018} constructed the 3D convex shape model of (9158) Plate. The rotation period and spin axis were also obtained to be 5.165 hr and $(119^\circ, -52^\circ)$ \citep{durech2018}. The NEATM results of diameter and geometric albedo from WISE observation are $4.734 \pm 0.125$ km and $0.314 \pm 0.075$ \citep{masiero2011}. In this study, we use 54 WISE/NEOWISE observations ($16 \times 4.6 \rm \mu m$, $16 \times 12.0 \rm \mu m$ and $22 \times 22.0 \rm \mu m$) to investigate the thermal parameters for this asteroid. As shown in Fig. \ref{chi2ga9158},  a low thermal inertia of $10_{-10}^{+18}$  $\rm J m^{-2} s^{-1/2} K^{-1}$ as well as a low roughness fraction of $0.30_{-0.30}^{+0.20}$ are obtained, with respect to a $\chi_{\rm min}^{2} = 4.902$. The geometric albedo is given to be $0.379_{-0.024}^{+0.026}$, and the diameter is $4.113_{-0.134}^{+0.137}$ km. Our result of the effective diameter is close to that of \citet{masiero2011}. Thermal light curves for (9158) Plate are displayed in Fig. \ref{w234thli9158}. As can be seen, our results provide a formally acceptable fit although the modeled fluxes seem to match the observations at W2 and W4 waveband rather than that of W3. Similarly, Fig. \ref{fmobsratio9158} shows that the value of Observation/ATPM ratios of  various wavelengths moves around 1.

\subsection{(12088) Macalintal}

\begin{figure}
        \includegraphics[width=9cm,height=4.8cm]{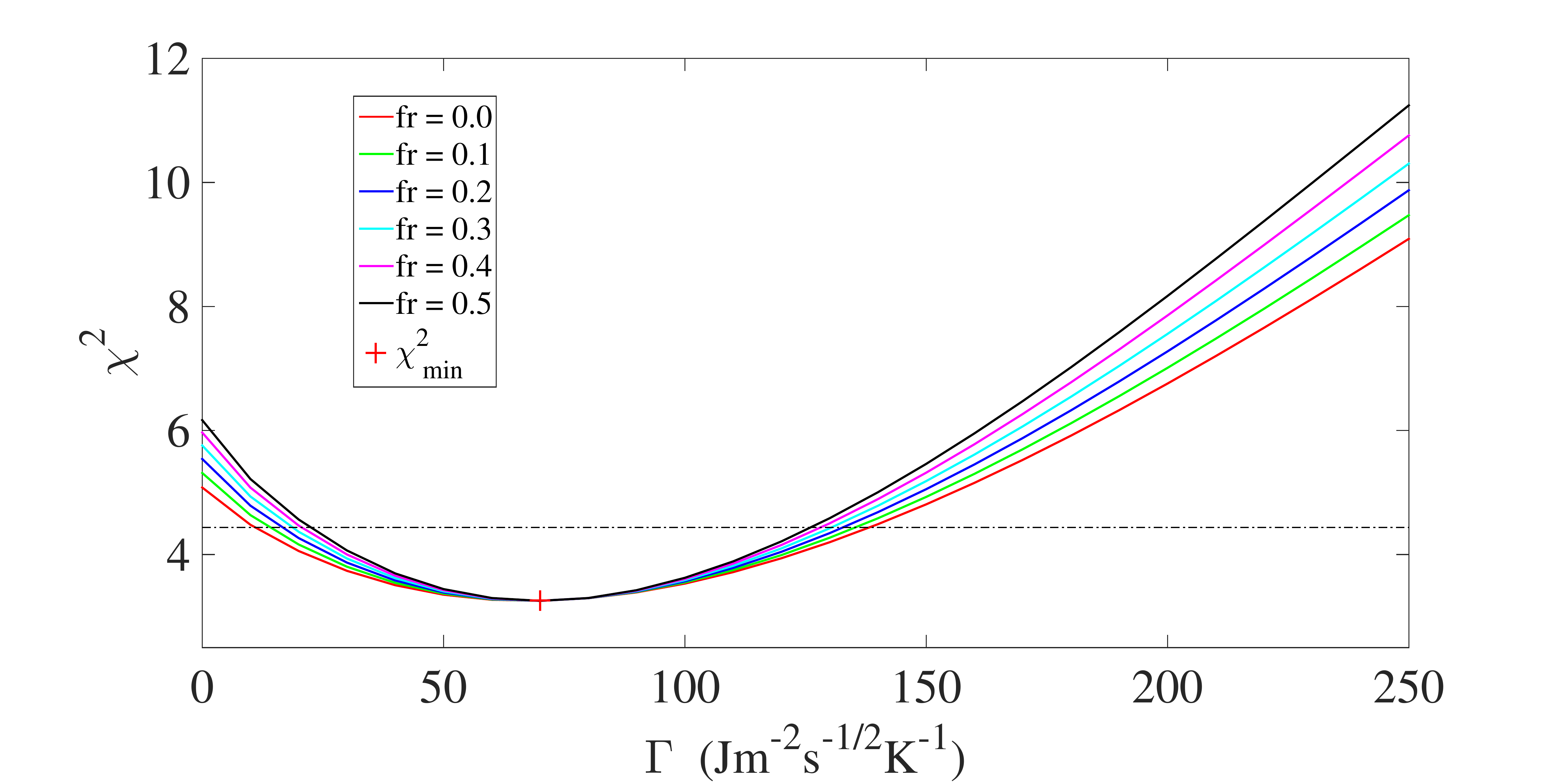}
    \caption{$\Gamma -\chi^2$ profile of (12088) Macalintal.}
    \label{chi2ga12088}
\end{figure}

\begin{figure}
        \includegraphics[width=9cm,height=10cm]{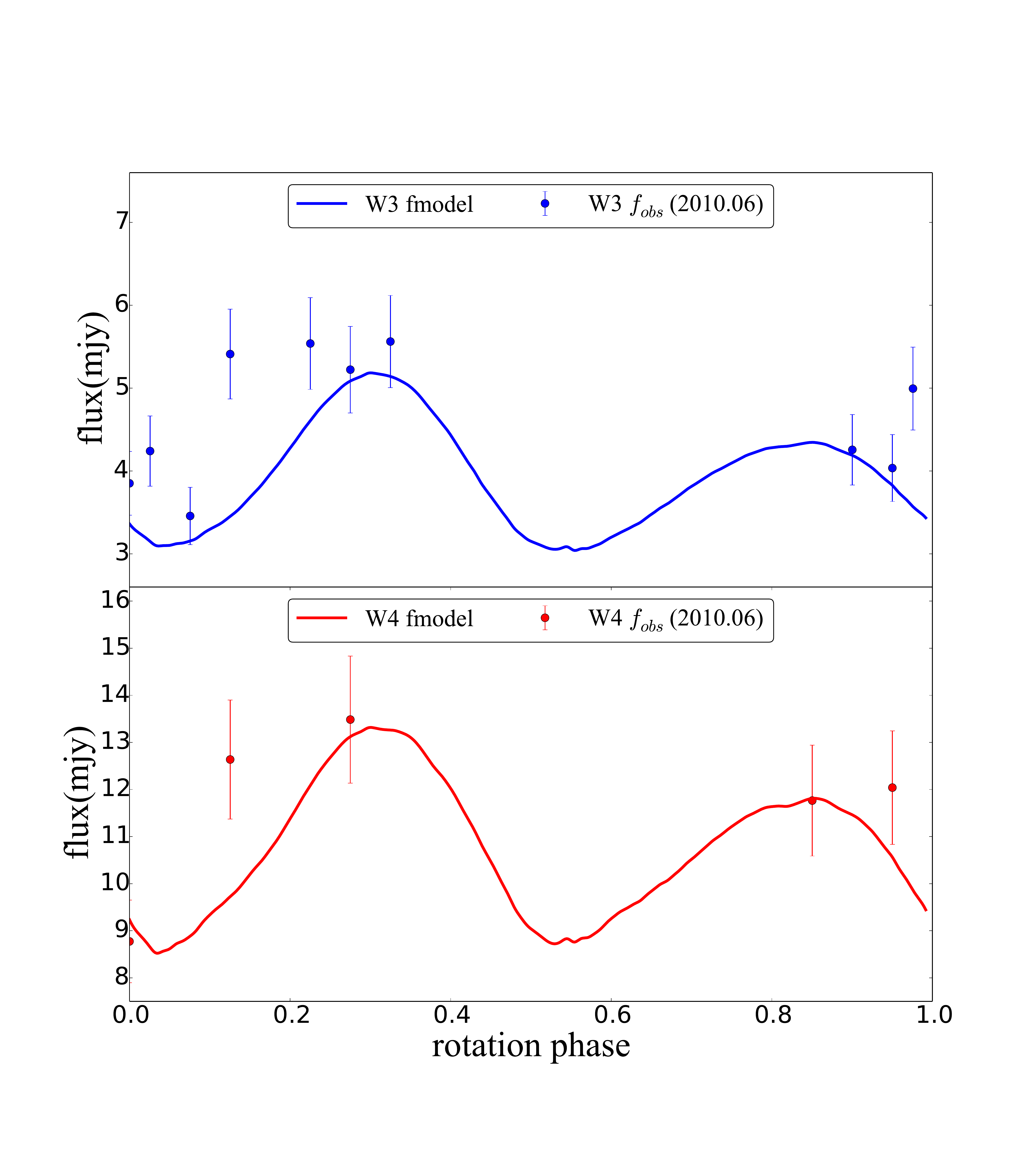}
    \caption{W3 and W4 thermal light curves of (12088) Macalintal.}
    \label{w34thli12088}
\end{figure}

\begin{figure}
        \includegraphics[width=9.5cm,height=6cm]{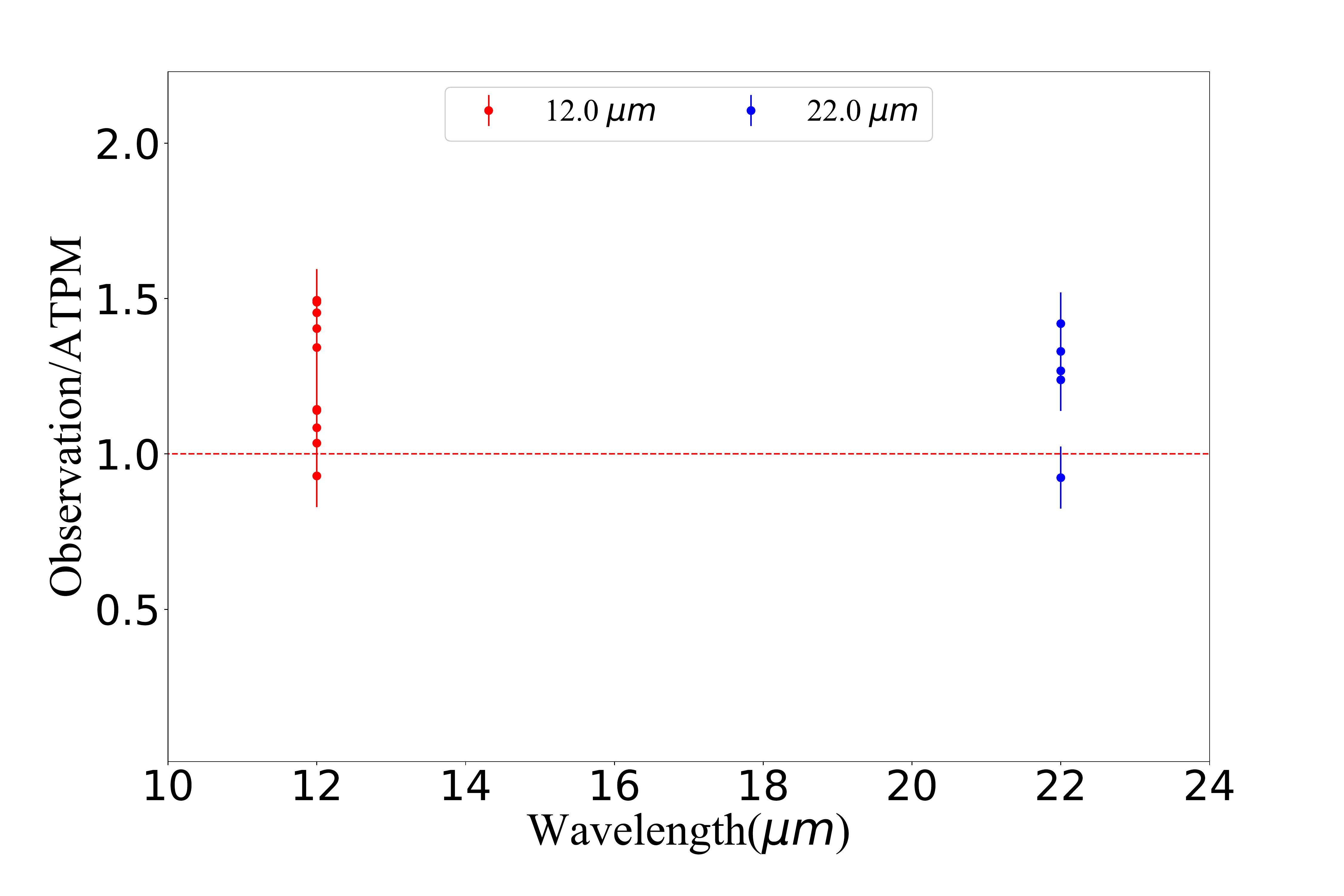}
    \caption{Observation/ATPM ratio of  (12088) Macalintal at two wavebands.}
    \label{fmobsratio12088}
\end{figure}

(12088) Macalintal was discovered by the Lincoln Observatory in 1998. In the SDSS-based taxonomic classification, it is a V-type asteroid \citep{Carvano2010}. \citet{durech2018} presented the rotation period of this asteroid to be 3.342 hr. Besides, using the WISE observation and NEATM, the geometric albedo and diameter of this asteroid are, respectively, 0.385 $\pm 0.097$ and $3.724 \pm 0.250$ km. In this work, we first collect the observations for our fitting procedure, but find that the fewer WISE data are available for this asteroid ($10 \times 12.0 \rm \mu m$ and $5 \times 22.0 \rm \mu m$). By performing the fitting,  we derive that the geometric albedo is $0.344_{-0.050}^{+0.120}$, the effective diameter $3.591_{-0.499}^{+0.293}$ km.  From Fig. \ref{chi2ga12088}, we can constrain the thermal inertia and roughness fraction to be  $70_{-53}^{+68}$  $\rm J m^{-2} s^{-1/2} K^{-1}$ and $0.00_{-0.00}^{+0.50}$, respectively. The uncertainties for $\Gamma$ and $f_{\rm r}$ are quite large because we simply adopt 15 observations during the fitting process. Using the derived $\Gamma$ and $f_{\rm r}$, thermal light curves and the observation/ATPM ratio for W3 and W4 wavebands are shown in Figs. \ref{w34thli12088} and \ref{fmobsratio12088}.  We can notice that the observed fluxes are generally larger than the theoretical results.

\subsection{(15032) Alexlevin}
\begin{figure}
        \includegraphics[width=9cm,height=4.8cm]{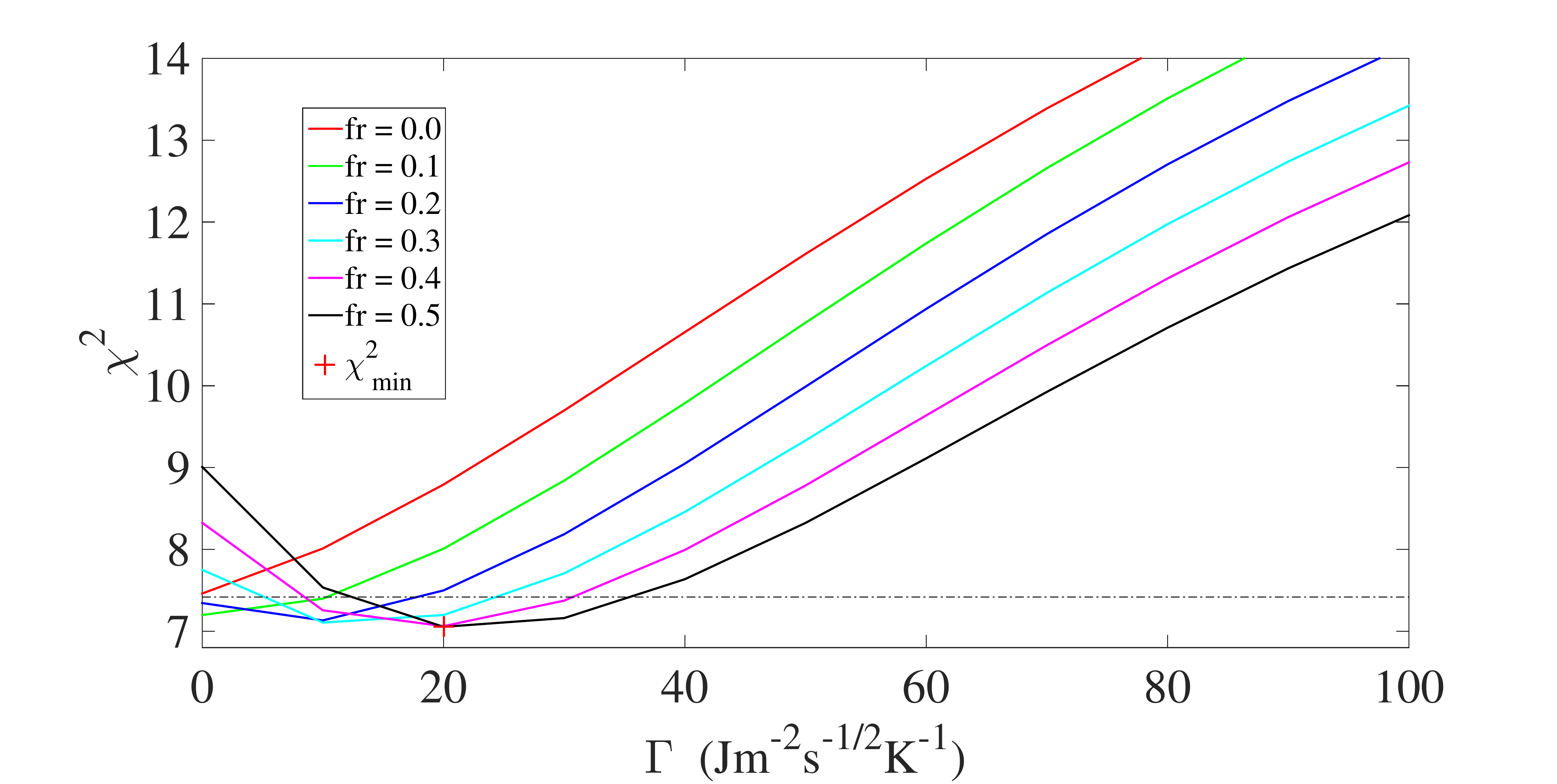}
    \caption{$\Gamma -\chi^2$ profile of (15032) Alexlevin.}
    \label{chi2ga15032}
\end{figure}

\begin{figure}
        \includegraphics[width=9cm,height=12cm]{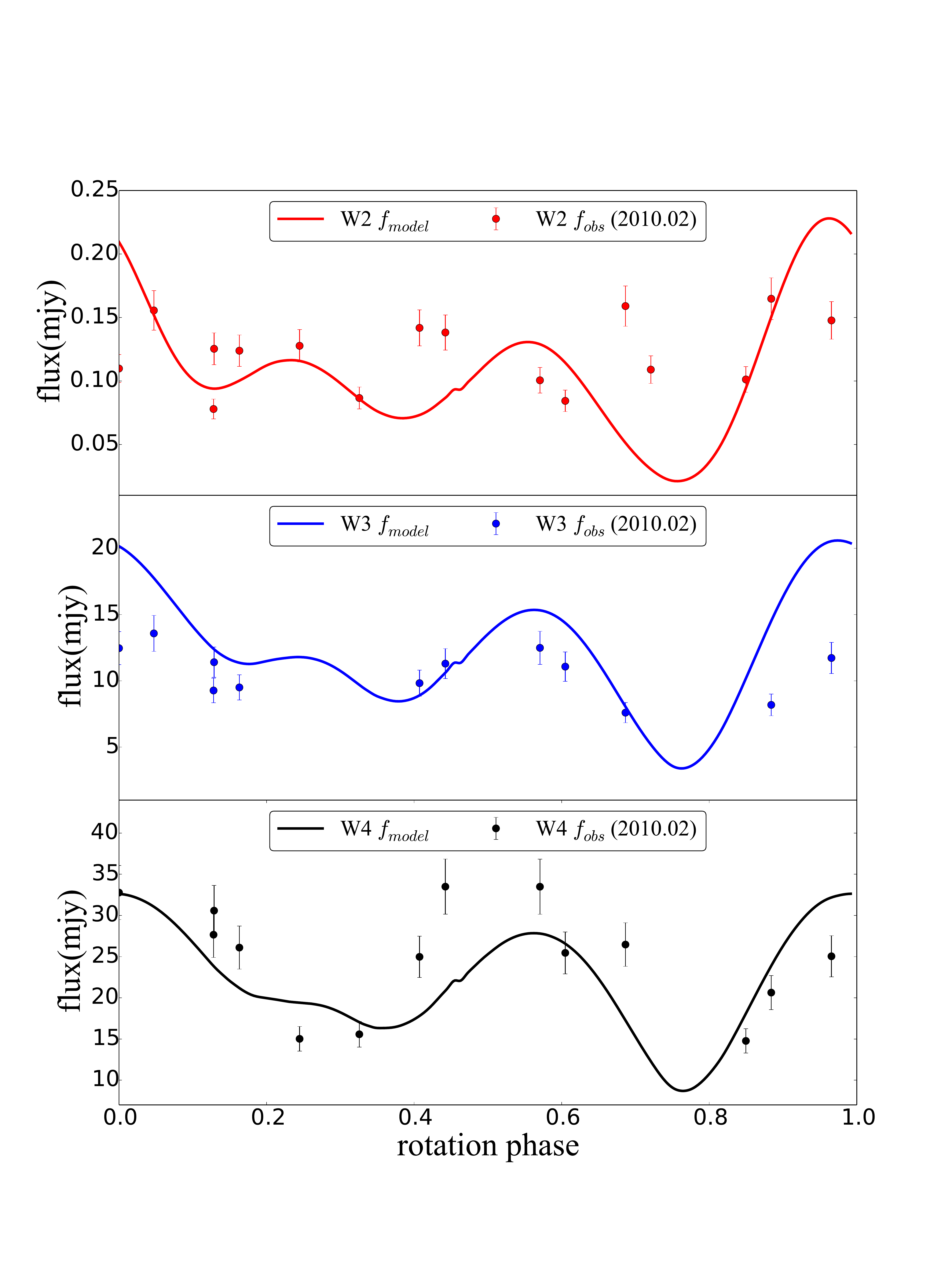}
    \caption{W2, W3 and W4 thermal light curves of (15032) Alexlevin.}
    \label{w234thli15032}
\end{figure}

\begin{figure}
        \includegraphics[width=9.5cm,height=6cm]{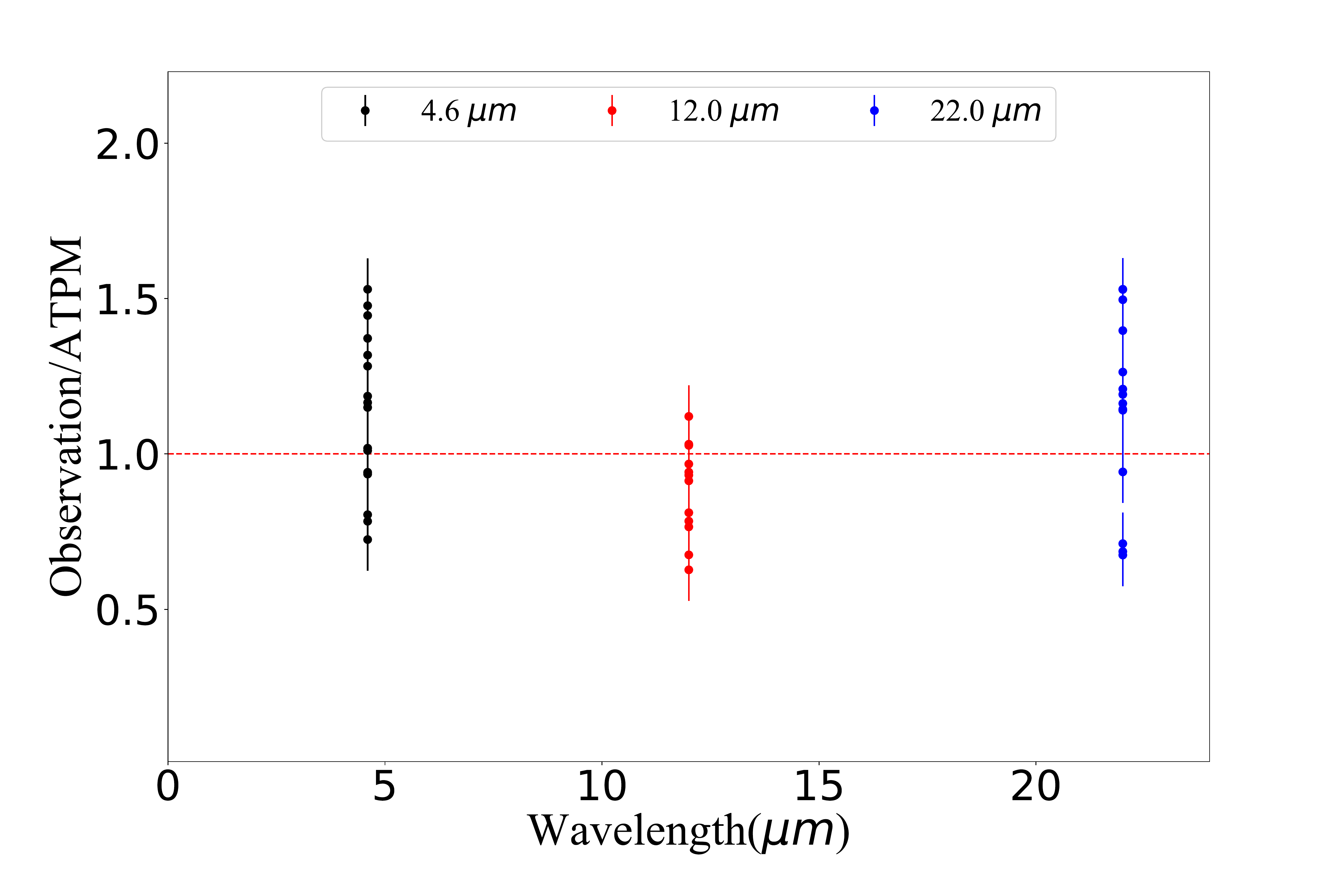}
    \caption{Observation/ATPM ratio of (15032) Alexlevin at three wavebands.}
    \label{fmobsratio15032}
\end{figure}

Asteroid (15032) Alexlevin was discovered in 1998. In the Moving Objects VISTA (MOVIS) catalogue and SDSS-based taxonomic classification, it is recogonized as a V-type asteroid \citep{Carvano2010,lic2017}. This asteroid has an orbital period of 3.66 yr. \citet{masiero2011} derived the diameter and albedo of (15032) Alexlevin to be $3.579 \pm 0.059$ km and $0.288 \pm 0.048$. In this work, we adopt the absolute magnitude of 14.5 from MPC and the rotation period of 4.406 hr from \citet{durech2018} in our ATPM model, and 42 WISE/NEOWISE observations ($16 \times 4.6 \rm \mu m$, $12 \times 12.0 \rm \mu m$ and $14 \times 22.0 \rm \mu m$) are utilized in the fitting. By comparing the theoretical flux and the observations, we further obtain the geometric albedo of $0.349_{-0.025}^{+0.024}$ and the effective diameter of $2.832_{-0.093}^{+0.154}$ km. The derived diameter is smaller than that of \citet{masiero2011}. As shown in Fig. \ref{chi2ga15032}, we can place constraints on the thermal inertia of asteroid (15032) Alexlevin to be $20_{-20}^{+15}$ $\rm J m^{-2} s^{-1/2} K^{-1}$ and a roughness fraction of $0.50_{-0.40}^{+0.00}$. The ratio between observed and theoretical fluxes for 3 wavebands are plotted in Fig. \ref{fmobsratio15032}.  In addition, the 3-bands thermal light curves of (15032) Alexlevin are exhibited in Fig. \ref{w234thli15032}. As can be seen, our theoretical flux can fit well with the observations.

\section{Discussion of the radiometric results}\label{discu}

In this work, we present the first attempt to determine the thermal parameters of 10 Vesta family asteroids by using ATPM and combined with the thermal infrared observations from IRAS, AKARI and WISE/NEOWISE. Our results are summarized in Table.\ref{resultsall}. All of the Vesta family asteroids have thermal inertia less than 100 $\rm J m^{-2} s^{-1/2} K^{-1}$ as well as relatively low roughness fractions, which may suggest that they have undergone a long time surface evolution process. It should be noticed that, among the 10 Vesta family members, (1906) Neaf, (2511) Patterson, (5111) Jacliff, (12088) Macalintal, and (15032) Alexlevin are V-type asteroids, they are also called "Vestoids", and the other 5 members are non-Vestoids. We obtain the mean value of $p_{\rm v}$ for the 5 Vestoids to be 0.328 and is very close to the median value ($0.362\pm0.100$) of V-type (Bus-Demeo taxonomy) asteroids in \citet{mainzer2011}. While for non-Vestoids, this value is 0.300. As mentioned above, the derived $p_{\rm v}$ for (63) Ausonia (Sa/Sw) and (556) Phyllis (S) are $0.180_{-0.009}^{+0.010}$ and $0.209_{-0.010}^{+0.010}$, respectively, which is similar to the median $p_{\rm v}$ of these two spectral types obtained by \citet{mainzer2011}. As for the SQp type asteroid (9158) Plate, we have derived the value of $p_{\rm v}$ to be $0.379_{-0.024}^{+0.026}$, which is inside the geometric albedo range ($0.062\sim0.617$) of SQp type asteroid obtained by \citet{mainzer2012}. Additionally, (3281) Maupertuis has a geometric albedo of $0.484_{-0.074}^{+0.051}$ and this value is within the range of V-type asteroids's $p_{\rm v}$ in \citet{mainzer2011} but is larger than their median value. While for (7001) Neother, the derived $p_{\rm v}$ is $0.241_{0.013}^{+0.034}$, which is comparable with the geometric albedo of S-type asteroids obtained by \citet{mainzer2011,mainzer2012}. Except for asteroid (63) Ausonia and (556) Phyllis, other Vesta family asteroids we studied have effective diameters smaller than 10 km. In the following, we will make a brief discussion on thermal nature for the Vesta family asteroids based on our derived results.

\makeatletter\def\@captype{table}\makeatother
\begin{table*}
        \centering
        \caption{Derived thermal characteristics of 10 Vesta family asteroids in this work}
        \label{resultsall}
        \begin{tabular}{lccccc|cc} % four columns, alignment for each
                \hline
        \specialrule{0em}{2.0pt}{2.0pt}
                Asteroid & $\Gamma$ ($\rm J m^{-2} s^{-1/2} K^{-1}$)&$ f_{\rm r}$&$D_{\rm eff}$ (km)&$p_{\rm v}$&$\chi^2_{\rm min}$&$p_{\rm v}^*$&$D_{\rm eff}^*$ (km)\\
                \hline
                \specialrule{0em}{2.0pt}{2.0pt}
                (63) Ausonia & $50_{-24}^{+12}$ & $0.50_{-0.30}^{+0.00}$ & $94.595_{-2.483}^{+2.343}$& $0.189_{-0.009}^{+0.010}$ &3.102 &$0.168 \pm 0.008$&$102.975 \pm 2.754$\\
                \specialrule{0em}{2.0pt}{2.0pt}
                (556) Phyllis& $30_{-11}^{+12}$ & $0.40_{-0.20}^{+0.10}$& $35.600_{-0.901}^{+0.883}$& $0.209_{-0.010}^{+0.010}$ &2.715&$0.185 \pm 0.011$& $37.810 \pm 1.100$\\
                \specialrule{0em}{2.0pt}{2.0pt}
                (1906) Neaf& $70_{-16}^{+19}$ &$0.50_{-0.20}^{+0.00}$  & $7.728_{-0.211}^{+0.407}$& $0.246_{-0.024}^{+0.014}$ &8.748&$0.228 \pm 0.047$&$8.057 \pm 0.083$\\
                \specialrule{0em}{2.0pt}{2.0pt}
                (2511) Patterson & $90_{-43}^{+58}$ & $0.00_{-0.00}^{+0.50}$ & $9.034_{-1.128}^{+0.997}$& $0.180_{-0.034}^{+0.055}$ &3.853&$0.287 \pm 0.039$&$7.849 \pm 0.174$\\
                \specialrule{0em}{2.0pt}{2.0pt}
                (3281) Maupertuis & $60_{-31}^{+58}$ & $0.50_{-0.30}^{+0.00}$ & $5.509_{-0.270}^{+0.447}$& $0.484_{-0.074}^{+0.051}$ &3.698&$0.489 \pm 0.020$&$5.482 \pm 0.043$\\
                \specialrule{0em}{2.0pt}{2.0pt}
                (5111) Jacliff & $0_{-0}^{+15}$ & $0.00_{-0.00}^{+0.40}$ & $ 5.302_{-0.397}^{+0.237}$ & $0.523_{-0.044}^{+0.088}$ &3.583&$0.425 \pm 0.039$&$6.447 \pm 0.129$\\
                \specialrule{0em}{2.0pt}{2.0pt}
                (7001) Neother&$20_{-20}^{+21}$ & $0.00_{-0.00}^{+0.40}$ & $5.923_{-0.378}^{+0.167}$& $0.241_{-0.013}^{+0.034}$ &3.851&$0.216 \pm 0.022$&$6.122 \pm 0.073$\\
                \specialrule{0em}{2.0pt}{2.0pt}
            (9158) Plate&$10_{-10}^{+19}$ & $0.30_{-0.30}^{+0.20}$ & $4.113_{-0.134}^{+0.137}$& $0.379_{-0.024}^{+0.026}$ &4.902&$0.314 \pm 0.075$&$4.734 \pm 0.125$\\
            \specialrule{0em}{2.0pt}{2.0pt}
            (12088) Macalintal&$70_{-53}^{+68}$ & $0.00_{-0.00}^{+0.50}$ & $3.591_{-0.499}^{+0.293}$& $0.340_{-0.052}^{+0.051}$ &3.253&$0.385 \pm 0.097$&$3.724 \pm 0.250$\\
            \specialrule{0em}{2.0pt}{2.0pt}
            (15032) Alexlevin&$20_{-20}^{+15}$ & $0.50_{-0.40}^{+0.00}$ & $2.832_{-0.093}^{+0.154}$& $0.349_{-0.035}^{+0.024}$ &7.054&$0.288 \pm 0.048$&$3.579 \pm 0.059$\\
                \hline
        \end{tabular}
        \begin{tablenotes}
    \footnotesize
    \item[]Note: All the results of thermal properties are in SI units, where $\Gamma$ is the thermal inertia, $ f_{\rm r}$ is the roughness fraction, $D_{\rm eff}$ is the effective diameter and $p_{\rm v}$ is the geometric albedo. $p_{\rm v}^*$ and $D_{\rm eff}^*$ represent the geometric albedo and effective diameter outcomes of \citet{mainzer2011}, \citet{masiero2012} and \citet{masiero2014}.
    \end{tablenotes}
\end{table*}

\begin{figure*}
        \includegraphics[width=20cm,height=12cm]{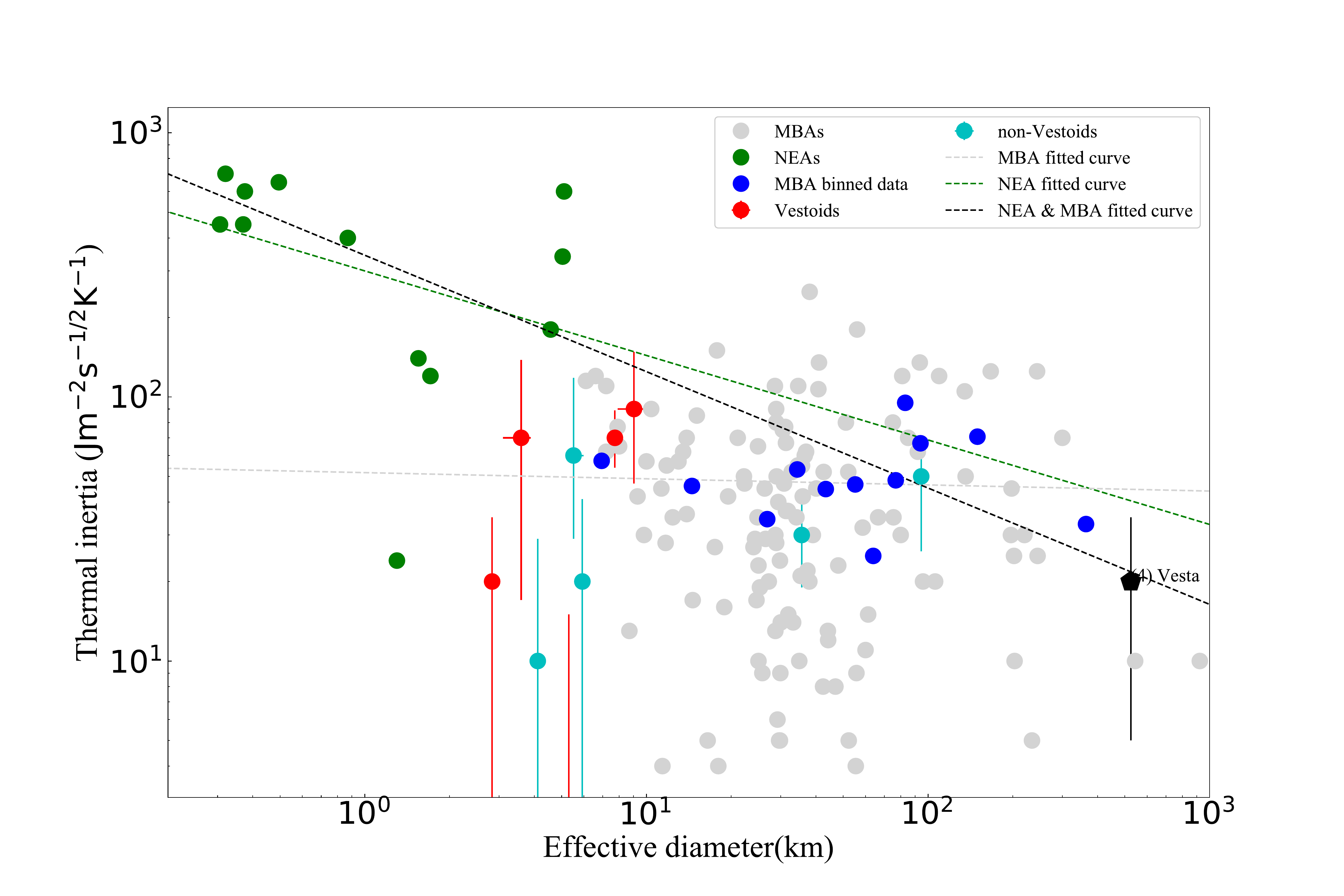}
    \caption{Thermal inertia as a function of effective diameter. The gray and green dots represent the results of MBAs and NEAs from \citet{delbo2015} and \citet{hanus2018}, the blue dots are the binned data of MBAs, while the red and cyan dots illustrate our results of Vestoids and non-Vestoids. The gray and green dashed lines show the relationship of $\Gamma-D_{\rm eff}$ by fitting for MBAs and NEAs, respectively. Note that the black dashed line shows that of the NEAs and MBAs (binned data), indicating that the thermal inertia decreases as the effective diameter increases.}
    \label{thdeff}
\end{figure*}

\begin{figure*}
        \includegraphics[width=18cm,height=12cm]{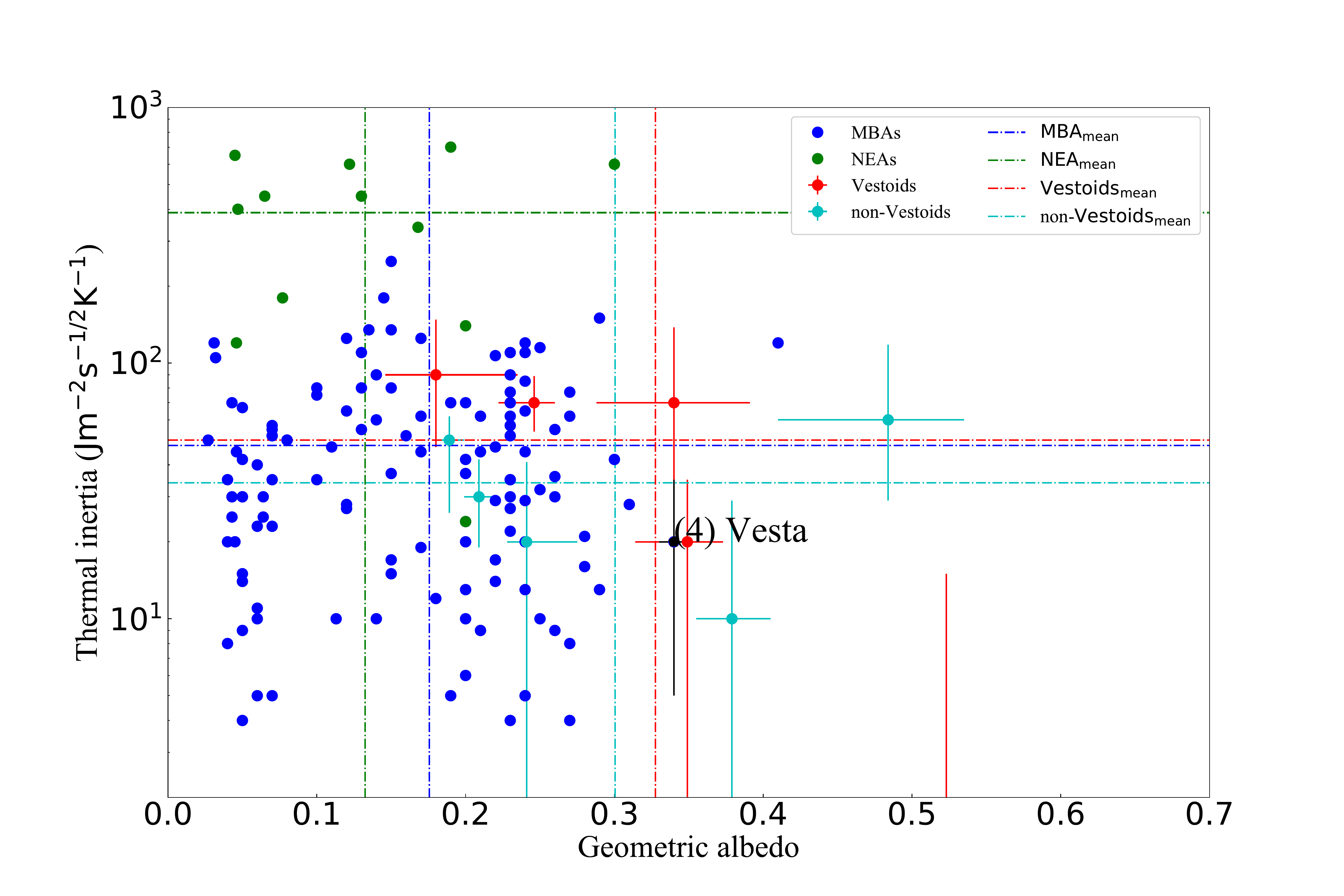}
    \caption{Thermal inertia as a function of geometric albedo. The blue and green dots represent the results of MBAs and NEAs of \citet{delbo2015} and \citet{hanus2018}, whereas the red and cyan dots with error bars indicate those of the Vestoids and non-Vestoids of this work.}
    \label{thpv}
\end{figure*}

\begin{figure*}
        \centering
        \includegraphics[scale=0.38]{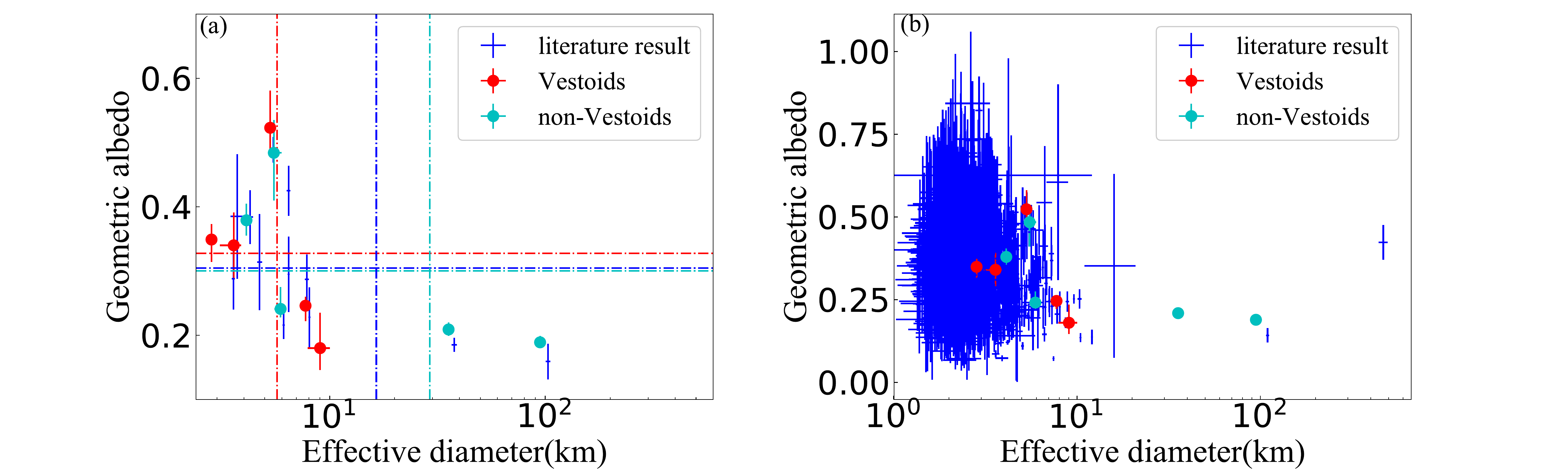}
    \caption{(a)The geometric albedo $p_{\rm v}$ with effective diameter $D_{\rm eff}$ for the Vesta family asteroids. Red circles with error bars represent our results, while the blue dots with error bars represent those of the literature. The horizontal and vertical dash lines stand for the mean value of geometric albedo and effective diameter, respectively. (b) The updated $p_{\rm v}$ and $D_{\rm eff}$ profile from \citet{masiero2011} for nearly 1000 asteroids (blue dots with error bars). Our results are given by red circles with error bars. }
    \label{pvdeffall}
\end{figure*}

\begin{figure}
        \includegraphics[width=9.5cm,height=6cm]{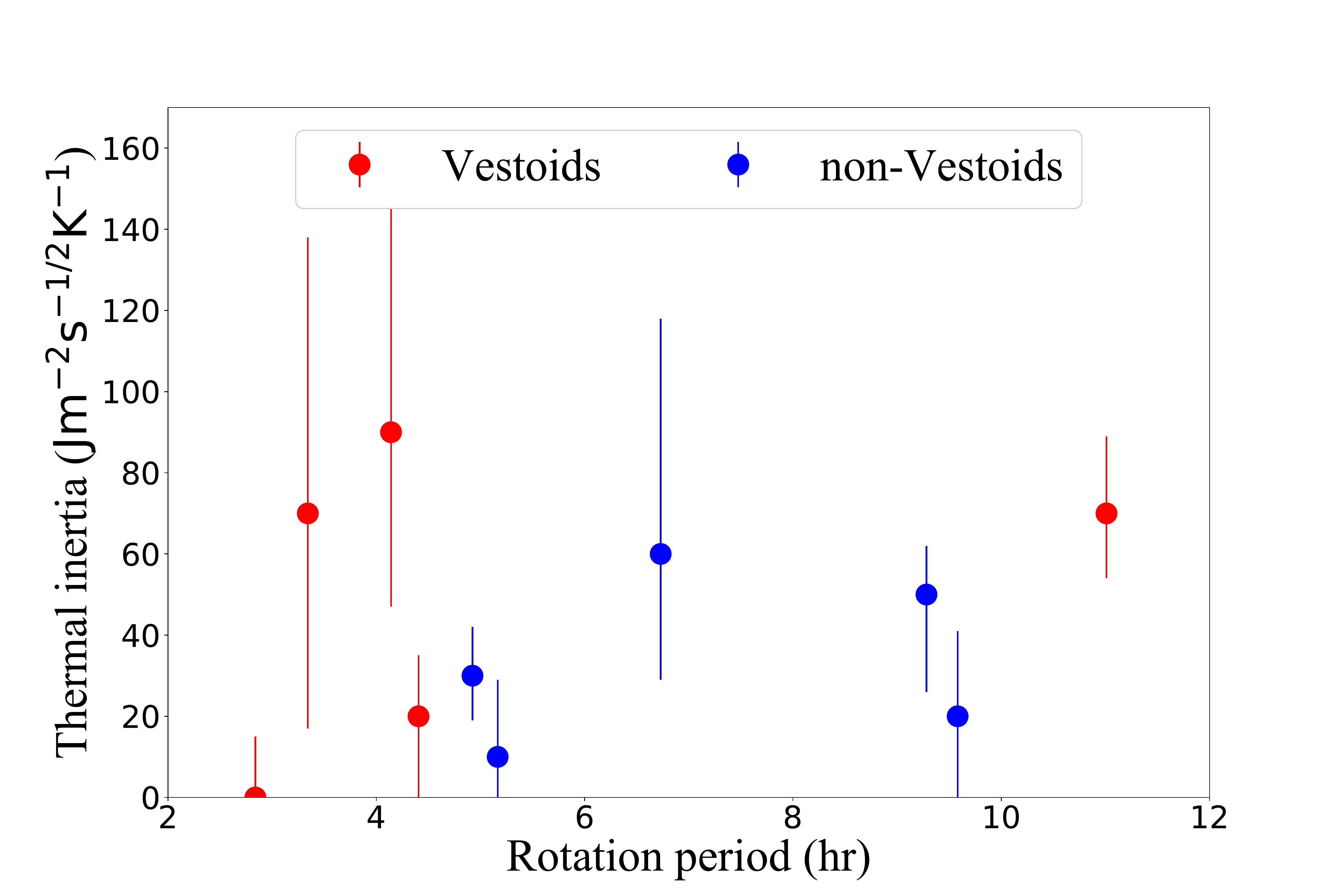}
    \caption{Thermal inertia as a function of rotation period. No expected increasing trend for 10 Vesta family members.}
    \label{throt}
\end{figure}

\begin{figure}
        \includegraphics[width=9.5cm,height=6cm]{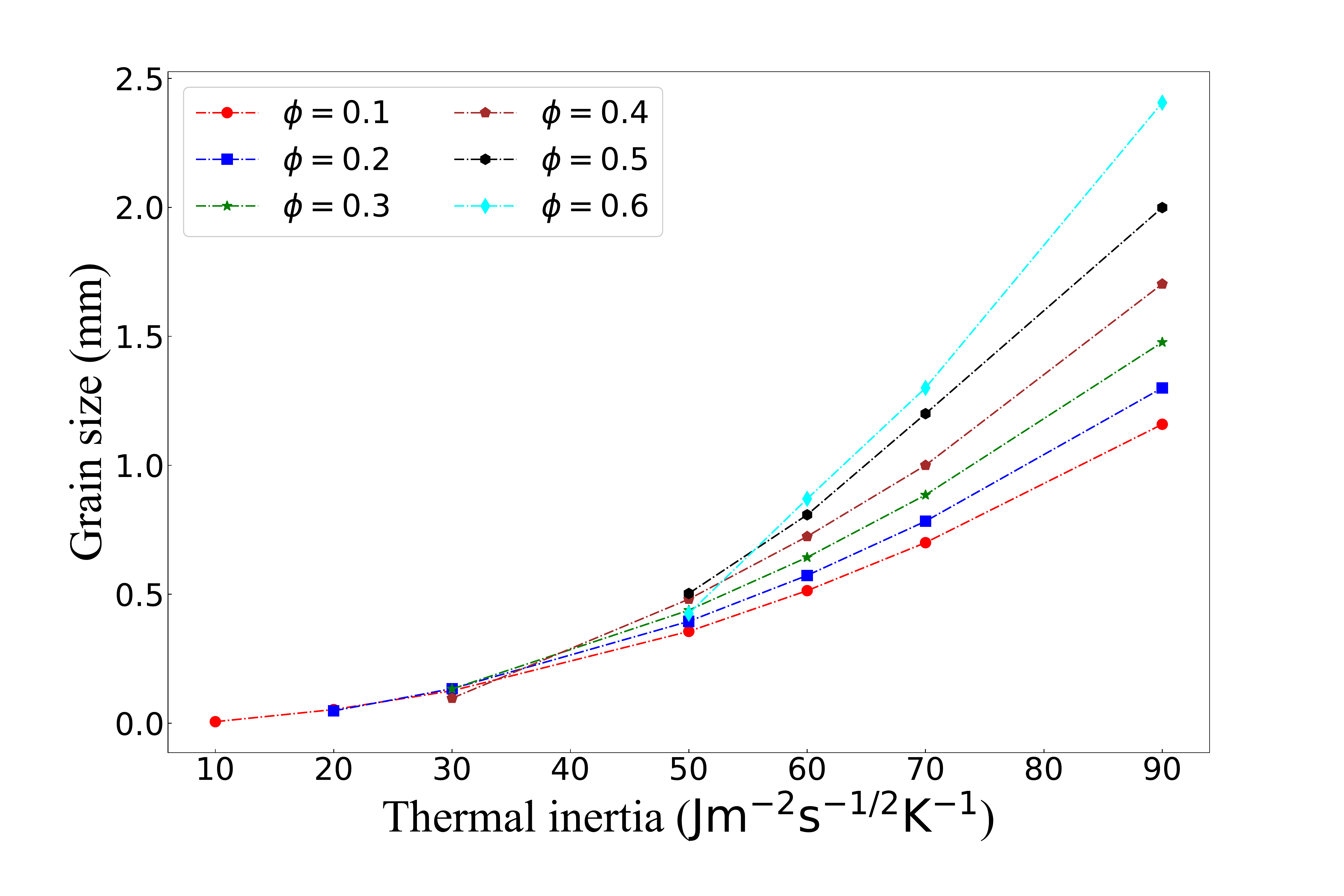}
    \caption{Regolith grain size of the 10 Vesta family asteroids as a function of thermal inertia, where different colors represent the different volume filling factor \citet{gundlach2013}. The values of grain size give an obvious increasing trend with the increasing thermal inertia.}
    \label{thgs}
\end{figure}

\subsection{Thermal inertia, effective diameter and geometric albedo}

\citet{delbo2007} investigated the relationship between thermal inertia and effective diameter and provided the power law formula $ \Gamma = d_0 D^{-\xi}$, where a linear regression gives the best-fitting parameters of $d_0$ and $\xi$ to be $300 \pm 47$ and $0.48 \pm 0.04$. Furthermore, \citet{delbo2008} showed the thermal inertia of main-belt asteroids by using IRAS data, and they obtained the values of $\rm \xi$ to be $1.4 \pm 0.2$ for MBAs and $0.32 \pm 0.09$ for NEAs, respectively. In addition, \citet{hanus2018} presented thermal parameters of $\sim$ 300 main-belt asteroids. Here, we combine our results of $\Gamma$ and $D_{\rm eff}$ with those of \citet{delbo2015} and \citet{hanus2018} to further explore the relationship between thermal inertia and effective diameter. In Fig. \ref{thdeff}, we present our results given by red (Vestoids) and cyan (non-Vestoids) dots with error bars, where the values of thermal inertia and effective diameter from \citet{delbo2015} and \citet{hanus2018} are shown in gray (for MBAs) and green (for NEAs) dots (in order to make the diagram more clearer, we do not plot their error bars for each value of $\Gamma$ and $D_{\rm eff}$). The green dashed line is fitted by using the value of $\rm \xi = 0.32$ for NEAs of \citet{delbo2008}. However, according to our results, it should be noteworthy that the small-sized MBAs can have low thermal inertia, thus the gap in km-sized and low thermal inertia area are filled. By fitting the results of $\rm \Gamma$ and $D_{\rm eff}$ for the main-belt asteroids from \citet{delbo2008} and \citet{hanus2018} as well as the Vesta family asteroids we have investigated in this work, then we obtain the value of $d_0$ and $\rm \xi$ to be 51.68 and 0.023, respectively. Moreover, we further explore the relationship of $\Gamma-D_{\rm eff}$ by means of the data from NEAs (green dots) and binned MBAs (blue squares), which is denoted by black dashed line with respect to $d_0$ and $\xi$ to be 344 and 0.441, respectively.  To obtain the binned data, we divide the main-belt asteroids into 12 intervals according to the size of diameter, then the average effective diameter and thermal inertia in each interval are calculated.  As shown in Fig. \ref{thdeff}, the gray dashed line is almost horizontal because of the very low value of $\rm \xi$ for MBAs when compared with that of \citet{delbo2008} for NEAs. Note that the slope of the black dashed line is a bit higher than that of the green dashed line by fitting the NEA results.

As shown in Eq. \ref{deffpv}, the effective diameter $ D_{\rm eff}$ and geometric albedo $\ p_{\rm v}$ is correlated with each other. Thus, Fig. \ref{thpv} shows thermal inertia as a function of geometric albedo for MBAs, NEAs and Vesta family asteroids (including Vestoids and non-Vestoids), represented by blue, green, red, and cyan dots, respectively. In Fig. \ref{thpv}, we further show the mean value of thermal inertia and geometric albedo for MBAs, NEAs, Vestoids and non-Vestoids by the dashed lines in the relevant colors.  From Fig. \ref{thpv}, we can see that the $p_{\rm v}$ of Vesta family members is relatively larger than that of other main-belt asteroids. Again, the average $\Gamma$ of the Vesta family asteroids here is very close to the average thermal inertia of all the MBAs in Fig. \ref{thpv}. The average $p_{\rm v}$ of the 5 Vestoids bears resemblance to that of (4) Vesta (see Fig. \ref{thpv}), which may indicate a close relationship between Vesta family asteroids and (4) Vesta.

Fig. \ref{pvdeffall} shows the profile of $D_{\rm eff}$ and $p_{\rm v}$ for 10 Vesta family asteroids (the left panel), where red and cyan circles with error bars represent our results whereas the blue dots with error bars indicate those of the literature \citep{mainzer2011, masiero2011,masiero2014}. To compare our results with the previous work, we again plot the mean value of effective diameter and geometric albedo, shown by dashed lines in different colors. As can be seen from the left panel of Fig. \ref{pvdeffall}, we observe that the $p_{\rm v}$ and $D_{\rm eff}$ of most Vesta family asteroids agree well with the earlier results \citep{mainzer2011, masiero2011,masiero2014} from NEATM model. Moreover, we show the $D_{\rm eff}$ - $p_{\rm v}$ profile for over 1000 main-belt asteroids using the data from \citet{masiero2011} (the right panel of Fig. \ref{pvdeffall}). From Fig. \ref{pvdeffall}, we infer that the geometric albedo may show a downward trend with the increasing effective diameter.

\begin{table}
        \centering
        \caption{Results of regolith grain size for 10 Vesta family asteroids with different volume filling factors.}
        \label{grainsizeresult}
        \begin{tabular}{lcc} % four columns, alignment for each
                \hline
                \specialrule{0em}{1pt}{1pt}
                Asteroid & $r_{\rm grain} (\rm mm)$&$\phi$\\
                \hline
                \specialrule{0em}{1pt}{1pt}
                (63) Ausonia  &$0.433_{-0.341}^{+0.310}$&$0.1\sim0.6$ \\
                \specialrule{0em}{1pt}{1pt}
                (556) Phyllis  &$0.082_{-0.067}^{+0.156}$&$0.1\sim0.3$\\
                \specialrule{0em}{1pt}{1pt}
                (1906) Neaf & $0.976_{-0.444}^{+0.974}$&$0.1\sim0.6$\\
                \specialrule{0em}{1pt}{1pt}
                (2511) Patterson &$1.673_{-1.314}^{+2.970}$&$0.1\sim0.6$\\
                \specialrule{0em}{1pt}{1pt}
                (3281) Maupertuis  &$0.687_{-0.577}^{+1.749}$&$0.1\sim0.6$\\
                \specialrule{0em}{1pt}{1pt}
                (5111) Jacliff  &N.A.&-\\
                \specialrule{0em}{1pt}{1pt}
                (7001) Neother &$0.052_{\rm -N.A.}^{+0.216}$&$0.1\sim0.2$\\
                \specialrule{0em}{1pt}{1pt}
            (9158) Plate &$0.006_{\rm -N.A.}^{+0.104}$&0.1\\
            \specialrule{0em}{1pt}{1pt}
            (12088) Macalintal &$0.976_{-0.973}^{+3.053}$&$0.1\sim0.6$\\
            \specialrule{0em}{1pt}{1pt}
            (15032) Alexlevin  &$0.052_{\rm -N.A.}^{+0.134}$&$0.1\sim0.2$\\
            \specialrule{0em}{1pt}{1pt}
                \hline
        \end{tabular}
        \begin{tablenotes}
    \footnotesize
    \item[]Note: The uncertainties of the grain sizes are determined by the errors of thermal inertia. N.A. appears in the lower limits of the grain sizes because for some asteroids the thermal inertias are too small to constrain the corresponding regolith grain sizes.
    \end{tablenotes}
\end{table}

\subsection{Thermal inertia and rotation period}

\citet{harris2016} developed an NEATM-based thermal inertia estimator and calculated the thermal inertia for roughly 50 asteroids provided by \citet{delbo2015}. Based on their results, \citet{harris2016} investigated the dependence of thermal inertia on asteroid rotation period and showed that for both MBAs and NEOs, $\Gamma$ has an increasing trend with the decreasing of spin rate. This is probably because for slowly rotating asteroids thermal wave penetrates much deeper into the subsurface than the fast rotators.  However, \citet{Marciniak2019} investigated 16 slow rotators with sizes ranging from $30 \sim 150$ km and found that for slowly rotating asteroid, there exists no obvious correlation between thermal inertia and rotation period. In this work, we obtain thermal inertia for 10 Vesta family members, which they may had suffered similar dynamical and thermal histories ever since their formation. Therefore, it is more reliable to investigate the relationship between spin rate and thermal inertia in the asteroid family. For 10 Vesta family asteroids, the rotation period ranges from $2.839 \sim 11.010$ hours, while the thermal inertia varies from $0 \sim 90$  $\rm J m^{-2} s^{-1/2} K^{-1} $. However, as shown in Fig. \ref{throt}, we do not find any obvious correlation between thermal inertia and rotation period, probably because of the limited population of asteroids. Thus it does not mean there exist no growing relationship between $\Gamma$ and $P_{\rm rot}$, future study with an abundant of Vesta family asteroids may reveal a clear correlation between $\Gamma$ and rotation period.

\subsection{Regolith Grain Size}

According to the method described by \citet{gundlach2013}, the thermal conductivity can be expressed by thermal inertia $\kappa = {\Gamma^2}/{(\phi \rho c)}$, where $\phi$ is the volume filling factor, $\rho$ is the bulk density and $c$ the specific heat capacity. Besides, $\kappa$ can also be regarded as a function of regolith grain size, according to the theoretical model developed for granular materials in vacuum \citep{gundlach2012}.  Based on our thermal inertia, we derive the regolith grain size of 10 Vesta family asteroids, where the results are shown in Fig. \ref{thgs}. We use various colors to denote the volume filling factors, which ranges from $0.1 \sim 0.6$ and the temperature is set to be 200 K. As can be seen, we obtain  eight values of thermal inertia for the Vesta family asteroids. Here when an asteroid has thermal inertia $\Gamma < 50$ $\rm Jm^{-2}s^{-1/2}K^{-1}$, there is no good agreement between the model and the derived thermal conductivities from thermal inertia measurements for some volume filling factors. Take (9158) Plate (with thermal inertia of 10 $\rm Jm^{-2}s^{-1/2}K^{-1}$) as an example, we simply have a regolith grain size of $0.006$ mm with a volume filling factor $\phi = 0.1$. While for the asteroids with $\Gamma<10$ $\rm Jm^{-2}s^{-1/2}K^{-1}$, the regolith grain sizes cannot be well constrained, but may be smaller than 0.006 mm.  Table. \ref{grainsizeresult} summarizes the major outcomes for the Vesta family asteroids. We also evaluate the lower and upper limits of the regolith grainsizes for these Vesta family members according to the errors of thermal inerita we obtained. As can be seen in the table, the lower limits of grain radius of (7001) Neother, (9158) Plate, (15032) Alexlevin are not constrained, because the lower limits of $\Gamma$ for these 3 asteroids are smaller than 10 $\rm Jm^{-2}s^{-1/2}K^{-1}$.

\section{Conclusions}
\label{conclu}
In conclusion, we investigate thermal properties of 10 Vesta family asteroids, including their thermal inertia, geometric albedo, effective diameter and roughness fraction. The average thermal inertia of these Vesta family asteroids is 42 $\rm Jm^{-2}s^{-1/2}K^{-1}$. For the V-type family members (Vestoids) we derive the average geometric albedos to be 0.328 while for the non-Vestoids family member, the average geometric albedo is 0.300. The mean value of $\Gamma$ is a bit larger than that of (4) Vesta \citep{delbo2015}, and the average $p_{\rm v}$ for both Vestoids and non-Vestoids are smaller than that of Vesta, but is similar to the mean value obtained by \citet{masiero2013} and \citet{mainzer2011,mainzer2012}. Moreover, we study the relationship between thermal inertia and effective diameter, as well as the relation between thermal inertia and rotation period. Considering both NEAs and MBAs, we place constraints on a new set of coefficients in the $\Gamma-D$ equation, which is slightly different from the result of \citet{delbo2015}. In addition, taking the published physical data for known Vesta family asteroids into consideration, we do not find the expected increasing trend between thermal inertia and rotation period. Moreover, since Vesta family asteroids are deemed as the fragments of the severe impact event, the wide range of geometric albedo of 10 Vesta family asteroids is in line with that of the surface of (4)Vesta. In the future work, we will address the thermophysical characteristics of more Vesta family asteroids and other families, which can help us better comprehend the formation, evolution and classification of diverse sorts of asteroid population in the main-belt region.

\noindent
\acknowledgments
This work is financially supported by the National Natural Science Foundation of China (Grant Nos. 11661161013, 11633009), CAS Interdisciplinary Innovation Team, the Strategic Priority Research Program on Space Science, the Chinese Academy of Sciences,
(Grant No. XDA15020302) and Foundation of Minor Planets of the Purple Mountain Observatory. This research has made use of the NASA/IPAC Infrared Science Archive, which is operated by the Jet Propulsion Laboratory, California Institute of Technology, under contract with the National Aeronautics and Space Administration. Research using WISE Release data is eligible for proposals to the NASA ROSES Astrophysics Data Analysis Program.

\appendix
\section{WISE observation flux}
We list the color-corrected WISE observational fluxes and uncertainties for 10 Vesta family members in this work.
\begin{table}[h]
%\small
        \renewcommand\tabcolsep{3.5pt}
        %\centering
        \caption{3-bands WISE/NEOWISE observed flux for the 10 Vesta family asteroids}
        \label{fobs}
        \begin{supertabular}{|c|ccccccc|} % four columns, alignment for each
                \hline

                Asteroid & Epoch   & $W2_{\rm obs}$  &$W2_{\rm err}$& $W3_{\rm obs}$ &$W3_{\rm err}$ & $W4_{\rm obs}$  &$W4_{\rm err}$ \\
                        & MJD & (mjy)& (mjy)&  (mjy)& (mjy)& (mjy)& (mjy)\\
                \hline
(63) Ausonia            &55276.20263    & 100.43000     &  10.04300     &12870.87000    &1287.08700     &19210.95000    &1921.09500     \\
&55276.20275    & 108.61000     &  10.86100     &15660.58000    &1566.05800     &20584.90000    &2058.49000     \\
&55276.33506    &  98.15000     &   9.81500     &10844.46000    &1084.44600     &17312.06000    &1731.20600     \\
&55276.46736    & 124.93000     &  12.49300     &14997.12000    &1499.71200     &20603.87000    &2060.38700     \\
&55276.59967    & 117.46000     &  11.74600     &13766.00000    &1376.60000     &20509.20000    &2050.92000     \\
&55276.73197    &  96.62000     &   9.66200     &11890.71000    &1189.07100     &17813.48000    &1781.34800     \\
&55276.79819    & 122.77000     &  12.27700     &17898.14000    &1789.81400     &25023.54000    &2502.35400     \\
&55276.86427    & 140.18000     &  14.01800     &17410.37000    &1741.03700     &22220.24000    &2222.02400     \\
&55276.93049    &  85.40000     &   8.54000     &10120.64000    &1012.06400     &17879.23000    &1787.92300     \\
&55276.99658    &   -   &   -   &15474.19000    &1547.41900     &22036.81000    &2203.68100     \\
&55277.06279    & 126.79000     &  12.67900     &14047.79000    &1404.77900     &20246.45000    &2024.64500     \\
&55277.12888    &  93.39000     &   9.33900     &12022.86000    &1202.28600     &19913.56000    &1991.35600     \\
&55277.19510    & 129.50000     &  12.95000     &16413.75000    &1641.37500     &22990.52000    &2299.05200     \\
&55277.32740    & 145.31000     &  14.53100     &11502.91000    &1150.29100     &17731.64000    &1773.16400     \\
&55277.45971    & 131.31000     &  13.13100     &14588.41000    &1458.84100     &20509.20000    &2050.92000     \\
&55277.59201    & 142.39000     &  14.23900     &17603.86000    &1760.38600     &24453.94000    &2445.39400     \\
&55277.72431    &   -   &   -   &11397.46000    &1139.74600     &19034.82000    &1903.48200     \\
&56982.11199    &  52.71000     &   5.27100     &   -   &   -   &   -   &   -   \\
&56982.24353    &  74.18000     &   7.41800     &   -   &   -   &   -   &   -   \\
&56982.37494    &  70.91000     &   7.09100     &   -   &   -   &   -   &   -   \\
&56982.50648    &  50.57000     &   5.05700     &   -   &   -   &   -   &   -   \\
&56982.57219    &  64.67000     &   6.46700     &   -   &   -   &   -   &   -   \\
&56982.63789    &  72.02000     &   7.20200     &   -   &   -   &   -   &   -   \\
&56982.70360    &  55.86000     &   5.58600     &   -   &   -   &   -   &   -   \\
&56982.70373    &  55.81000     &   5.58100     &   -   &   -   &   -   &   -   \\
&56982.76943    &  71.69000     &   7.16900     &   -   &   -   &   -   &   -   \\
&56982.83514    &  70.58000     &   7.05800     &   -   &   -   &   -   &   -   \\
&56982.96655    &  66.17000     &   6.61700     &   -   &   -   &   -   &   -   \\
&56983.09809    &  56.22000     &   5.62200     &   -   &   -   &   -   &   -   \\
&56983.22951    &  68.15000     &   6.81500     &   -   &   -   &   -   &   -   \\
&56983.36105    &  68.40000     &   6.84000     &   -   &   -   &   -   &   -   \\
&57291.87510    &  26.15000     &   2.61500     &   -   &   -   &   -   &   -   \\
&57292.26871    &  27.97000     &   2.79700     &   -   &   -   &   -   &   -   \\
&57446.06743    &  40.58000     &   4.05800     &   -   &   -   &   -   &   -   \\
&57446.19859    &  33.23000     &   3.32300     &   -   &   -   &   -   &   -   \\
                 \hline
        \end{supertabular}
\end{table}

\begin{table}[!h]
%\small
        \renewcommand\tabcolsep{2.8pt}
        %\centering
        \caption{\textbf{Continued}}
        \label{fobsc1}
        \begin{supertabular}{|l|ccccccc|}
                \hline

                Asteroid & Epoch &  $W2_{\rm obs}$  &$W2_{\rm err}$& $W3_{\rm obs}$ &$W3_{\rm err}$ & $W4_{\rm obs}$  &$W4_{\rm err}$ \\
                        & MJD & (mjy)& (mjy)&  (mjy)& (mjy)& (mjy)& (mjy)\\
                \hline
                &57446.46078    &  40.32000     &   4.03200     &   -   &   -   &   -   &   -   \\
        &57446.52623    &  32.35000     &   3.23500     &   -   &   -   &   -   &   -   \\
        &57446.59181    &  35.31000     &   3.53100     &   -   &   -   &   -   &   -   \\
        &57446.65739    &  34.41000     &   3.44100     &   -   &   -   &   -   &   -   \\
        &57446.78842    &  37.11000     &   3.71100     &   -   &   -   &   -   &   -   \\
        &57446.91958    &  30.81000     &   3.08100     &   -   &   -   &   -   &   -   \\
        &57922.93303    &  92.79000     &   9.27900     &   -   &   -   &   -   &   -   \\
        &57923.19496    & 120.42000     &  12.04200     &   -   &   -   &   -   &   -   \\
        &57923.32587    & 101.83000     &  10.18300     &   -   &   -   &   -   &   -   \\
        \hline
(556) Phyllis&55203.22130       &   8.22397     &   0.82240     &1012.06380     & 101.20638     &2429.67849     & 242.96785     \\
&55364.87970    &   7.11020     &   0.71102     & 873.38935     &  87.33894     &2077.53720     & 207.75372     \\
&55365.01200    &   6.37209     &   0.63721     & 802.43136     &  80.24314     &1905.23613     & 190.52361     \\
&55365.14430    &   5.92491     &   0.59249     & 809.85616     &  80.98562     &2049.03232     & 204.90323     \\
&55365.27661    &   6.37209     &   0.63721     & 780.56288     &  78.05629     &1819.48638     & 181.94864     \\
&55365.34270    &   6.25578     &   0.62558     & 896.20609     &  89.62061     &2106.43862     & 210.64386     \\
&55365.40891    &   6.80271     &   0.68027     & 783.44388     &  78.34439     &2045.26134     & 204.52613     \\
&55365.47500    &   5.67390     &   0.56739     & 703.40895     &  70.34090     &1729.61247     & 172.96125     \\
&55365.54121    &   6.15860     &   0.61586     & 784.88837     &  78.48884     &1953.20941     & 195.32094     \\
&55365.60730    &   6.38972     &   0.63897     & 800.95458     &  80.09546     &1953.20941     & 195.32094     \\
&55365.67352    &   5.61671     &   0.56167     & 719.79345     &  71.97935     &   -   &   -   \\
&55365.73961    &   6.44290     &   0.64429     & 793.61139     &  79.36114     &1898.22989     & 189.82299     \\
&55365.87191    &   6.64786     &   0.66479     & 956.77075     &  95.67708     &2335.33296     & 233.53330     \\
&55366.00422    &   6.35450     &   0.63545     & 669.89678     &  66.98968     &1561.52452     & 156.15245     \\
&55366.13652    &   6.62951     &   0.66295     & 801.69263     &  80.16926     &1865.30051     & 186.53005     \\
&55367.92256    &   6.57479     &   0.65748     & 818.85667     &  81.88567     &2039.61787     & 203.96179     \\
&55368.05486    &   5.75283     &   0.57528     & 746.80616     &  74.68062     &1868.73968     & 186.87397     \\
&55368.18716    &   5.54986     &   0.55499     & 745.43176     &  74.54318     &1987.69071     & 198.76907     \\
&55368.31947    &   5.86518     &   0.58652     & 702.11442     &  70.21144     &1794.52209     & 179.45221     \\
&55368.38556    &   6.05176     &   0.60518     & 881.47072     &  88.14707     &1976.73662     & 197.67366     \\
&55368.45177    &   6.70937     &   0.67094     & 806.87803     &  80.68780     &2007.93111     & 200.79311     \\
&55368.51786    &   5.46867     &   0.54687     & 684.86960     &  68.48696     &1654.81256     & 165.48126     \\
&55368.58407    &   5.90856     &   0.59086     & 746.11864     &  74.61186     &1851.60698     & 185.16070     \\
&55368.65016    &   6.10776     &   0.61078     & 778.40908     &  77.84091     &1898.22989     & 189.82299     \\
&55368.71625    &   5.44354     &   0.54435     & 753.71627     &  75.37163     &1814.46589     & 181.44659     \\
&55368.71638    &   5.19376     &   0.51938     & 719.79345     &  71.97935     &1938.87052     & 193.88705     \\
&55368.78247    &   6.08530     &   0.60853     & 744.05989     &  74.40599     &1905.23613     & 190.52361     \\
&55368.91477    &   5.75283     &   0.57528     & 767.72890     &  76.77289     &2002.39065     & 200.23907     \\
&55369.04708    &   5.23217     &   0.52322     & 669.28007     &  66.92801     &1651.76709     & 165.17671     \\
&55369.17938    &   6.18133     &   0.61813     & 786.33552     &  78.63355     &1886.03092     & 188.60309     \\
&56672.80351    &   8.00722     &   0.80072     &   -   &   -   &   -   &   -   \\
&56672.93531    &   6.65398     &   0.66540     &   -   &   -   &   -   &   -   \\
&56673.06698    &   8.24673     &   0.82467     &   -   &   -   &   -   &   -   \\
&56673.19877    &   7.78183     &   0.77818     &   -   &   -   &   -   &   -   \\
&56673.26461    &   7.56975     &   0.75698     &   -   &   -   &   -   &   -   \\
&56673.26473    &   7.51418     &   0.75142     &   -   &   -   &   -   &   -   \\
&56673.33057    &   7.83216     &   0.78322     &   -   &   -   &   -   &   -   \\
&56673.39640    &   8.18619     &   0.81862     &   -   &   -   &   -   &   -   \\
&56673.46236    &   6.84040     &   0.68404     &   -   &   -   &   -   &   -   \\

                 \hline
        \end{supertabular}
\end{table}

\begin{table}[!h]
%\small
        \renewcommand\tabcolsep{2.8pt}
        %\centering
        \caption{\textbf{Continued}}
        \label{fobsc1}
        \begin{supertabular}{|l|ccccccc|}
                \hline

                Asteroid & Epoch &  $W2_{\rm obs}$  &$W2_{\rm err}$& $W3_{\rm obs}$ &$W3_{\rm err}$ & $W4_{\rm obs}$  &$W4_{\rm err}$ \\
                        & MJD & (mjy)& (mjy)&  (mjy)& (mjy)& (mjy)& (mjy)\\
                \hline
&56673.52819    &   8.11114     &   0.81111     &   -   &   -   &   -   &   -   \\
&56673.59402    &   8.06644     &   0.80664     &   -   &   -   &   -   &   -   \\
&56673.65998    &   6.75899     &   0.67590     &   -   &   -   &   -   &   -   \\
&56673.79178    &   7.90463     &   0.79046     &   -   &   -   &   -   &   -   \\
&56673.92345    &   8.01460     &   0.80146     &   -   &   -   &   -   &   -   \\
&56674.05524    &   8.46217     &   0.84622     &   -   &   -   &   -   &   -   \\
&56681.03859    &   8.22397     &   0.82240     &   -   &   -   &   -   &   -   \\
&56843.60027    &   3.94350     &   0.39435     &   -   &   -   &   -   &   -   \\
&56849.12647    &   3.85373     &   0.38537     &   -   &   -   &   -   &   -   \\
&56849.25801    &   4.74987     &   0.47499     &   -   &   -   &   -   &   -   \\
&56849.38955    &   4.32794     &   0.43279     &   -   &   -   &   -   &   -   \\
&56849.52121    &   4.21388     &   0.42139     &   -   &   -   &   -   &   -   \\
&56849.65275    &   4.09904     &   0.40990     &   -   &   -   &   -   &   -   \\
&56849.71846    &   4.49864     &   0.44986     &   -   &   -   &   -   &   -   \\
&56849.71859    &   4.47798     &   0.44780     &   -   &   -   &   -   &   -   \\
&56849.78429    &   4.77179     &   0.47718     &   -   &   -   &   -   &   -   \\
&56849.85013    &   3.72119     &   0.37212     &   -   &   -   &   -   &   -   \\
&56849.91583    &   4.20612     &   0.42061     &   -   &   -   &   -   &   -   \\
&56849.98167    &   4.62042     &   0.46204     &   -   &   -   &   -   &   -   \\
&56850.11321    &   4.14841     &   0.41484     &   -   &   -   &   -   &   -   \\
&56850.24487    &   3.88939     &   0.38894     &   -   &   -   &   -   &   -   \\
&56850.37641    &   3.87509     &   0.38751     &   -   &   -   &   -   &   -   \\
&56850.50795    &   4.57385     &   0.45739     &   -   &   -   &   -   &   -   \\
&57140.64850    &   2.10030     &   0.21003     &   -   &   -   &   -   &   -   \\
&57310.66880    &   3.40943     &   0.34094     &   -   &   -   &   -   &   -   \\
&57310.73438    &   2.23605     &   0.22361     &   -   &   -   &   -   &   -   \\
&57632.64094    &   7.96309     &   0.79631     &   -   &   -   &   -   &   -   \\
&57632.77197    &  10.60429     &   1.06043     &   -   &   -   &   -   &   -   \\
&57632.96845    &   9.11762     &   0.91176     &   -   &   -   &   -   &   -   \\
        &57795.90567    &  13.25201     &   1.32520     &   -   &   -   &   -   &   -   \\
        &57796.16761    &  14.39732     &   1.43973     &   -   &   -   &   -   &   -   \\
        &57796.36409    &  10.31530     &   1.03153     &   -   &   -   &   -   &   -   \\
        &57796.42954    &  14.18671     &   1.41867     &   -   &   -   &   -   &   -   \\
        &57796.56058    &   7.37024     &   0.73702     &   -   &   -   &   -   &   -   \\
        &57796.62603    &  12.39030     &   1.23903     &   -   &   -   &   -   &   -   \\
        &57796.88796    &  14.43716     &   1.44372     &   -   &   -   &   -   &   -   \\
                \hline
(1906) Neaf& 55255.85786        &   0.42100     &    0.04210    &   54.10132    &    5.41013    &  135.75280    &   13.57528    \\
& 55255.99017   &   0.22135     &    0.02214    &   21.12554    &    2.11255    &   62.22269    &    6.22227    \\
& 55256.12247   &   0.36131     &    0.03613    &   53.06497    &    5.30650    &  144.12815    &   14.41282    \\
& 55256.25478   &   0.25298     &    0.02530    &   31.71085    &    3.17109    &   84.40124    &    8.44012    \\
& 55256.32099   &   0.45111     &    0.04511    &   57.01746    &    5.70175    &  144.39389    &   14.43939    \\
& 55256.38708   &   0.25368     &    0.02537    &   41.00227    &    4.10023    &  117.04391    &   11.70439    \\
& 55256.38721   &   0.22237     &    0.02224    &   39.95838    &    3.99584    &  105.86412    &   10.58641    \\
& 55256.45330   &   -   &    -  &   23.20641    &    2.32064    &   65.15516    &    6.51552    \\
& 55256.51951   &   0.33472     &    0.03347    &   43.09312    &    4.30931    &  111.87906    &   11.18791    \\
& 55256.58560   &   0.28098     &    0.02810    &   51.61880    &    5.16188    &  132.90729    &   13.29073    \\
& 55256.65182   &   0.23982     &    0.02398    &   29.70352    &    2.97035    &   79.35034    &    7.93503    \\

                 \hline
        \end{supertabular}
\end{table}

\begin{table}[!h]
%\small
        \renewcommand\tabcolsep{2.8pt}
        %\centering
        \caption{\textbf{Continued}}
        \label{fobsc1}
        \begin{supertabular}{|l|ccccccc|}
                \hline

                Asteroid & Epoch &  $W2_{\rm obs}$  &$W2_{\rm err}$& $W3_{\rm obs}$ &$W3_{\rm err}$ & $W4_{\rm obs}$  &$W4_{\rm err}$ \\
                        & MJD & (mjy)& (mjy)&  (mjy)& (mjy)& (mjy)& (mjy)\\
                \hline
& 55256.71790   &   0.17926     &    0.01793    &   28.97396    &    2.89740    &   85.65423    &    8.56542    \\
& 55256.71803   &   0.29153     &    0.02915    &   33.66727    &    3.36673    &   93.14270    &    9.31427    \\
& 55256.85034   &   0.26102     &    0.02610    &   37.49798    &    3.74980    &  102.69488    &   10.26949    \\
& 55256.98264   &   0.32590     &    0.03259    &   44.66917    &    4.46692    &  111.05773    &   11.10577    \\
& 55257.11495   &   0.20393     &    0.02039    &   29.08090    &    2.90809    &   66.24434    &    6.62443    \\
& 57573.15646   &   0.38751     &    0.03875    &    -  &    -  &    -  &    -  \\
& 57573.28749   &   0.31760     &    0.03176    &    -  &    -  &    -  &    -  \\
& 57573.54955   &   0.34728     &    0.03473    &    -  &    -  &    -  &    -  \\
& 57573.28749   &   0.31760     &    0.03176    &    -  &    -  &    -  &    -  \\
& 57573.54955   &   0.34728     &    0.03473    &    -  &    -  &    -  &    -  \\
& 57573.68058   &   0.76187     &    0.07619    &    -  &    -  &    -  &    -  \\
& 57573.74616   &   0.32380     &    0.03238    &    -  &    -  &    -  &    -  \\
& 57573.81161   &   0.33969     &    0.03397    &    -  &    -  &    -  &    -  \\
& 57573.87719   &   0.60295     &    0.06030    &    -  &    -  &    -  &    -  \\
& 57574.00822   &   0.32440     &    0.03244    &    -  &    -  &    -  &    -  \\
& 57574.07367   &   0.35341     &    0.03534    &    -  &    -  &    -  &    -  \\
& 57574.13925   &   0.73229     &    0.07323    &    -  &    -  &    -  &    -  \\
& 57574.40131   &   0.55345     &    0.05535    &    -  &    -  &    -  &    -  \\
& 57574.53234   &   0.40094     &    0.04009    &    -  &    -  &    -  &    -  \\
& 57745.21349   &   -   &    -  &    -  &    -  &    -  &    -  \\
& 57745.34452   &   -   &    -  &    -  &    -  &    -  &    -  \\
& 57745.47556   &   0.55961     &    0.05596    &    -  &    -  &    -  &    -  \\
& 57745.60659   &   -   &    -  &    -  &    -  &    -  &    -  \\
& 57745.67204   &   -   &    -  &    -  &    -  &    -  &    -  \\
& 57745.80307   &   -   &    -  &    -  &    -  &    -  &    -  \\
& 57745.86852   &   -   &    -  &    -  &    -  &    -  &    -  \\
& 57745.93397   &   0.65446     &    0.06545    &    -  &    -  &    -  &    -  \\
& 57745.99955   &   0.48919     &    0.04892    &    -  &    -  &    -  &    -  \\
\hline
(2511) Patterson& 55243.02056   &    -  &    -  &   23.81266    &    2.38127    &   75.43082    &    7.54308    \\
& 55243.15286   &    -  &    -  &   28.28837    &    2.82884    &   82.70825    &    8.27083    \\
& 55243.28517   &    -  &    -  &   19.46292    &    1.94629    &   58.49905    &    5.84990    \\
& 55243.41747   &    -  &    -  &   31.79859    &    3.17986    &   85.02542    &    8.50254    \\
& 55243.54990   &    -  &    -  &    -  &    -  &   49.01723    &    4.90172    \\
& 55243.61599   &    -  &    -  &   27.90024    &    2.79002    &   88.29758    &    8.82976    \\
& 55243.68221   &    -  &    -  &   32.00427    &    3.20043    &   89.03255    &    8.90325    \\
& 55243.74830   &    -  &    -  &   24.61546    &    2.46155    &   72.70241    &    7.27024    \\
& 55243.81451   &    -  &    -  &    -  &    -  &   53.15557    &    5.31556    \\
& 55243.88073   &    -  &    -  &   24.12170    &    2.41217    &   73.91776    &    7.39178    \\
& 55244.01303   &    -  &    -  &   25.02695    &    2.50270    &   78.26162    &    7.82616    \\
& 55244.14534   &    -  &    -  &   19.26672    &    1.92667    &   62.45235    &    6.24524    \\
& 55244.27764   &    -  &    -  &   28.97396    &    2.89740    &   80.23222    &    8.02322    \\
\hline
(3281) Maupertuis&55329.95339   &   0.19984     &   0.01998     &  29.29597     &   2.92960     &  75.98866     &   7.59887\\
&55329.95352    &   0.27611     &   0.02761     &  25.68076     &   2.56808     &  65.57659     &   6.55766\\
&55330.08570    &   0.22839     &   0.02284     &  21.91835     &   2.19184     &  65.93998     &   6.59400\\
&55330.08583    &   -   &   -   &  19.57077     &   1.95708     &  45.03486     &   4.50349\\
&55330.21800    &   -   &   -   &  20.45540     &   2.04554     &  48.92702     &   4.89270\\
                 \hline
        \end{supertabular}
\end{table}

\begin{table}[!h]
%\small
        \renewcommand\tabcolsep{2.8pt}
        %\centering
        \caption{\textbf{Continued}}
        \label{fobsc1}
        \begin{supertabular}{|l|ccccccc|}
                \hline

                Asteroid & Epoch &  $W2_{\rm obs}$  &$W2_{\rm err}$& $W3_{\rm obs}$ &$W3_{\rm err}$ & $W4_{\rm obs}$  &$W4_{\rm err}$ \\
                        & MJD & (mjy)& (mjy)&  (mjy)& (mjy)& (mjy)& (mjy)\\
                \hline
&55330.35031    &   0.22839     &   0.02284     &   -   &   -   &   -   &   -\\
&55330.61491    &   0.19314     &   0.01931     &  19.97137     &   1.99714     &  50.52994     &   5.05299\\
&55330.61504    &   0.20487     &   0.02049     &   -   &   -   &  51.99347     &   5.19935\\
&55330.74722    &   0.18059     &   0.01806     &  21.65748     &   2.16575     &  68.85711     &   6.88571\\
&55330.74735    &   0.23894     &   0.02389     &  19.77006     &   1.97701     &  62.22269     &   6.22227\\
&55330.81343    &   0.26223     &   0.02622     &   -   &   -   &   -   &   -\\
&55330.87952    &   0.26296     &   0.02630     &  27.24005     &   2.72401     &  75.22268     &   7.52227\\
&55330.87965    &   0.21532     &   0.02153     &  27.08993     &   2.70899     &  69.36634     &   6.93663\\
&55330.94574    &   0.28516     &   0.02852     &   -   &   -   &   -   &   -\\
&55331.07804    &   0.23264     &   0.02326     &  27.77205     &   2.77721     &  71.70490     &   7.17049\\
&55331.21034    &   0.27157     &   0.02716     &  20.79734     &   2.07973     &  55.66072     &   5.56607\\
&55331.34265    &   0.20715     &   0.02072     &  19.82476     &   1.98248     &  46.94062     &   4.69406\\
&56867.67765    &   0.26368     &   0.02637     &   -   &   -   &   -   &   -\\
&56867.80919    &   0.31672     &   0.03167     &   -   &   -   &   -   &   -\\
&56867.94073    &   0.31760     &   0.03176     &   -   &   -   &   -   &   -\\
\hline
(5111) Jacliff&55239.84411      &   0.27057     &   0.02706     &  28.89401     &   2.88940     &  77.54411     &   7.75441     \\
&55239.97654    &   0.14135     &   0.01414     &  23.83460     &   2.38346     &  61.70904     &   6.17090     \\
&55240.10885    &   0.20318     &   0.02032     &  22.71998     &   2.27200     &  58.60691     &   5.86069     \\
&55240.24115    &   0.21852     &   0.02185     &  25.96617     &   2.59662     &  79.93718     &   7.99372     \\
&55240.37346    &   -   &   -   &  28.73478     &   2.87348     &  85.26068     &   8.52607     \\
&55240.50589    &   -   &   -   &  23.05727     &   2.30573     &  53.10664     &   5.31066     \\
&55240.63819    &   0.13737     &   0.01374     &  19.78827     &   1.97883     &  53.54868     &   5.35487     \\
&55240.70428    &   0.18650     &   0.01865     &  23.68143     &   2.36814     &  64.79609     &   6.47961     \\
&55240.83671    &   0.28489     &   0.02849     &  30.20007     &   3.02001     &  68.73039     &   6.87304     \\
&55240.96902    &   0.20187     &   0.02019     &  24.84322     &   2.48432     &  68.22583     &   6.82258     \\
&55241.10132    &   -   &   -   &  21.26218     &   2.12622     &  61.53877     &   6.15388     \\
&57579.18234    &   0.35017     &   0.03502     &   -   &   -   &   -   &   -   \\
&57579.31337    &   0.24813     &   0.02481     &   -   &   -   &   -   &   -   \\
&57579.44440    &   0.37418     &   0.03742     &   -   &   -   &   -   &   -   \\
&57579.57543    &   0.43359     &   0.04336     &   -   &   -   &   -   &   -   \\
&57579.70646    &   0.37453     &   0.03745     &   -   &   -   &   -   &   -   \\
&57579.77191    &   0.41676     &   0.04168     &   -   &   -   &   -   &   -   \\
&57579.77204    &   0.32083     &   0.03208     &   -   &   -   &   -   &   -   \\
&57579.83749    &   0.29942     &   0.02994     &   -   &   -   &   -   &   -   \\
&57579.90294    &   0.31180     &   0.03118     &   -   &   -   &   -   &   -   \\
&57579.96852    &   0.37384     &   0.03738     &   -   &   -   &   -   &   -   \\
&57580.03397    &   0.30583     &   0.03058     &   -   &   -   &   -   &   -   \\
&57580.03410    &   0.35635     &   0.03564     &   -   &   -   &   -   &   -   \\
&57580.09955    &   0.37834     &   0.03783     &   -   &   -   &   -   &   -   \\
&57580.16500    &   0.38079     &   0.03808     &   -   &   -   &   -   &   -   \\
&57580.16513    &   0.34600     &   0.03460     &   -   &   -   &   -   &   -   \\
&57580.29603    &   0.42061     &   0.04206     &   -   &   -   &   -   &   -   \\
                 \hline
        \end{supertabular}
\end{table}

\begin{table}[!h]
%\small
        \renewcommand\tabcolsep{2.8pt}
        %\centering
        \caption{\textbf{Continued}}
        \label{fobsc1}
        \begin{supertabular}{|l|ccccccc|}
                \hline

                Asteroid & Epoch &  $W2_{\rm obs}$  &$W2_{\rm err}$& $W3_{\rm obs}$ &$W3_{\rm err}$ & $W4_{\rm obs}$  &$W4_{\rm err}$ \\
                        & MJD & (mjy)& (mjy)&  (mjy)& (mjy)& (mjy)& (mjy)\\
                \hline
&57580.29616    &   0.46589     &   0.04659     &   -   &   -   &   -   &   -   \\
&57580.42706    &   0.30753     &   0.03075     &   -   &   -   &   -   &   -   \\
&57580.42719    &   0.38431     &   0.03843     &   -   &   -   &   -   &   -   \\
&57580.55809    &   0.35899     &   0.03590     &   -   &   -   &   -   &   -   \\
&57749.40470    &   0.39182     &   0.03918     &   -   &   -   &   -   &   -   \\
&57749.40483    &   0.31731     &   0.03173     &   -   &   -   &   -   &   -   \\
&57749.53574    &   0.40877     &   0.04088     &   -   &   -   &   -   &   -   \\
&57749.79780    &   0.41446     &   0.04145     &   -   &   -   &   -   &   -   \\
&57749.99428    &   0.27484     &   0.02748     &   -   &   -   &   -   &   -   \\
\hline
(7001) Neother&55319.36945      &   -   &   -   &  37.25700     &   3.72570     &   -   &   -   \\
&55319.50175    &   0.26637     &   0.02664     &  24.52494     &   2.45249     &  54.09396     &   5.40940     \\
&55319.63406    &   -   &   -   &  23.14237     &   2.31424     &  56.95722     &   5.69572     \\
&55319.76636    &   -   &   -   &  37.56712     &   3.75671     &  85.65423     &   8.56542     \\
&55319.89867    &   -   &   -   &  24.01087     &   2.40109     &  55.76334     &   5.57633     \\
&55319.96488    &   -   &   -   &  34.13563     &   3.41356     &   -   &   -   \\
&55320.03097    &   -   &   -   &  22.90909     &   2.29091     &  58.87743     &   5.88774     \\
&55320.09719    &   0.25158     &   0.02516     &   -   &   -   &  52.23346     &   5.22335     \\
&55320.16328    &   -   &   -   &  35.12432     &   3.51243     &   -   &   -   \\
&55320.22949    &   0.16154     &   0.01615     &   -   &   -   &  60.30437     &   6.03044     \\
&55320.29558    &   0.23631     &   0.02363     &   -   &   -   &  54.19370     &   5.41937     \\
&55320.36180    &   -   &   -   &  33.35861     &   3.33586     &  80.23222     &   8.02322     \\
&55320.49410    &   -   &   -   &   -   &   -   &  50.02059     &   5.00206     \\
&55320.62640    &   -   &   -   &  22.76187     &   2.27619     &  62.33742     &   6.23374     \\
&55320.75871    &   -   &   -   &  32.68947     &   3.26895     &   -   &   -   \\
&57631.72614    &   0.53590     &   0.05359     &   -   &   -   &   -   &   -   \\
&57632.11923    &   0.68404     &   0.06840     &   -   &   -   &   -   &   -   \\
&57632.31571    &   0.65687     &   0.06569     &   -   &   -   &   -   &   -   \\
&57632.51220    &   0.58275     &   0.05828     &   -   &   -   &   -   &   -   \\
&57632.51232    &   0.59031     &   0.05903     &   -   &   -   &   -   &   -   \\
&57632.70881    &   0.62443     &   0.06244     &   -   &   -   &   -   &   -   \\
&57632.97087    &   0.50336     &   0.05034     &   -   &   -   &   -   &   -   \\
&57633.36383    &   0.54788     &   0.05479     &   -   &   -   &   -   &   -   \\
&57633.36396    &   0.47806     &   0.04781     &   -   &   -   &   -   &   -   \\
&57801.40529    &   0.63370     &   0.06337     &   -   &   -   &   -   &   -   \\
&57801.66723    &   0.55601     &   0.05560     &   -   &   -   &   -   &   -   \\
&57801.73281    &   0.75698     &   0.07570     &   -   &   -   &   -   &   -   \\
&57801.79826    &   0.71430     &   0.07143     &   -   &   -   &   -   &   -   \\
&57801.99474    &   0.59687     &   0.05969     &   -   &   -   &   -   &   -   \\
\hline
(9158) Plate&55218.40459        &   -   &   -   &   -   &   -   &  34.35175     &   3.43518     \\
&55218.53690    &   0.17341     &   0.01734     &   -   &   -   &  25.39503     &   2.53950     \\
&55218.66920    &   0.18875     &   0.01888     &  14.42806     &   1.44281     &  34.47854     &   3.44785     \\
&55218.80164    &   0.19910     &   0.01991     &  17.53912     &   1.75391     &  46.29657     &   4.62966     \\
&55218.93394    &   -   &   -   &  15.19175     &   1.51918     &  46.33923     &   4.63392     \\
&55219.00003    &   0.14437     &   0.01444     &  16.85807     &   1.68581     &  39.00756     &   3.90076     \\
&55219.00016    &   -   &   -   &  17.23487     &   1.72349     &  44.86925     &   4.48693     \\
&55219.06624    &   -   &   -   &   -   &   -   &  26.83792     &   2.68379     \\
&55219.13246    &   0.19837     &   0.01984     &  17.89814     &   1.78981     &  44.99340     &   4.49934     \\
&55219.26476    &   0.19457     &   0.01946     &   -   &   -   &  45.28442     &   4.52844     \\
&55219.39707    &   -   &   -   &   -   &   -   &  27.94788     &   2.79479     \\
                 \hline
        \end{supertabular}
\end{table}

\begin{table}[!h]
%\small
        \renewcommand\tabcolsep{2.8pt}
        %\centering
        \caption{\textbf{Continued}}
        \label{fobsc1}
        \begin{supertabular}{|l|ccccccc|}
                \hline

                Asteroid & Epoch &  $W2_{\rm obs}$  &$W2_{\rm err}$& $W3_{\rm obs}$ &$W3_{\rm err}$ & $W4_{\rm obs}$  &$W4_{\rm err}$ \\
                        & MJD & (mjy)& (mjy)&  (mjy)& (mjy)& (mjy)& (mjy)\\
                \hline
&55221.05132    &   0.18962     &   0.01896     &  16.04013     &   1.60401     &   -   &   -   \\
&55221.18362    &   -   &   -   &  15.34645     &   1.53465     &  35.54257     &   3.55426     \\
&55221.18375    &   0.15815     &   0.01582     &  17.06113     &   1.70611     &  36.50470     &   3.65047     \\
&55221.31605    &   0.18009     &   0.01801     &   -   &   -   &   -   &   -   \\
&55221.44836    &   0.18771     &   0.01877     &   -   &   -   &  36.77467     &   3.67747     \\
&55221.51457    &   0.14252     &   0.01425     &  15.57428     &   1.55743     &  23.43953     &   2.34395     \\
&55221.58066    &   -   &   -   &  16.30827     &   1.63083     &  39.95295     &   3.99530     \\
&55221.64688    &   0.17696     &   0.01770     &   -   &   -   &   -   &   -   \\
&55221.71309    &   -   &   -   &  16.82705     &   1.68271     &  46.76800     &   4.67680     \\
&55221.77918    &   0.10315     &   0.01032     &  12.87087     &   1.28709     &  31.94102     &   3.19410     \\
&55221.91149    &   -   &   -   &  14.54815     &   1.45481     &  37.11494     &   3.71149     \\
&55221.91161    &   0.17631     &   0.01763     &  16.61146     &   1.66115     &  42.10647     &   4.21065     \\
&55222.04392    &   0.17040     &   0.01704     &  14.85963     &   1.48596     &  44.45789     &   4.44579     \\
&55222.17622    &   0.18599     &   0.01860     &   -   &   -   &  37.52743     &   3.75274     \\
\hline
(12088) Macalintal&55375.95934  &   -   &   -   &   3.85130     &   0.38513     &   8.77301     &   0.87730     \\
&55376.09165    &   -   &   -   &   4.04025     &   0.40403     &  12.05449     &   1.20545     \\
&55376.22395    &   -   &   -   &   4.26587     &   0.42659     &   -   &   -   \\
&55376.35625    &   -   &   -   &   -   &   -   &  11.81269     &   1.18127     \\
&55376.42234    &   -   &   -   &   5.58738     &   0.55874     &   -   &   -   \\
&55376.55464    &   -   &   -   &   5.25301     &   0.52530     &  13.56278     &   1.35628     \\
&55376.68695    &   -   &   -   &   5.57709     &   0.55771     &   -   &   -   \\
&55376.95155    &   -   &   -   &   5.46522     &   0.54652     &  12.76288     &   1.27629     \\
&55377.08386    &   -   &   -   &   3.49629     &   0.34963     &   -   &   -   \\
&55377.21616    &   -   &   -   &   4.29346     &   0.42935     &   -   &   -   \\
&55377.34847    &   -   &   -   &   5.06300     &   0.50630     &   -   &   -   \\
\hline
(15032) Alexlevin&55243.02082   &   0.10982     &   0.01098     &  12.46258     &   1.24626     &  32.77546     &   3.27755     \\
&55243.15312    &   0.10871     &   0.01087     &   -   &   -   &   -   &   -   \\
&55243.28555    &   0.13762     &   0.01376     &  11.25143     &   1.12514     &  33.35407     &   3.33541     \\
&55243.41785    &   0.12311     &   0.01231     &   9.44513     &   0.94451     &  25.93874     &   2.59387     \\
&55243.55016    &   0.16349     &   0.01635     &   8.12097     &   0.81210     &  20.47146     &   2.04715     \\
&55243.61637    &   0.12656     &   0.01266     &   -   &   -   &  14.88500     &   1.48850     \\
&55243.68246    &   0.08354     &   0.00835     &  10.96498     &   1.09650     &  25.20860     &   2.52086     \\
&55243.74868    &   0.14611     &   0.01461     &  11.59866     &   1.15987     &  24.77130     &   2.47713     \\
&55243.81477    &   0.08556     &   0.00856     &   -   &   -   &  15.38682     &   1.53868     \\
&55243.88098    &   0.15699     &   0.01570     &   7.50253     &   0.75025     &  26.13057     &   2.61306     \\
&55243.94720    &   0.15342     &   0.01534     &  13.39084     &   1.33908     &   -   &   -   \\
&55244.01329    &   0.13979     &   0.01398     &   9.68295     &   0.96830     &  24.61211     &   2.46121     \\
&55244.14559    &   0.07668     &   0.00767     &   9.12027     &   0.91203     &  27.21127     &   2.72113     \\
&55244.14572    &   0.12333     &   0.01233     &  11.22039     &   1.12204     &  30.08492     &   3.00849     \\
&55244.27802    &   0.09933     &   0.00993     &   -   &   -   &  14.49268     &   1.44927     \\
&55244.41033    &   0.09851     &   0.00985     &  12.23511     &   1.22351     &  32.80566     &   3.28057     \\
\hline
\end{supertabular}
\end{table}

%%%%%%
%%%%%%
%% This command is needed to show the entire author+affilation list when
%% the collaboration and author truncation commands are used.  It has to
%% go at the end of the manuscript.
%\allauthors
%
%% Include this line if you are using the \added, \replaced, \deleted
%% commands to see a summary list of all changes at the end of the article.
%\listofchanges


\newpage
\begin{thebibliography}{}

\bibitem[Al{\'\i}-Lagoa et al.(2018)]{alilagoa2018_akari}
Al{\'\i}-Lagoa, V., M{\"u}ller, T.~G., Usui, F., et al.\ 2018, A\&A, 612, A85

\bibitem[Binzel \& Xu(1993)]{binzel1993}
Binzel, R.~P., \& Xu, S.\ 1993, Science, 260, 186

\bibitem[\protect\citeauthoryear{Bottke et al.}{2005}]{bottke2005}
Bottke, W.~F., Durda, D.~D., Nesvorn{\'y}, D., et al.\ 2005, Icarus, 175, 111

\bibitem[\protect\citeauthoryear{Burbine et al.}{2001}]{burb2001}
Burbine, T. H., Buchanan, P. C., Binzel, R. P., et al. 2001, M\&PS, 36, 761

\bibitem[Bus \& Binzel(2002)]{bus_binzel2002}
Bus, S.~J., \& Binzel, R.~P.\ 2002, Icarus, 158, 146

\bibitem[\protect\citeauthoryear{Carruba et al.}{2005}]{carr2005}
Carruba, V., Michtchenko, T. A., Roig, F., et al. 2005, A\&A, 441,819

\bibitem[Carvano et al.(2010)]{Carvano2010}
Carvano, J.~M., Hasselmann, P.~H., Lazzaro, D., et al.\ 2010, A\&A, 510, A43

\bibitem[Cruikshank et al.(1991)]{cruikshank1991}
Cruikshank, D.~P., Tholen, D.~J., Hartmann, W.~K., et al.\ 1991, Icarus, 89, 1


\bibitem[\protect\citeauthoryear{Delbo  et  al.}{2015}]{delbo2015}
Delbo, M., Mueller, M., Emery, J. P., et al. 2015, Asteroid IV, Univ. Arizona Press, Tucson, p. 107

\bibitem[\protect\citeauthoryear{Delbo et al.}{2007}]{delbo2007}
Delbo, M., dell'Oro, A., Harris, A. W., et al. 2007, Icarus, 190, 236

\bibitem[\protect\citeauthoryear{Delbo \& Tanga}{2008}]{delbo2008}
Delbo, M., Tanga, P., 2009, P\&SS, 57, 259

\bibitem[DeMeo et al.(2009)]{Demeo2009}
DeMeo, F.~E., Binzel, R.~P., Slivan, S.~M., et al.\ 2009, Icarus, 202, 160

%%\bibitem[\protect\citeauthoryear{De Sanctis et al.}{2011}]{sanc2011}
%%De Sanctis, M. C., Migliorini, A., Luzia, J. F., et al.  2011, A\&A, 533, %%A77

\bibitem[\protect\citeauthoryear{De Sanctis et al.}{2012}]{sanctis2012}
De Sanctis, M. C., et al. 2012, Sci, 336, 697

\bibitem[\protect\citeauthoryear{\v{D}urech et al.}{2018}]{durech2018}
\v Durech, J., Hanu\v s, J., Al\'i Lagoa, V., 2018, A\&A, 617, A57

\bibitem[\protect\citeauthoryear{\v{D}urech et al.}{2016}]{durech2016}
\v Durech, J., Hanu\v s, J., Oszkiewicz, D., Van\v co, R. 2016, A\&A, 587A, 48

\bibitem[\protect\citeauthoryear{\v{D}urech et al.}{2017}]{durech2017}
\v Durech, J., Hanu\v s, J., Ali-Lagoa, V. 2017, DPS, 4911027

%%\bibitem[\protect\citeauthoryear{Erik}{1997}]{erik1997}
%%Erik, A. 1997, M\&PS, 32, 965

\bibitem[\protect\citeauthoryear{Fulvio et al.}{2012}]{ful2012}
Fulvio D., Brunetto R., Vernazza P., Strazzulla G., 2012, A\&A, 537, L11

\bibitem[\protect\citeauthoryear{Grav et al.}{2012}]{grav2012}
Grav, T., et al., 2012, ApJ, 744, 197

\bibitem[\protect\citeauthoryear{Gundlach et al.}{2012}]{gundlach2012}
Gundlach, B., Bulm, J. 2012, Icarus, 219, 618

\bibitem[\protect\citeauthoryear{Gundlach et al.}{2013}]{gundlach2013}
Gundlach, B., Bulm, J. 2013, Icarus, 223, 479

\bibitem[\protect\citeauthoryear{Hanu\v{s} et al.}{2016}]{hanus2016}
Hanu\v s, J., et al. 2016, A\&A, 592A, 34

\bibitem[\protect\citeauthoryear{Hanu\v{s} et al.}{2018}]{hanus2018}
Hanu\v s, J., Delbo, M., \v Durech, J., Al\'i-Lagoa, V. 2018, Icarus, 309, 297

\bibitem[\protect\citeauthoryear{Hardersen et al.}{2014}]{harder2014}
Hardersen, P. S., Reddy, V., Roberts, R., Mainzer, A., 2014, Icarus, 242, 269

\bibitem[\protect\citeauthoryear{Harris \& Drube}{2016}]{harris2016}
Harris, A. W., Drube, L., 2016, ApJ, 832, 127

\bibitem[\protect\citeauthoryear{Hasegawa et al.}{2014a}]{hasegawa2014}
Hasegawa, S., et al., 2014a, PASJ, 66, 54

\bibitem[\protect\citeauthoryear{Hasegawa et al.}{2014b}]{hasegawa2014b}
Hasegawa, S., Miyasaka, S., Tokimasa, N., et al.\ 2014b, PASJ, 66, 89

\bibitem[\protect\citeauthoryear{Jiang et al.}{2019}]{jiang2019}
Jiang, H. X., Yu, L. L., Ji, J. H.,  2019, AJ, 158, 205

\bibitem[\protect\citeauthoryear{Kaasalainen \& Torppa}{2001}]{kaa2001}
Kaasalainen, M., Torppa, J., 2001, Icarus, 153,24

\bibitem[\protect\citeauthoryear{Lagerros}{1996}]{lag1996}
Lagerros, J. S. V., 1996, A\&A, 310, 1011

\bibitem[\protect\citeauthoryear{Licandro et al.}{2017}]{lic2017}
Licandro, J., Popesu, M., Morate, D., de Le\'on J., 2017, A\&A, 600, A126

\bibitem[Mainzer et al.(2011)]{mainzer2011}
Mainzer, A., Grav, T., Masiero, J., et al.\ 2011, ApJ, 741, 90

\bibitem[Mainzer et al.(2012)]{mainzer2012}
Mainzer, A., Masiero, J., Grav, T., et al.\ 2012, ApJ, 745, 7

\bibitem[Mandler \& Elkins-Tanton(2013)]{mandler2013}
Mandler, B.~E., \& Elkins-Tanton, L.~T.\ 2013, Meteoritics and Planetary Science, 48, 2333

\bibitem[\protect\citeauthoryear{Marchi et al.}{2012}]{Marchi2012}
Marchi, S., et al, 2012, Sci, 336, 690

\bibitem[\protect\citeauthoryear{Marciniak et al.}{2007}]{Marciniak2007}
Marciniak, A., et al, 2007, A\&A 473, 633

\bibitem[\protect\citeauthoryear{Marzari et al.}{1996}]{marz1996}
Marzari, F., Cellino, A., Davis, D. R., et al. 1996, A\&A, 316, 248

\bibitem[\protect\citeauthoryear{Marzari et al.}{1995}]{marz1995}
Marzari, F., Davis, D. R.,  Vanzani, V., 1995, Icarus, 113, 168

\bibitem[\protect\citeauthoryear{Marciniak et al.}{2019}]{Marciniak2019}
Marciniak, A., et al., 2019, A\&A, 625, 139

\bibitem[\protect\citeauthoryear{Masiero et al.}{2011}]{masiero2011}
Masiero, J. R., et al., 2011, ApJ, 741, 68

\bibitem[\protect\citeauthoryear{Masiero et al.}{2012}]{masiero2012}
Masiero, J. R., Mainzer, A. K., Grav, T., et al. 2012, ApJ, 759, 8

\bibitem[\protect\citeauthoryear{Masiero et al.}{2013}]{masiero2013}
Masiero, J. R., Mainzer, A. K., Bauer, J. M., et al. 2013, ApJ, 770, 7

\bibitem[\protect\citeauthoryear{Masiero et al.}{2014}]{masiero2014}
Masiero, J. R., Grav, T., Mainzer, A. K., et al. 2014, ApJ, 791, 121

%%\bibitem[\protect\citeauthoryear{Michel et al.}{2015}]{michel2015}
%%Michel, P., Richardson, D. C., Durda, D. D., et al. 2015, Asteroid %%\uppercase\expandafter{\romannumeral4}, Univ. Arizona Press, Tucson, p.341

\bibitem[\protect\citeauthoryear{Migliorini et al.}{1997}]{mig1997}
Migliorini, F., Morbidelli, A., Zappal\`a, V., et al. 1997, M\&PS, 32, 903

\bibitem[Myhrvold(2018)]{myhrovold2018}
Myhrvold, N.\ 2018, Icarus, 303, 91

\bibitem[Moskovitz et al.(2010)]{Moskovitz2010}
Moskovitz, N.~A., Willman, M., Burbine, T.~H., et al.\ 2010, Icarus, 208, 773

\bibitem[\protect\citeauthoryear{Nesvorn\'y et al.}{2008}]{nesv2008}
Nesvorn\'y, D.,Roig, F., Gladman, et al. 2008, Icarus, 193, 85

\bibitem[\protect\citeauthoryear{Nesvorn\'y et al.}{2015}]{nesv2015}
Nesvorn\'y, D., Bro\v z, M., Carruba, V., 2015, \uppercase\expandafter{\romannumeral4}, Univ. Arizona Press, Tucson,  p.297

\bibitem[\protect\citeauthoryear{Press et al.}{2007}]{press2007}
Press, W. H., et al. 2007, Numerical Recipes, Cambridge University Press, New York, p. 815

\bibitem[\protect\citeauthoryear{Reddy et al.}{2012}]{reddy2012}
Reddy, V., Nathues, A., Le Corre L., et al. 2012, Science, 336, 700

\bibitem[\protect\citeauthoryear{Reddy et al.}{2013}]{reddy2013}
Reddy, V., Li, J. Y., Le corre L., et al. 2013, Icarus, 226, 1103

\bibitem[\protect\citeauthoryear{Rozitis \& Green}{2011}]{roz2011}
Rozitis, B., Green, S. F. 2011, MNRAS, 415, 2042

\bibitem[\protect\citeauthoryear{Russell et al.}{2012}]{rus2012}
Russell, C. T., et al. 2012, Science, 336, 684

\bibitem[\protect\citeauthoryear{Spencer et al.}{1989}]{spe1989}
Spencer, J. R., Lebofsky, L. A., Sykes, M. V., 1989, Icarus, 78, 337

\bibitem[\protect\citeauthoryear{Schenk et al.}{2012}]{schenk2012}
Schenk, P., et al., 2012, Science, 336, 694

\bibitem[\protect\citeauthoryear{Tanga et al.}{2003}]{tanga2003}
Tanga, P., Hestroffer, D., Cellino, A., et al. 2003, A\&A, 401, 733

\bibitem[\protect\citeauthoryear{Tedesco et al.}{2004}]{tedesco2004}
Tedesco, E. F., Noah, P. V., Noah, M., et al. 2004, PDSS, 12

\bibitem[Tholen(1984)]{tholen1984}
Tholen, D.~J.\ 1984, Ph.D. Thesis

\bibitem[\protect\citeauthoryear{Thomas et al.}{1997a}]{tho1997}
Thomas, P., Binzel, R. P., Gaffey, M. J., et al. 1997, Icarus, 128, 88

\bibitem[Usui et al.(2011)]{usui2011_akari}
Usui, F., Kuroda, D., M{\"u}ller, T.~G., et al.\ 2011, PASJ, 63, 1117

\bibitem[\protect\citeauthoryear{Warner et al.}{2009}]{warner2009}
Warner, B. D., Harris, A. W., Pravec, P. 2009, Icarus, 202, 134

\bibitem[\protect\citeauthoryear{Wright et al.}{2010}]{wright2010}
Wright, E. L. et al. 2010, AJ, 140, 1868

\bibitem[Xu et al.(1995)]{xu1995_icarus} Xu, S., Binzel, R.~P., Burbine, T.~H., et al.\ 1995, Icarus, 115, 1

\bibitem[\protect\citeauthoryear{Yu et al.}{2017}]{yu2017}
Yu, L. L., Yang, B., Ji, J. H., Ip, W.H. 2017, MNRAS, 472, 2388

\bibitem[\protect\citeauthoryear{Zappal\`a et al.}{1983}]{zapp1983}
Zappal\`a, V., Scaltriti, F., Di Martino, M. 1983, Icarus, 56, 325

\bibitem[\protect\citeauthoryear{Zappal\`a et al.}{1990}]{zapp1990}
Zappal\`a, V., Cellino, A., Farinella, P., Knezevic, Z., 1990, AJ, 100, 2030

\bibitem[\protect\citeauthoryear{Zappal\`a et al.}{1995}]{zapp1995}
Zappal\`a, V., Bendjoya, P., Cellino, A., et al. 1995, Icarus, 116, 291

%%\bibitem[\protect\citeauthoryear{Zellner et al.}{1995}]{zellner1995}
%%Zellner, B., Storrs, A., Wells, E. N., 1995, Lunar Planet Sci, XXVI, 1553

\end{thebibliography}
\end{document}